\documentclass{article}
\usepackage[arxiv]{icml2025}

\makeatletter
\@ifpackageloaded{natbib}{
  \AtBeginDocument{\setcitestyle{numbers,square,sort&compress}}
}{}
\makeatother

\usepackage{hyperref}

\usepackage{mathtools}
\usepackage{amsthm}
\usepackage{xspace}
\usepackage[capitalize,noabbrev]{cleveref}
\usepackage[utf8]{inputenc} %
\usepackage[T1]{fontenc}    %
\usepackage{url}            %
\usepackage{booktabs}       %
\usepackage{amsmath, amssymb, amsfonts}       %
\usepackage{xcolor}
\usepackage{nicefrac}       %
\usepackage{microtype}      %
\usepackage{fancyhdr}       %
\usepackage{graphicx}       %
\usepackage{siunitx}
\graphicspath{{media/}}     %
\usepackage{enumitem}
\usepackage{tikz}
\usetikzlibrary{positioning, calc, shapes, arrows.meta, backgrounds, spy}
\usepackage{subcaption}
\usepackage{mwe}
\usepackage{booktabs} %
\usepackage[most]{tcolorbox}
\usepackage{varwidth}
\usepackage{adjustbox}

\newcommand{\R}{\mathbb{R}}

\newcommand{\E}{\mathop{\mathbb{E}}}
\newcommand{\identity}{\mathbb{I}}
\newcommand{\loss}{\mathcal{L}}
\newcommand{\lossSDS}{\loss_{\mathrm{SDS}}}
\newcommand{\optParams}{\boldsymbol{\theta}}
\newcommand{\optParamsDomain}{\Theta}
\newcommand{\renderFunc}{\mathbf{g}}
\newcommand{\render}{\mathbf{x}}
\newcommand{\renderDomain}{\mathcal{X}}
\newcommand{\noise}{\boldsymbol{\epsilon}}
\newcommand{\timestep}{{t'}}
\newcommand{\timestepMin}{\timestep_{\!\!\textnormal{min}}}
\newcommand{\timestepMax}{\timestep_{\!\!\textnormal{max}}}
\newcommand{\signalScale}{\alpha}
\newcommand{\noiseScale}{\sigma}
\newcommand{\noisedRender}{\mathbf{z}}
\newcommand{\diffusionParams}{\boldsymbol{\phi}}
\newcommand{\noisePrediction}{\noise_{\diffusionParams}}
\newcommand{\prompt}{\boldsymbol{p}} %
\newcommand{\camera}{\boldsymbol{c}}
\newcommand{\cameraDomain}{\mathcal{C}}
\newcommand{\timestepWeight}{\omega(\timestep)}
\newcommand{\audioSampleIndex}{t}
\newcommand{\totalAudioSamples}{T}
\newcommand{\audioRenderFunc}{\renderFunc_{{\textnormal{audio}}}}
\newcommand{\audioRender}{\render_{{\textnormal{audio}}}}
\newcommand{\audioRenderDomain}{\R^{2 \times \totalAudioSamples}}
\newcommand{\frequency}{\lambda}
\newcommand{\decay}{d}
\newcommand{\amplitude}{a}
\newcommand{\impulse}{\mathbf{I}}
\newcommand{\impulseObj}{\impulse_{\text{obj}}}
\newcommand{\impulseReverb}{\impulse_{\text{reverb}}}
\newcommand{\impulseImpact}{\impulse_{\text{impact}}}
\newcommand{\numImpulseComponent}{N}
\newcommand{\impulseComponentIndex}{n}
\newcommand{\bandpassFunc}{F}
\newcommand{\targetAudio}{\mathbf{m}}
\newcommand{\sourceIndex}{k}
\newcommand{\totalNumSources}{K}
\newcommand{\encodeFunc}{\textnormal{enc}}
\newcommand{\decodeFunc}{\textnormal{dec}}
\newcommand{\latentRender}{\mathbf{h}}
\newcommand{\denoisedNoisedLatent}{\hat{\noisedRender}}
\newcommand{\noisedDenoisedRender}{\hat{\render}}
\newcommand{\stableAudioLatentDomain}{\R^{64 \times \nicefrac{T}{2048}}}
\newcommand{\lossRecons}{\loss_{\textnormal{rec}}}

\newcommand{\STFTIndex}{m}
\newcommand{\totalNumSTFT}{M}
\newcommand{\STFT}{\mathcal{S}}
\newcommand{\lossScaleSourceSep}{\gamma}
\newcommand{\guidanceScale}{\tau}
\newcommand{\guidanceNoisePrediction}{\hat{\noise}_{\diffusionParams}}

\newcommand{\update}{\mathbf{u}}
\newcommand{\specRender}{\mathbf{s}}
\newcommand{\noisedDenoisedSpecRender}{\hat{\specRender}}
\newcommand{\updateSDS}{\update_{\textnormal{SDS}}}
\newcommand{\figref}[1]{Fig.~\ref{#1}}
\newcommand{\secref}[1]{Sec.~\ref{#1}}
\newcommand{\fmMatrix}{\mathbf{A}}
\newcommand{\fmMatrixSize}{V}
\newcommand{\fmMatrixIndex}{v}
\newcommand{\fmState}{\mathbf{u}}
\newcommand{\fmFrequency}{\omega}
\newcommand{\fmAttackDecayA}{\alpha}
\newcommand{\fmAttackDecayB}{\delta}
\newcommand{\fmAttackDecayFunc}{f}
\newcommand{\updateSourceSep}{\update_{\textnormal{Sep}}}

\DeclareMathOperator*{\argmin}{arg\,min}

\definecolor{nvidiagreen}{RGB}{118, 185, 0}
\definecolor{darkgreen}{rgb}{0,0.8,0}
\definecolor{param}{HTML}{61988E}
\definecolor{inwave}{HTML}{28587B}
\definecolor{upwave}{HTML}{3A0842}

\theoremstyle{plain}

\theoremstyle{definition}

\theoremstyle{remark}

\usepackage[textsize=tiny]{todonotes}

\icmltitlerunning{Audio-SDS: Source Separation, Synthesis, and Beyond}

\begin{document}

\twocolumn[
\icmltitle{Score Distillation Sampling for Audio: \\
           Source Separation, Synthesis, and Beyond}

\icmlsetsymbol{equal}{*}

\vspace{-0.015\textheight}
\begin{icmlauthorlist}
\icmlauthor{\hspace{0.15\textwidth}}{}
\icmlauthor{Jessie Richter-Powell$^{\textnormal{1 2}}$}{}%
\icmlauthor{Antonio Torralba$^{\textnormal{1 2}}$}{}
\icmlauthor{Jonathan Lorraine$^{\textnormal{1}}$}{}
\icmlauthor{\hspace{0.15\textwidth}}{}
\icmlauthor{\hspace{0.35\textwidth}}{}
\icmlauthor{$^{\textnormal{1}}$NVIDIA \hspace{0.03\textwidth} $^{\textnormal{2}}$MIT}{}
\icmlauthor{\hspace{0.35\textwidth}}{}
\icmlauthor{Project Website: \href{https://research.nvidia.com/labs/toronto-ai/Audio-SDS/}{{\color{nvidiagreen}research.nvidia.com/labs/toronto-ai/Audio-SDS/}}}{}
\end{icmlauthorlist}

\icmlkeywords{Machine Learning, ICML, Audio}

\vskip 0.15in
]

\begin{abstract}

    We introduce Audio-SDS, a generalization of Score Distillation Sampling (SDS) to text-conditioned audio diffusion models. While SDS was initially designed for text-to-3D generation using image diffusion, its core idea of distilling a powerful generative prior into a separate parametric representation extends to the audio domain. Leveraging a single pretrained model, Audio-SDS enables a broad range of tasks without requiring specialized datasets. In particular, we demonstrate how Audio-SDS can guide physically informed impact sound simulations, calibrate FM-synthesis parameters, and perform prompt-specified source separation. Our findings illustrate the versatility of distillation-based methods across modalities and establish a robust foundation for future work using generative priors in audio tasks.

\end{abstract}

\vspace{-0.03\textheight}
\section{Introduction}\label{sec:introduction}
\vspace{-0.005\textheight}
    Sound is a fundamental medium for interacting with the world and the people within it.
    Audio data, however, poses distinct challenges: it is naturally high-frequency -- requiring dense sampling, exhibits long-range temporal structure, 
    and is often noisy or constrained by copyright and licensing concerns.
    Consequently, image diffusion model applications have matured rapidly, while audio has lagged behind. 
    However, applications in music production, interactive VR/AR systems, sound effect design, and source separation demand robust audio learning methods that can adapt to novel tasks and constraints. 
    Leveraging pretrained audio diffusion models \citep{liu2023audioldm, evans2024stable} presents a promising way to bypass the need for datasets curated to each individual task, thereby expanding the range of machine learning to new audio use cases.
    
    Diffusion-based audio models have demonstrated remarkable fidelity in speech synthesis, music, and foley sound generation \citep{kongdiffwave, engel2020ddsp}. However, they focus on sampling in their modality (usually given some conditioning input), providing limited support for explicit parameter optimization.
    Tasks such as physically informed impact sound generation or prompt-driven source separation require a mixture of structural constraints and interpretability beyond what standard sampling pipelines can offer. At the same time, Score Distillation Sampling (SDS) has enabled powerful capabilities in text-to-3D pipelines \citep{pooledreamfusion}, but its potential across other domains has yet to be realized. 
    A direct adaptation of SDS to audio is an attractive solution, allowing the optimization of parametric audio representations using the learned distribution of a pretrained audio diffusion model.
    
    This paper introduces \textbf{Audio-SDS}, extending SDS to large-scale text-to-audio diffusion. At each optimization step, we encode a rendered audio signal (e.g., from a synthesizer or physical simulation), add noise, and then have the diffusion model predict the noise. By backpropagating the difference between predicted and actual noise, we nudge the underlying audio parameters to align slightly more with the text-conditioned distribution of the pretrained model. This framework unifies data-driven priors with explicit parametrizations (oscillators, impact simulations, or multi-source mixtures), enabling tasks that previously required separate, specialized pipelines. Empirically, we show that Audio-SDS produces perceptually convincing and semantically faithful results while maintaining controllability through physics-based or user-defined parameters.
    
\begin{figure*}[!htbp]
  \centering

  \begin{subfigure}{.65\textwidth}
    \centering
    \vspace{-0.05\textheight}
    \caption{Audio-SDS Framework: By reusing one large pretrained audio diffusion model, our method improves performance on diverse audio tasks.}
    \vspace{0.01\textheight}
    \includegraphics[width=\linewidth]{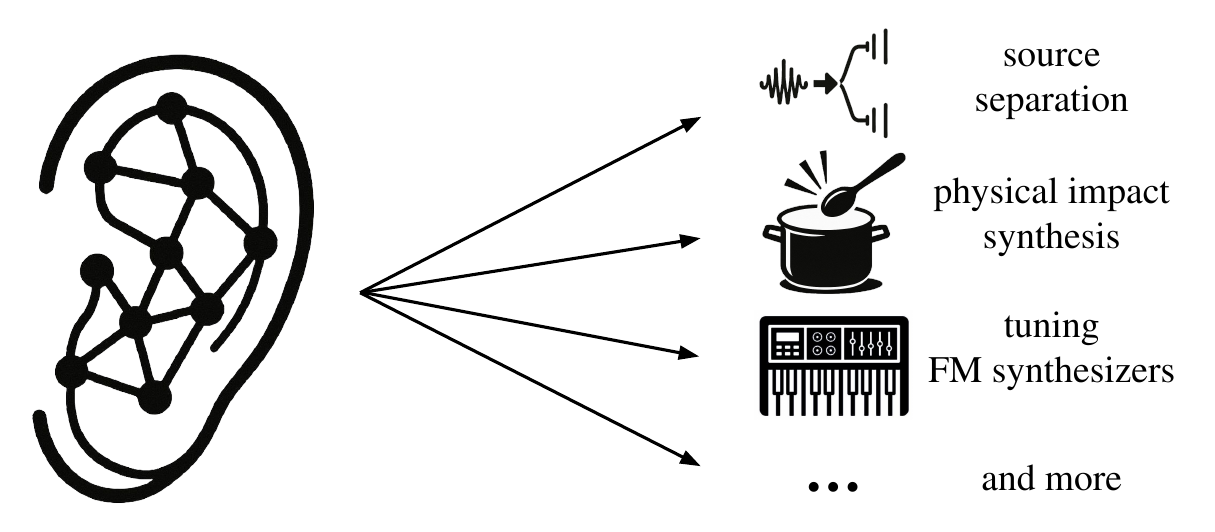}
    \label{fig:left}
  \end{subfigure}
  
  \begin{subfigure}{.99\textwidth}
    \centering
    \vspace{0.0\textheight}
    \caption{Audio-SDS Update}
    \vspace{-0.00\textheight}
    \begin{adjustbox}{scale=0.75}
    \begin{tikzpicture}[node distance=1cm, every node/.style={align=center}]
        \tikzstyle{input} = [rectangle, rounded corners, minimum width=1cm, minimum height=1cm, text centered, draw=black, dashed]
        \tikzstyle{process} = [rectangle, rounded corners, minimum width=1cm, minimum height=1cm, text centered, draw=black]
        \tikzstyle{arrow} = [thick,->,>=stealth]
    
        \node (audio_sim) [input] {{\color{cyan}optimizable}\\{\color{cyan}audio parameters $\optParams$}};

        \node (prompt) [process, below of=audio_sim, yshift=-1.3cm] {{\color{darkgreen}user-provided}\\{\color{darkgreen}text-prompt}};
        
        \node (render) [right of=audio_sim, xshift=5.5cm] {\includegraphics[trim={2.5cm 1.85cm 4.7cm .95cm}, clip, width=2cm]{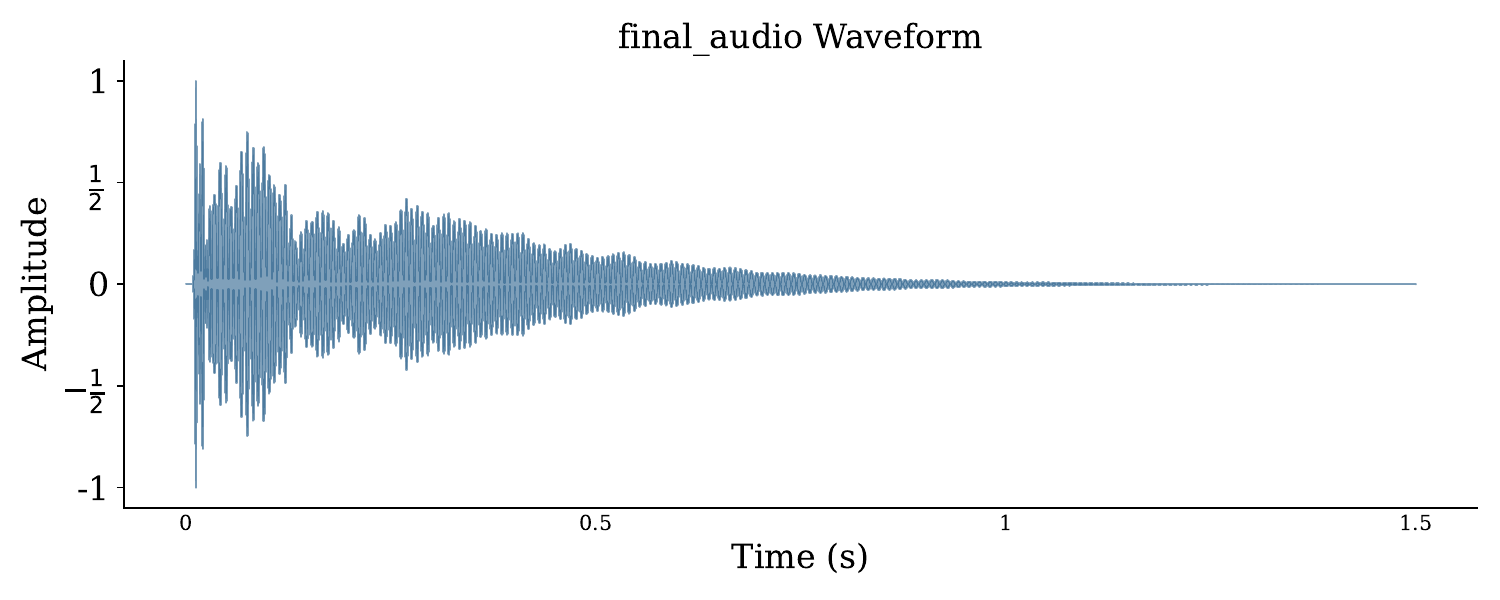}};
        \node[above=of render, yshift=-1.07cm, minimum width=0.25\linewidth, align=center] {differentiably\\rendered audio $\renderFunc(\optParams)$};
        
        \node (denoise) [process, below of=render, yshift=-1.3cm] {Add noise then denoise,\\ with diffusion model,\\conditioned on prompt};

        \node (update_signal) [below of=denoise, yshift=-1.3cm] {\includegraphics[trim={2.5cm 1.85cm 4.7cm .95cm}, clip, width=2cm]{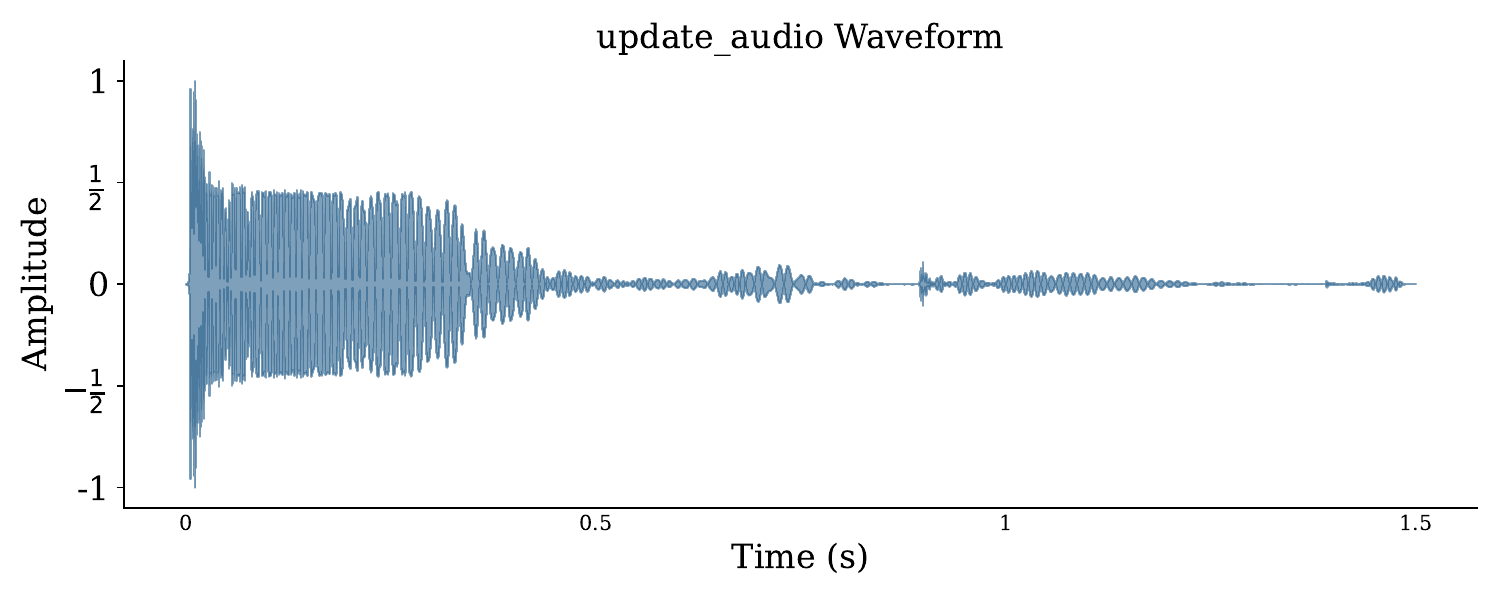}};
        \node[below=of update_signal, yshift=1.04cm, minimum width=0.25\linewidth, align=center] {``higher-probability''\\updated audio};
        \node (calc_update) [process, below of=audio_sim, yshift=-3.6cm] {audio update is expected \\ ``updated $-$ rendered audio''};

        \draw [arrow] (render) -- (denoise);
        \draw [arrow] (denoise) -- (update_signal);
        \draw [arrow] (update_signal) -- (calc_update);

        \draw [arrow] ($(audio_sim.east) + (0.0cm, 0.14cm)$) -- ($(render.west) + (0.0cm, 0.14cm)$);
        \draw [arrow, dashed] ($(render.west) + (0.0cm, -0.14cm)$) -- ($(audio_sim.east) + (0.0cm, -0.14cm)$);
        
        \draw [arrow] ($(render.south west) + (0.0cm, 0.2cm)$) -- ($(calc_update.north east) + (-0.2cm, 0.0cm)$);
        \draw [arrow, dashed] ($(calc_update.north east) + (0.0cm, -0.2cm)$) -- ($(render.south west) + (0.2cm, 0.0cm)$);

        \draw [arrow] (prompt.east) -- ++(2.0cm,0.0cm) 
            to[out=0, in=180] ++(0.65cm, 0.3cm)
            to[out=0, in=180] ++(0.65cm, -0.3cm)
            -- ($(denoise.west) + (0.0cm, 0.0cm)$);
    \end{tikzpicture}
    \end{adjustbox}
    
    \label{fig:middle}
  \end{subfigure}

  \begin{subfigure}{.95\textwidth}
    \centering
    \vspace{0.0\textheight}
    \caption{Audio-SDS Tasks}
    \vspace{0.01\textheight}
    \includegraphics[width=\linewidth]{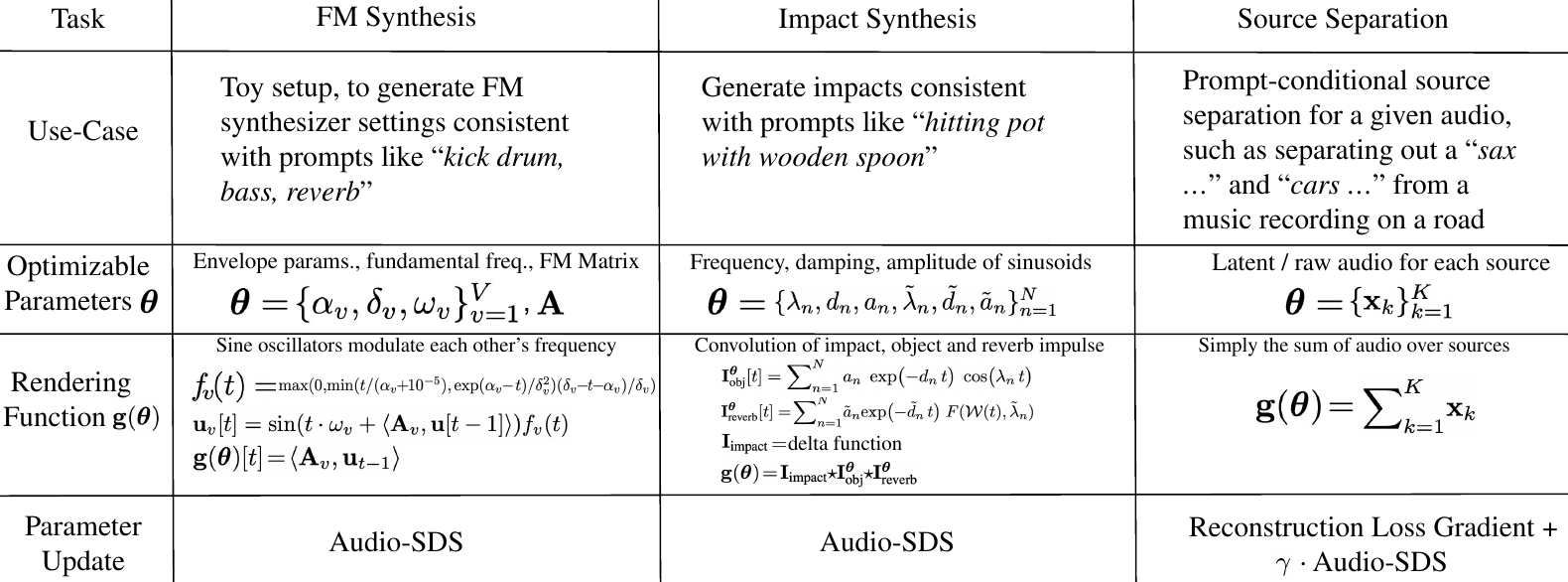}
    \label{fig:right}
  \end{subfigure}

  \vspace{-0.5em}
  \caption{We show an overview of our Audio-SDS framework on the top, our update in the middle, and our different tasks on the bottom. Our framework is further detailed in \secref{sec:method}. A more detailed overview of the update is in \figref{fig:audio_sds_overview_detailed}, highlighting our modifications to the SDS update. Our tasks are outlined further in \secref{sec:audio_sds_applications}, with results in \secref{sec:experiments}.}
  \label{fig:audio_sds_overview}
\end{figure*}

    Our contributions are:
    \vspace{-0.005\textheight}
    \begin{itemize}[leftmargin=*, nosep]
        \item \textbf{Unified Framework for Audio Tasks.} We introduce Audio-SDS (see \figref{fig:audio_sds_overview}), which uses a single, pretrained text-to-audio diffusion model for diverse audio tasks, distilling semantic updates from the frozen diffusion prior into parameters, unifying tasks ranging from synthesis to editing without requiring task-specific datasets.
    
        \item \textbf{Improving SDS For Audio.} First, a decoder-based SDS (\secref{sec:avoiding_instabilities}) circumvents instabilities we found differentiating through latent encoders. Second, we use multistep denoising (\secref{sec:method_multi_step_audio_sds}) for a more stable guidance signal. Third, our multiscale spectrogram emphasis (\secref{sec:method_spectrogram_updates}) better captures transient and high-frequency details, which are crucial for audio realism.
    
        \item \textbf{Demonstrations on Concrete Audio Tasks.} We use Audio-SDS to tune (i) an FM synthesizer for prompt-based tuning (\figref{fig:fm_and_impact_synthesis_overview_qualitative}), (ii) a physically informed impact model, tuning parameters to match text descriptions (\figref{fig:diff_impact_overview}), and -- most notably -- (iii) prompt-guided source separation for disentangling audio (\figref{fig:source_separation_overview}).
    \end{itemize}

\newpage

\section{Background}\label{sec:background}
    We begin by providing relevant background on diffusion models in \secref{sec:background_diffusion_models}, Score Distillation Sampling (SDS) in \secref{sec:background_sds}, audio generative modeling in \secref{sec:background_audio_generation}, and selected problems of interest in audio processing in \secref{sec:background_audio_applications}.

    \subsection{Contextualizing Diffusion Models}\label{sec:background_diffusion_models}
        Diffusion models have emerged as a powerful class of generative models, achieving remarkable success in the image domain~\cite{ho2020denoising} and more recently in high-resolution latent space generation~\cite{rombach2022high}. They operate by learning to reverse a forward noising process, where clean samples are incrementally perturbed by noise, with a parametrized denoising network trained to remove noise in small steps. These models have demonstrated strong mode coverage and have produced high-fidelity samples across various tasks. However, audio signals pose distinct challenges due to their potentially long durations, phase coherence requirements, and high sampling rates~\cite{kongdiffwave, zhang2023survey}.
        In this work, we aim to adapt techniques originally developed for image-based diffusion -- specifically Score Distillation Sampling (SDS) -- to address the unique aspects of audio generation.

    \subsection{Score Distillation Sampling (SDS)}\label{sec:background_sds}
        \textbf{Intuition:}
            SDS -- generalized to audio in \secref{sec:audio_sds_method} -- was first introduced in text-to-3D generation (DreamFusion), where updates from a pretrained image diffusion model drive 3D object optimization~\cite{pooledreamfusion,liu2023zero}. The diffusion model's score function acts as a guidance signal, effectively ``distilling'' its domain knowledge. Applying this guidance to a new representation (e.g., a 3D asset in the original use case or audio parameters here) with iterative updates aligns its renderings with samples the diffusion model deems likely. We repurpose SDS for audio generation by leveraging pretrained audio diffusion models, allowing us to impose strong priors -- learned from large-scale datasets -- on the generated output.

        \textbf{Technical Details:}
            Let $\renderFunc\!:\! \optParamsDomain \!\times\! \cameraDomain \!\to\! \renderDomain$ be our differentiable renderer, where $\optParams \!\in\! \optParamsDomain$ are our trainable parameters, $\camera \!\in\! \cameraDomain$ are our sampled parameters, and $\renderFunc(\optParams) \!=\! \render \!\in\! \renderDomain$ is a render in the diffusion model modality (e.g., RGB image or audio waveform). The sampled parameters $\camera$ include the camera location, lighting, and more when using SDS for text-to-3D. Let $\noisePrediction(\noisedRender,\timestep,\prompt)$ be the model noise prediction function, with noised rendering $\noisedRender$ at timestep $\timestep$ with prompt $\prompt$ and (frozen) diffusion parameters $\diffusionParams$. 
            In practice, we use classifier-free guidance (CFG, \citep{ho2022classifier}), with a guidance parameter $\guidanceScale$, and $\guidanceNoisePrediction(\noisedRender, \timestep\!, \prompt) \!=\! (1 \!+\! \guidanceScale)\noisePrediction(\noisedRender, \timestep\!, \prompt) \!-\! \guidanceScale \noisePrediction(\noisedRender, \timestep)$ to enable higher quality generation.
            We sample a random timestep and noise $\noise \!\sim\! \mathcal{N}(0,\identity)$ to form the noised rendering $\noisedRender$, where $\timestepMin \!\approx\! 0$ and $\timestepMin \!\approx\! 1$ are low and high noise scales, respectively:
            \begin{equation}
            \mathclap{
                \noisedRender(\optParams, \camera) = \signalScale_\timestep \renderFunc(\optParams, \!\camera) + \noiseScale_\timestep \noise
            }
            \end{equation}
            where $\signalScale_\timestep$ and $\noiseScale_\timestep$ are the signal and noise scale, respectively, which come from a standard diffusion schedule, with the noised render $\noisedRender$'s dependence on the diffusion timestep $\timestep$ and noise sample $\noise$ suppressed. The SDS loss is:
            \begin{equation}\label{eq:loss_sds_k}
            \mathclap{
              \lossSDS(\optParams; \prompt) = \E_{\timestep\!\!,\noise,\camera} [ \timestepWeight || \guidanceNoisePrediction ( \noisedRender(\optParams, \!\camera), \!\timestep, \prompt ) - \noise ||^{2} ]
            }
            \end{equation}
            with $\timestepWeight$ being a time-dependent weight.
            Taking the gradient via the chain rule gives:
            \begin{equation}
            \mathclap{
                \!\!\!\nabla_{\!\optParams}\!\lossSDS(\!\optParams;\!\prompt\!) \!\!=\!\!\!\!
              \E_{\timestep\!\!,\noise,\camera} \!\![
                \timestepWeight \! (
                  \!\guidanceNoisePrediction \!(\! \noisedRender(\!\optParams\!, \!\camera\!)\!, \!\timestep\!, \prompt \!) \!-\! \noise
                )\!
                \mathbf{J}_{\!\guidanceNoisePrediction}\!\!(\!\noisedRender\!)
                \!\nabla_{\!\optParams} \noisedRender(\!\optParams\!,\! \camera)\!
              ]
            }
            \end{equation}
            where $\nabla_{\!\optParams} \noisedRender(\optParams, \camera) = \signalScale_\timestep \nabla_{\!\optParams} \renderFunc(\optParams, \camera)$.
            Crucially, the SDS update $\updateSDS$ of \citet{pooledreamfusion} takes $\nabla_{\optParams}\lossSDS(\optParams;\prompt)$ and avoids differentiating through the pretrained diffusion model by approximating the Jacobian of the CFG noise prediction $\mathbf{J}_{\!\guidanceNoisePrediction}\!(\noisedRender)$ as the identity:
            \begin{equation}\label{eq:sds_update}
            \mathclap{
              \updateSDS(\optParams;\prompt) \!=\!
              \E_{\timestep\!\!,\noise,\camera} \!\![
                \timestepWeight(
                  \!\guidanceNoisePrediction \!( \! \noisedRender(\!\optParams\!, \!\camera)\!, \!\timestep\!, \prompt \!) \!-\! \noise
                )
                \!\nabla_{\!\optParams} \noisedRender(\!\optParams\!,\! \camera)
              ]
            }
            \end{equation}
            Intuitively, as shown in \figref{fig:middle}, this update ``nudges'' the parameters $\optParams$ so that the rendered audio $\render \!=\! \renderFunc(\optParams, \camera)$ becomes more likely under the text-conditioned distribution of the diffusion model at -- e.g., in text-to-3D -- camera location $\camera$.

    \subsection{Related Audio Generative Approaches}\label{sec:background_audio_generation}
        \textbf{Audio Representations:}
            An audio waveform is a time-domain representation of sound pressure (amplitude) variations, capturing the signal over time and serving as a fundamental input or output for generative modeling tasks. Audio waveforms can also be converted into spectrograms to explicitly capture frequency content, as leveraged in \secref{sec:method_spectrogram_updates}.

        \textbf{Audio Generation Paradigms:}
            Autoregressive models \citep{van2016wavenet} have led to highly expressive synthesis but at considerable computational cost. Adversarial \citep{kumar2019melgan, kong2020hifi, donahueadversarial}, and flow-based approaches \citep{prenger2019waveglow, valle2019flowtron, shih2022generative} reduce inference time but sometimes struggle with training stability or mode collapse.

        \textbf{Diffusion Models for Audio Generation:}
            Diffusion-based methods have garnered attention for various tasks such as text-to-speech~\cite{kongdiffwave, chen2021wavegrad}. By iteratively denoising samples, these methods provide strong sample diversity and quality. However, they typically rely on directly training a diffusion model on large, domain-specific datasets. In contrast, our SDS-based approach enables task-specific guidance without retraining from scratch, leveraging updates from pretrained models to shape audio outputs. This reduces data requirements and offers more flexible conditioning options than many existing diffusion frameworks.
            
        \textbf{Latent Diffusion Models for Audio and Stable Audio Open:}
            Recent work has shown that using a latent representation -- with input compressed via a learned VAE -- dramatically improves efficiency, enabling text-conditioned frameworks such as AudioLDM~\citep{liu2023audioldm}, Audio Flamingo~\citep{kong2024audioflamingo}, and Stable Audio Open~\citep{evans2024stable}. 
            Notably, Stable Audio Open operates on the stereo (2-channel) input waveforms $\render \!\in\! \audioRenderDomain$. For a $\SI{10}{\second}$ audio file with $\num{44.1}$kHz sample rate, $\totalAudioSamples \!=\! \num{441000}$. An encoder $\encodeFunc$ embeds the waveforms into the latent space: $\latentRender \!=\! \encodeFunc(\render) \!\in\! \stableAudioLatentDomain$, with repeated downsampling leading to an overall compression of $64\times$.

    \subsection{Data Driven Audio Tasks}\label{sec:background_audio_applications}
        Most prior audio generation works focus on speech or music, leaving non-speech sound effects (SFX) relatively underexplored~\citep{salamon2014dataset, gemmeke2017audio}.
        Unlike speech or music, many SFX demand rich transient dynamics (e.g., impacts, collisions) that are difficult to capture in conventional data-driven models, particularly when clean, large-scale datasets are scarce~\cite{farnell2010designing}. Meanwhile, SDS demonstrates how updates derived from pretrained diffusion models can guide other parametrized representations (e.g., 3D objects). Here, we bring this concept to the audio domain, showing how latent audio diffusion can inform physically interpretable or semantically guided source separation and generation.  In particular, we demonstrate SDS on three audio problems that benefit from both parametric modeling and semantic constraints:
        
        \textbf{FM-Synthesizer.}
            A classic technique where oscillators modulate each other’s frequencies to produce rich timbres~\citep{chowning1973synthesis}.  We treat the FM parameters as optimizable variables showing how SDS can guide the synthesized waveform to match user-specified prompts (e.g., ``\texttt{a warm bass note}'') via semantic alignment. 
        
        \textbf{Impact Synthesis.}
            Physically motivated audio models simulate collisions using modal resonances or decaying oscillators~\citep{obrien2002synthesizing, langlois2014eigenmode}.  We incorporate these differentiable physics-based components and use SDS to tune their parameters (e.g., frequencies, dampings) so the output aligns with text prompts like ``\texttt{sharp metallic clang}.''
        
        \textbf{Source Separation.}
            A long-standing challenge in audio, source separation decomposes a mixed signal into distinct parts (e.g., instruments or environmental sounds)~\citep{jansson2017singing, stoller2018wave}.  Rather than relying on isolated source data, we constrain each source to be plausible under a pretrained diffusion model, leveraging user prompts such as ``\texttt{saxophone}'' or ``\texttt{traffic noise}'' to guide decomposition.

\section{Approach: Audio-SDS and Applications}\label{sec:method}
    We include a glossary summarizing our notation in App. Table~\ref{tab:TableOfNotation}.
    First, we cover applying Score Distillation Sampling (SDS) to audio diffusion models in \secref{sec:audio_sds_method}.
    Then, we detail our proposed applications of Audio-SDS in \secref{sec:audio_sds_applications}.
    
    \begin{figure*}[ht]
        \vspace{-0.01\textheight}
        \centering
        \begin{adjustbox}{scale=0.90}
        \begin{tikzpicture}[node distance=1cm, every node/.style={align=center}]
            \tikzstyle{input} = [rectangle, rounded corners, minimum width=1cm, minimum height=1cm, text centered, draw=black, dashed]
            \tikzstyle{process} = [rectangle, rounded corners, minimum width=1cm, minimum height=1cm, text centered, draw=black]
            \tikzstyle{arrow} = [thick,->,>=stealth]
        
            \node (audio_sim) [input] {{\color{cyan}\footnotesize{Audio parameters $\optParams$}}};
            
            \node (render) [right of=audio_sim, xshift=3cm] {\includegraphics[trim={2.5cm 1.85cm 4.7cm .95cm}, clip, width=2cm]{images/audio_sds_overview/final_audio_waveform.pdf}};
            \node[above=of render, yshift=-1.47cm, minimum width=0.25\linewidth, align=center] {{\color{blue}\tiny{Rendered audio}}\\{\color{blue}$\render = \renderFunc({\color{cyan}\optParams})$}};
            
            \node (encode) [process, right of=render, xshift=3cm] {\footnotesize{Encoded latent}\\ $\latentRender = \encodeFunc_{\diffusionParams}(\render)$};
            \node (noised_audio) [process, right of=encode, xshift=3cm] {\footnotesize{Noised latent}\\$\noisedRender = \textnormal{noise}(\latentRender, \timestep, \noise)$};
            
            \node (stable_audio) [input, below of=encode, yshift=-1.cm] {{\color{darkgreen}\footnotesize{Audio diffusion model}}\\{\color{darkgreen}\footnotesize{with frozen parameters $\diffusionParams$}}};%

            \node (prompt) [input, right of=stable_audio, xshift=4cm] {{\color{darkgreen}\footnotesize{Prompt $\prompt$}}};

            \node (noise) [process, right of=stable_audio, xshift=2cm] {\tiny{Sample noise}\\\tiny{$\noise \sim \mathcal{N}(0, \identity)$}\\\tiny{and noise scale}\\\tiny{$\timestep \!\!\sim\! \mathcal{U}[0, 1]$}};
            
            \node (denoise) [process, below of=noised_audio, yshift=-3cm] {{\color{red}\footnotesize{Denoised latent}}\\{\color{red}$\denoisedNoisedLatent = \textnormal{denoise}_{\diffusionParams}(\noisedRender, \prompt)$}};

            \node (update_signal) [left of=denoise, xshift=-3cm] {\includegraphics[trim={2.5cm 1.85cm 4.7cm .95cm}, clip, width=2cm]{images/audio_sds_overview/update_audio_waveform.pdf}};
            \node[below=of update_signal, yshift=1.34cm, minimum width=0.25\linewidth, align=center] {{\color{red}\tiny{Decoded Latent}}\\{\color{red}$\noisedDenoisedRender = \decodeFunc_{\diffusionParams}(\denoisedNoisedLatent)$}};
            
            \node (spec_update_signal) [left of=update_signal, xshift=-3cm] {\includegraphics[trim={2.5cm 1.7cm 4.7cm .95cm}, clip, width=2cm]{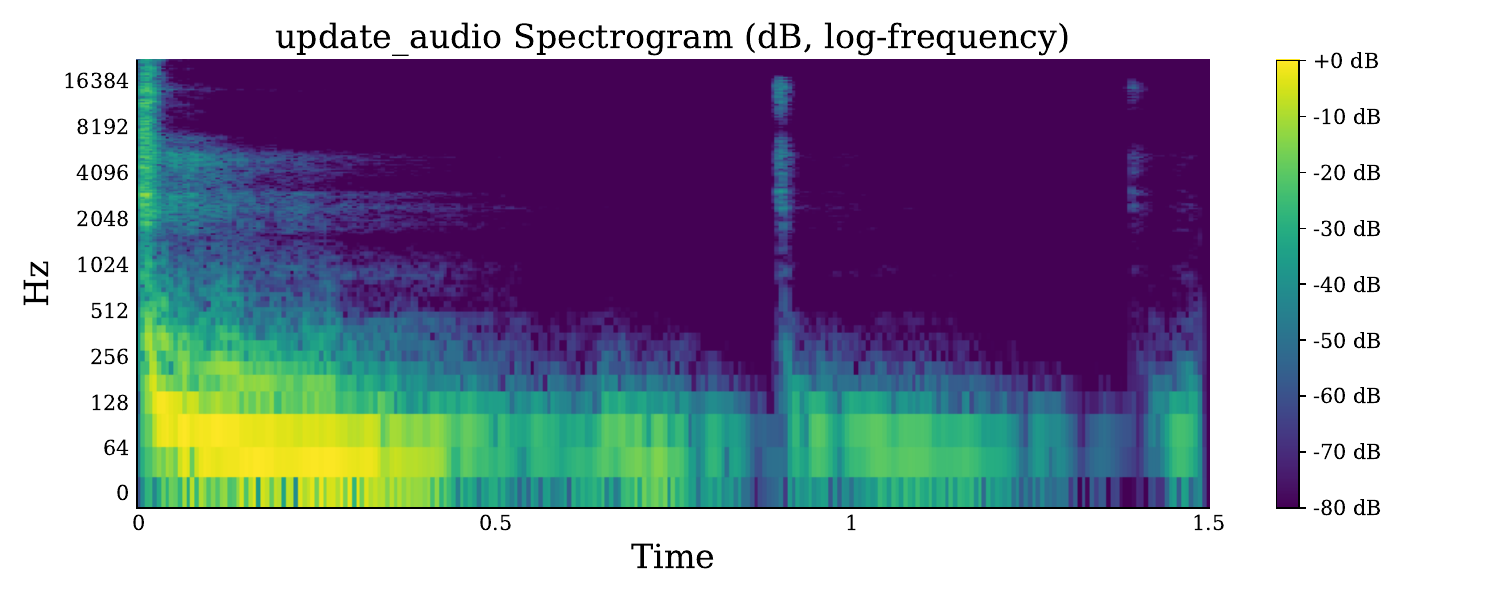}};
            \node[below=of spec_update_signal, yshift=1.2cm, minimum width=0.25\linewidth, align=center] {{\color{red}\footnotesize{$\noisedDenoisedSpecRender = \sum\STFT_{\STFTIndex}(\noisedDenoisedRender)$}}};%
            
            \node (spec_render) [below of=render, yshift=-1.25cm] {\includegraphics[trim={2.5cm 1.7cm 4.7cm .95cm}, clip, width=2cm]{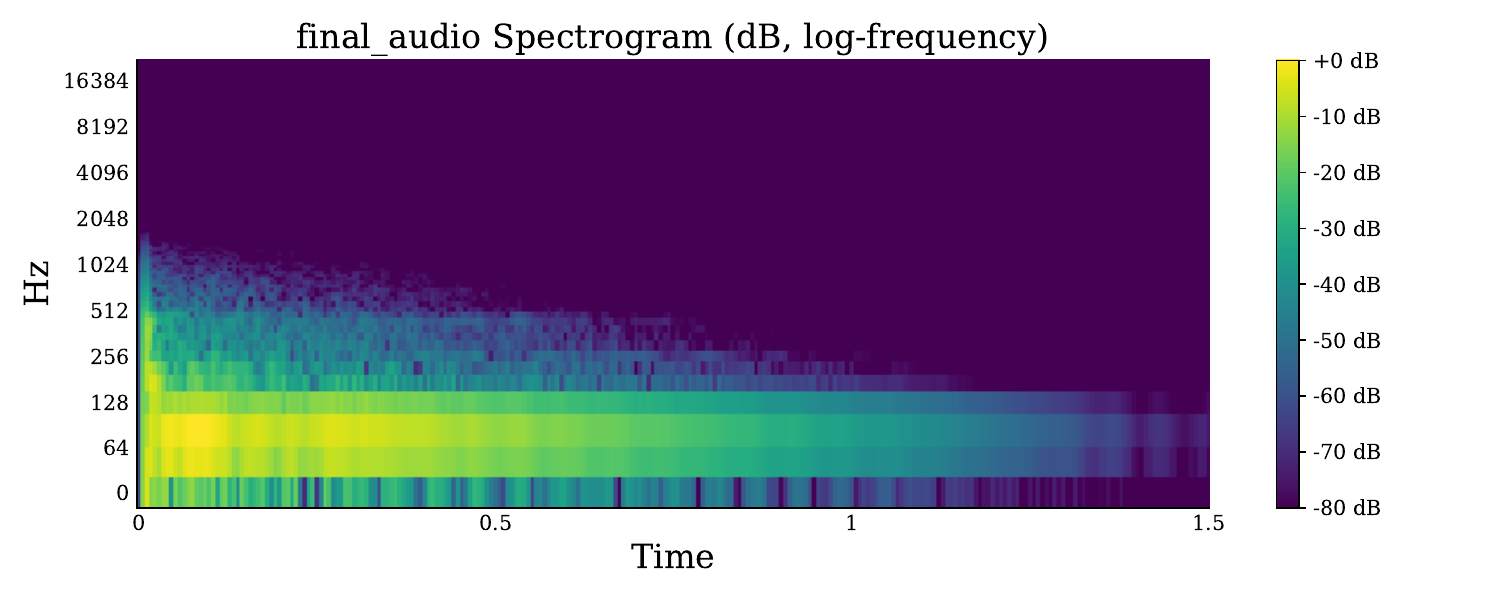}};
            \node[below=of spec_render, yshift=1.2cm, minimum width=0.25\linewidth, align=center] {{\color{red}\tiny{Multiscale spectrogram}}\\{\color{red}\footnotesize{$\specRender = \sum\STFT_{\STFTIndex}(\render)$}}};
            
            \node (calc_update) [process, left of=spec_update_signal, xshift=-3cm] {\tiny{Get update direction $\mathbf{d}^{\STFT\!, \decodeFunc}$ from Eq.\eqref{eq:spectrogram_decoder_sds_update}}\\\tiny{$
                  \mathbf{d}^{\STFT\!, \decodeFunc} = \E_{\timestep\!\!,\noise} \!\big[\noisedDenoisedSpecRender\big] \!-\! \specRender$}};

            \node (spec_anchor) [left of=spec_render] {};
        
            \draw [arrow] (audio_sim) -- (render);
            \draw [arrow] (render) -- (encode);
            \draw [arrow] (encode) -- (noised_audio);
            \draw [arrow] (noised_audio) -- (denoise);
            \draw [arrow] (denoise) -- (update_signal);
            \draw [arrow] (update_signal) -- (spec_update_signal);
            \draw [arrow] (spec_update_signal) -- (calc_update);
            \draw [arrow] (render) -- (spec_render);
            \draw [arrow] (spec_anchor) to  (calc_update) ;
            \draw [arrow] [thick, dashed,->,>=stealth] (calc_update) -- node[left] {\footnotesize{Update parameters {\color{cyan}$\optParams$}}\ \\\footnotesize{via Decoder-SDS}\\\footnotesize{in Eqs. \eqref{eq:decoder_sds_update}/\eqref{eq:spectrogram_decoder_sds_update}:}\\$\updateSDS^{\STFT\!, \decodeFunc} = \mathbf{d}^{\STFT\!, \decodeFunc} \,\nabla_{\!{\color{cyan}\optParams}}\specRender$} (audio_sim);

            \draw [arrow] (stable_audio) --  (encode);
            \draw [arrow] (stable_audio) --  (noised_audio);
            \draw [arrow] (stable_audio) --  (denoise);
            \draw [arrow] (stable_audio) --  (update_signal);

            \draw [arrow] (noise) --  (noised_audio);

            \draw [arrow] (prompt) --  (denoise);
        \end{tikzpicture}
        \end{adjustbox}
        \vspace{-0.023\textheight}
        \caption{
            Overview of our Audio-SDS method, marrying the Score Distillation Sampling (SDS)~\citep{pooledreamfusion} with an audio diffusion model in a robust framework for various audio tasks.
            SDS (see \secref{sec:background_sds}) -- originally developed for text-to-3D generation -- computes an update for rendered data $\render$ in the diffusion models modality (e.g., image or audio), then propagates that update through a differentiable simulation $\renderFunc$ to update parameters $\optParams$. %
            Intuitively, this nudges the render parameters to make it ``more likely'' under our conditioning, here the text prompt $\prompt$.
            Adapting SDS to audio, we propose three modifications {\color{red}shown in red}:
            (a) Decoder-SDS to circumvent differentiating through the encoder (\secref{sec:avoiding_instabilities}), (b) a spectrogram space update for perceptual details (\secref{sec:method_spectrogram_updates}), and (c) multi-step denoising for improved stability \secref{sec:method_multi_step_audio_sds}. We demonstrate three tasks, which, concretely, are choices of the rendering function $\renderFunc$ {\color{blue} shown in blue}: (1) tuning an FM synthesizer, (2) tuning physical impact synthesis, and (3) source separation -- see \secref{sec:audio_sds_applications}.
            Here, we show results for impact synthesis midway during training using the prompt $\prompt \!=\! $ ``\texttt{kick drum, bass, reverb}''.
            Optimizable parameters are {\color{cyan}shown in cyan}, and the user-specified prompt and choice of diffusion model are {\color{darkgreen}shown in green}.
        }
        \label{fig:audio_sds_overview_detailed}
        \vspace{-0.02\textheight}
    \end{figure*}
    
    \subsection{Using SDS on Audio Diffusion Models}\label{sec:audio_sds_method}
        \figref{fig:audio_sds_overview_detailed} details how we apply SDS to audio diffusion models. At its core, our methodology is similar to DreamFusion~\citep{pooledreamfusion}. Let $\audioRenderFunc\!:\! \optParamsDomain \!\to\! \audioRenderDomain$ denote the audio rendering function generating a stereo (2-channel) audio buffer output $\audioRender \!=\! \audioRenderFunc(\optParams)$ -- abbreviated to just $\render$ from here on out -- where $\totalAudioSamples$ is the total number of audio samples. We suppress the sampled parameter $\camera$ in the rendering function $\renderFunc$ (e.g., camera orientation in text-to-3D), as it is left unused in our current audio formulations.

        In practice, we found three modifications helpful for Audio-SDS:
        (i) \textbf{Circumventing Encoder-Decoder Instabilities} (Sec.~\ref{sec:avoiding_instabilities}), by avoiding differentiating through the encoder, 
        (ii) \textbf{Spectrogram Emphasis} (Sec.~\ref{sec:method_spectrogram_updates}), which places additional weight on transient and high-frequency details during updates, and 
        (iii) \textbf{Multi-step Denoising} (Sec.~\ref{sec:method_multi_step_audio_sds}), which refines the noised latent code over multiple partial denoising steps, improving perceptual fidelity.

        Synthesizers (\secref{sec:method_synthesizer}), physical simulations (\secref{sec:method_diff_impact}), or multiple audio buffers (\secref{sec:method_source_separation}) corresponding to separate sources could all serve as $\audioRenderFunc$ depending on the application, with $\optParams$ changing appropriately depending on the context. While DreamFusion introduced inverse rendering for 3D shapes via image diffusion, we show an analogous process for ``inverse audio synthesis'' using text-to-audio diffusion. Crucially, regardless of the specific parametrization, the same audio diffusion model can be used to update them to better align with the given prompt. Like the other Stability models, Stable Audio Open~\citep{evans2024stable} runs on a latent diffusion process, with encoder $\encodeFunc$ and decoder $\decodeFunc$.

        \subsubsection{Audio Diffusion Encoder Instabilities}\label{sec:avoiding_instabilities} 
            Score Distillation Sampling (SDS) has traditionally been compatible with latent diffusion models \cite{xie2024latte3d}. Latent audio diffusion models, however, specifically require computing gradients through the encoding process $\encodeFunc$, which maps the rendered audio $\render$ to its latent representation $\latentRender = \encodeFunc(\render)$. SDS updates then encourage $\latentRender$ to match the diffusion model's prediction by minimizing the discrepancy between the true $\noise$ and the predicted $\noisePrediction$.
            
            However, we observed instabilities and artifacts when differentiating through the encoder of a latent audio diffusion model, which can cause unreliable gradient estimates, hampering the optimization process and degrading the quality of the synthesized audio.
            To mitigate this, we use an approach circumventing the need to differentiate the encoder at all. Instead of directly comparing predicted noise with true added noise in latent space, we compute the discrepancy in the audio domain after decoding the denoised latent. In symbols, our modified decoder-SDS update is then:
            
            \begin{equation}\label{eq:decoder_sds_update}
            \mathclap{
                \updateSDS^{\decodeFunc}(\optParams;\prompt) \!=\! 
                 \Big( \E_{\timestep\!,\noise} \![ \noisedDenoisedRender_{\diffusionParams}(\optParams, \timestep, \noise, \prompt)] - \render(\optParams)\Big)\nabla_{\!\optParams}\render(\optParams)
            }
            \end{equation}
            Or, succinctly, $\updateSDS^{\decodeFunc} \!=\! (\E[\noisedDenoisedRender] \!-\! \render)\nabla_{\!\optParams}\render$, where:
            \begin{equation}
            \mathclap{
                \!\!\!\noisedDenoisedRender_{\diffusionParams}\!(\!\optParams,\! \timestep\!,\! \noise,\! \prompt\!) \!=\! \decodeFunc_{\diffusionParams}\!(\!\textnormal{denoise}_{\diffusionParams}\!(\! \textnormal{noise}(\!\encodeFunc_{\diffusionParams}\!(\render(\optParams)\!), \timestep\!,\! \noise\!)\!,\! \prompt\!)\!)
            }
            \end{equation}
            represents the result after denoising the noised latent and decoding. 
            Empirical evaluations confirm this significantly enhances the stability of SDS for audio diffusion models. The resulting audio maintains high fidelity and aligns closely with the text prompts, while also benefiting from a performance boost by omitting the encoder's derivatives. This is an audio-modality analog of the image-space updates of SparseFusion~\citep{zhou2023sparsefusion} and HIFA~\citep{zhuhifa}.
        
        \subsubsection{Spectrogram Emphasis for Audio-SDS}\label{sec:method_spectrogram_updates}
            While na\"ive $\ell_2$ losses in the time domain suffice for many diffusion-based pipelines, they can underweight transients and high-frequency details critical to perceived audio quality. To address this, we include a multiscale spectrogram residual. Given a rendered waveform $\render$ and an encoded-noised-denoised-decoded sample $\noisedDenoisedRender$, we compute a collection of Short-Time Fourier Transform (STFT) magnitudes $\STFT_{\!\STFTIndex}(\cdot)$ over $\STFTIndex \!=\! 1,\dots,\totalNumSTFT$ different window sizes (see App. \secref{sec:app_experimental_details}). Our final SDS update is:
            \begin{equation}\label{eq:spectrogram_decoder_sds_update}
            \mathclap{
            \footnotesize{
                \updateSDS^{\STFT\!, \decodeFunc}\!(\!\optParams;\!\prompt\!) \!=\!\! \sum_{\STFTIndex} 
                  \! \!\Big(\!\E_{\timestep\!,\noise} \!\!\!\big[\!\STFT_{\!\STFTIndex}\!(\!\noisedDenoisedRender_{\!\diffusionParams}(\!\optParams,\! \timestep\!,\! \noise,\! \prompt\!)\!)\!\big] \!-\! \STFT_{\!\STFTIndex}\!(\!\render(\!\optParams\!)\!)\!\!\Big) \!\nabla_{\!\!\optParams} \STFT_{\!\STFTIndex}\!(\!\render(\!\optParams\!)\!)
            }
            }
            \end{equation}
            Or, succinctly, $\updateSDS^{\STFT\!, \decodeFunc} \!\!=\! (\E[\noisedDenoisedSpecRender] \!-\! \specRender)\!\nabla_{\!\optParams}\specRender$, where $\noisedDenoisedSpecRender \!=\! \sum_{\STFTIndex=1}^{\totalNumSTFT}\!\STFT_{\!\STFTIndex}(\noisedDenoisedRender)$ and $\specRender \!=\! \sum_{\STFTIndex=1}^{\totalNumSTFT}\!\STFT_{\!\STFTIndex}(\render)$.
            Intuitively, spectrogram emphasis aligns better with perceptual similarity, as it compares frequency differences over each STFT window, emphasizing transients and higher-frequency information. 
            However, using a single STFT size is limited as frequency and temporal localization are opposed when using the Fourier transform \cite{smith2003digital}.
            Longer windows better resolve frequencies in the signal at the expense of temporal resolution. Alternatively, shorter windows increase temporal discrimination with worse frequency resolution. Multiscale spectrograms avoid this tradeoff by averaging varying window sizes, as in \cite{langlois2014inverse}, which also has the benefit of smoothing artifacts caused by the window boundaries.

        \subsubsection{Multistep Denoising for Enhanced SDS}\label{sec:method_multi_step_audio_sds}
            Standard SDS relies on backpropagating the residual between added noise and the noise predicted by one application of the denoiser. However, we found audio waveforms often benefit from multiple partial denoising steps per iteration, analogous to the SDEdit approach \citep{meng2021sdedit} in image diffusion, ensuring updates align more closely with the distribution of the diffusion model. In each iteration, we:
            (1) Sample $\timestep, \noise$, adding $\noise$ to the latent $\latentRender(\optParams)$ to get noised latent $\noisedRender$.
            (2) Denoise the noised latent to get $\denoisedNoisedLatent$ by running a handful of steps of the diffusion sampler, e.g., $2$--$10$.
            (3) Compute the SDS update as before using the multistep denoised example instead of single-step.
            Specifically, the denoising function is a partial sampling chain using a DDIM-style sampler, starting from the sampled time $\timestep$ back to $\num{0}$. %

    \vspace{-0.01\textheight}
    \subsection{Proposed Audio-SDS Applications}\label{sec:audio_sds_applications}
    \vspace{-0.01\textheight}
        One of our key objectives is to showcase how a \emph{single} pretrained text-to-audio diffusion model can tackle multiple audio tasks -- from tuning synthesizer parameters to physically informed impact modeling and prompt-driven source separation. While each application has well-established baselines, our emphasis here is on \emph{generality}: we demonstrate that Audio-SDS can flexibly guide \textbf{any} differentiable audio parametrization to align with high-level textual descriptions, all without needing specialized retraining. If the simulator output is mono and in $\R^{\totalAudioSamples}$, we duplicate it across channels to $\audioRenderDomain$ for Stable Audio Open's stereo input.

        \begin{figure}[t]
            \vspace{-0.01\textheight}
            \centering
            \begin{tikzpicture}
            \centering
                \node (img11){\includegraphics[trim={0.0cm 0.0cm 0.0cm 3.1cm}, clip, width=.95\linewidth]{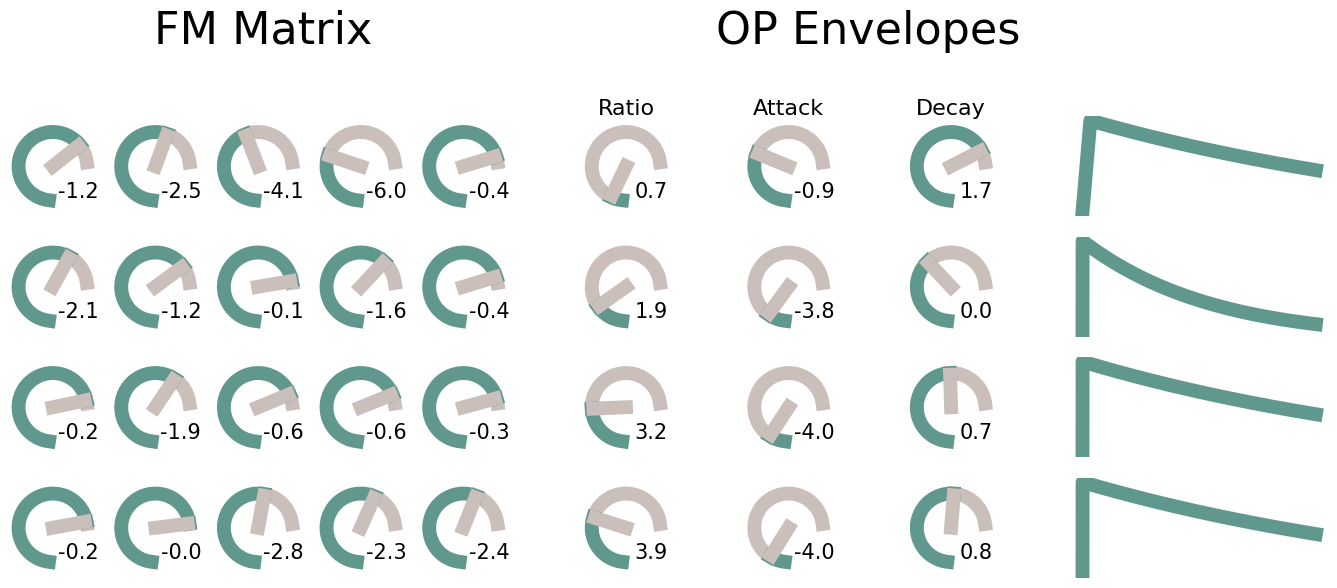}};

                \node[above=of img11, node distance=0cm, xshift=0.25cm, yshift=-1.15cm,  font=\color{black}]{\hspace{0.03\textwidth}\tiny{FM Matrix $\fmMatrix$ \hspace{0.065\textwidth} Ratio $\fmFrequency$ \hspace{0.005\textwidth} Attack $\fmAttackDecayA$ \hspace{0.005\textwidth} Decay $\fmAttackDecayB$ \hspace{0.016\textwidth} Operator Envelopes}};
            \end{tikzpicture}
            \vspace{-0.035\textheight}
            \caption{
                \textbf{FM Synthesis Overview.}
                We visualize the optimizable parameters $\optParams$ as dials for an FM user interface, including the FM matrix $\fmMatrix$ and each operator's ratio, attack, and decay $\{\fmFrequency_\fmMatrixIndex, \fmAttackDecayA_\fmMatrixIndex, \fmAttackDecayB_\fmMatrixIndex \}_{\fmMatrixIndex=1}^{\fmMatrixSize}$, where $\fmMatrixSize = 4$, at the end of optimization for ``\texttt{kick drum, bass, reverb}''. The optimized output (\href{https://drive.google.com/file/d/182ecHVg32w641PLL3XFoQHKBn8Z_BdJW/view?usp=sharing}{{\color{nvidiagreen}listen here}}) provides {\color{darkgreen} $+\num{0.13}$ CLAP} over the initialization (\href{https://drive.google.com/file/d/11JMwLdoQDhML08GbuKjlaCIjamVM1eDa/view?usp=sharing}{{\color{nvidiagreen}listen here}}) showing improved prompt alignment, with more results in \figref{fig:fm_and_impact_synthesis_overview_qualitative}.
            }
            \label{fig:fm_synthesis_overview}
            \vspace{-0.02\textheight}
        \end{figure}

        \vspace{-0.01\textheight}
        \subsubsection{FM Synthesizer}\label{sec:method_synthesizer}
        \vspace{-0.01\textheight}
            As a toy example and proof of concept, we deploy a rudimentary Frequency Modulation (FM) synthesizer, overviewed in \figref{fig:fm_and_impact_synthesis_overview_qualitative}. Simple sine oscillators run in parallel and interact by modulating each other's frequency -- hence the name \emph{frequency modulation}.
            This oscillation's strength is calculated by maintaining an \emph{FM} matrix, a (time-independent) $\fmMatrixSize \times (\fmMatrixSize+1)$ matrix $\fmMatrix$, denoting the strength of interactions between pairs. For each timestep $\audioSampleIndex$, each oscillator state $\fmState \in \R^{\fmMatrixSize \times \totalAudioSamples}$ is computed according to:\vspace{-0.005\textheight}
            \begin{equation}\label{eq:fm_state}
            \mathclap{
                \smash{
                \fmState_\fmMatrixIndex[\audioSampleIndex] = \sin( \audioSampleIndex \cdot \fmFrequency_\fmMatrixIndex + \langle \fmMatrix_\fmMatrixIndex, \fmState[\audioSampleIndex-1] \rangle) \fmAttackDecayFunc_\fmMatrixIndex(\audioSampleIndex)
                }
            }\vspace{-0.005\textheight}
            \end{equation}
            the rendered output $\smash{\render \!\in\! \R^{\totalAudioSamples}}$ is  $\smash{\render[\audioSampleIndex] \!=\! \renderFunc(\optParams)[\audioSampleIndex] \!=\! \langle \fmMatrix_\fmMatrixIndex, \fmState_{\audioSampleIndex-1} \rangle}$ where the optimizable parameters $\optParams$ are the FM matrix $\fmMatrix$ and envelope parameters $\{\fmFrequency_\fmMatrixIndex, \fmAttackDecayA_\fmMatrixIndex, \fmAttackDecayB_\fmMatrixIndex\}_{\fmMatrixIndex=1}^{\fmMatrixSize}$. Above, $\fmFrequency_{\fmMatrixIndex}$ is the fundamental frequency of the input note, and $\fmAttackDecayFunc_{\!\fmMatrixIndex\!}(\audioSampleIndex) \!=\! \max(0,\! \min( \audioSampleIndex/(\fmAttackDecayA_\fmMatrixIndex \!+\! 10^{-5}), \exp(\fmAttackDecayA_\fmMatrixIndex \!- \audioSampleIndex)/\fmAttackDecayB_\fmMatrixIndex^2)(\fmAttackDecayB_\fmMatrixIndex \!-\! \audioSampleIndex \!-\! \fmAttackDecayA_\fmMatrixIndex)/\fmAttackDecayB_\fmMatrixIndex)$ is an envelope controlling attack and decay for the $\fmMatrixIndex$th oscillator. While our setup modulates the phase of each oscillator, this is practically equivalent to frequency modulation \cite{schure1955frequency}, but preferred for numerical stability.
        
        \begin{figure}[t]
            \vspace{-0.015\textheight}
            \centering

            \begin{tikzpicture}
                \centering
                \node (img11){\includegraphics[trim={2.5cm 1.75cm 4.7cm 1.0cm}, clip, width=.4\linewidth]{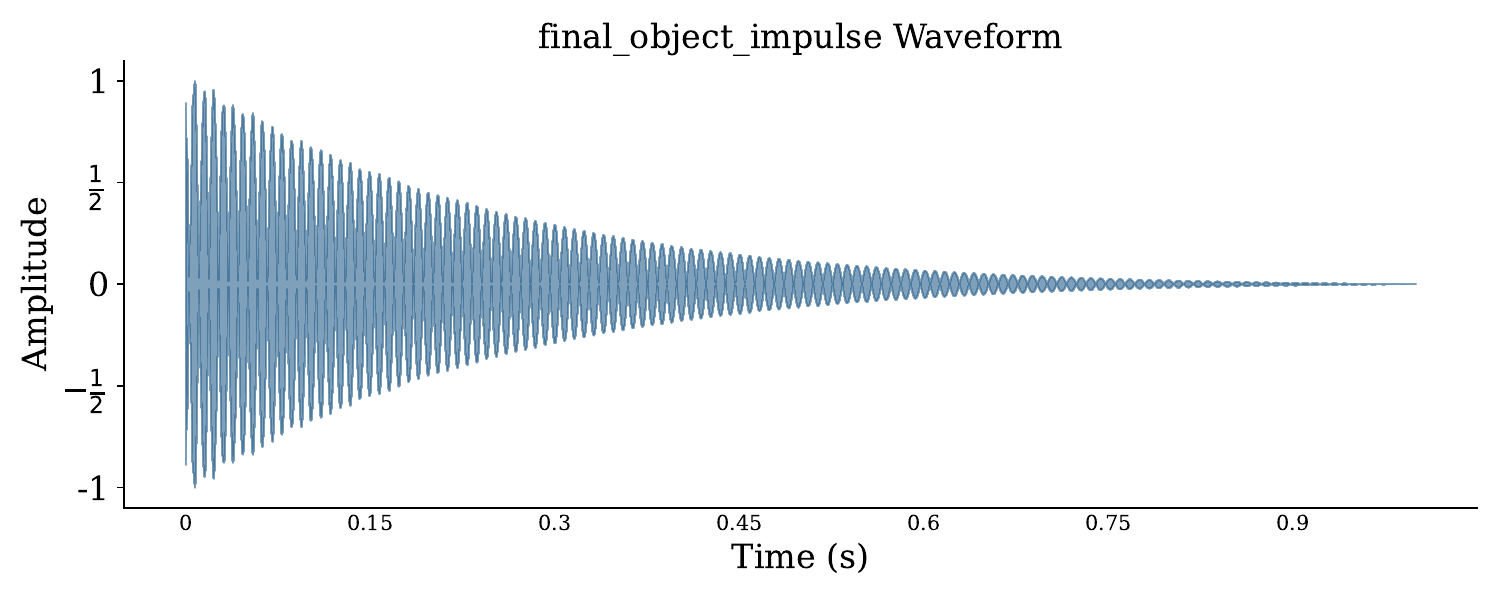}};
                \node (img21)[below=of img11, yshift=0.7cm]{\includegraphics[trim={2.5cm 1.75cm 4.7cm 1.0cm}, clip, width=.4\linewidth]{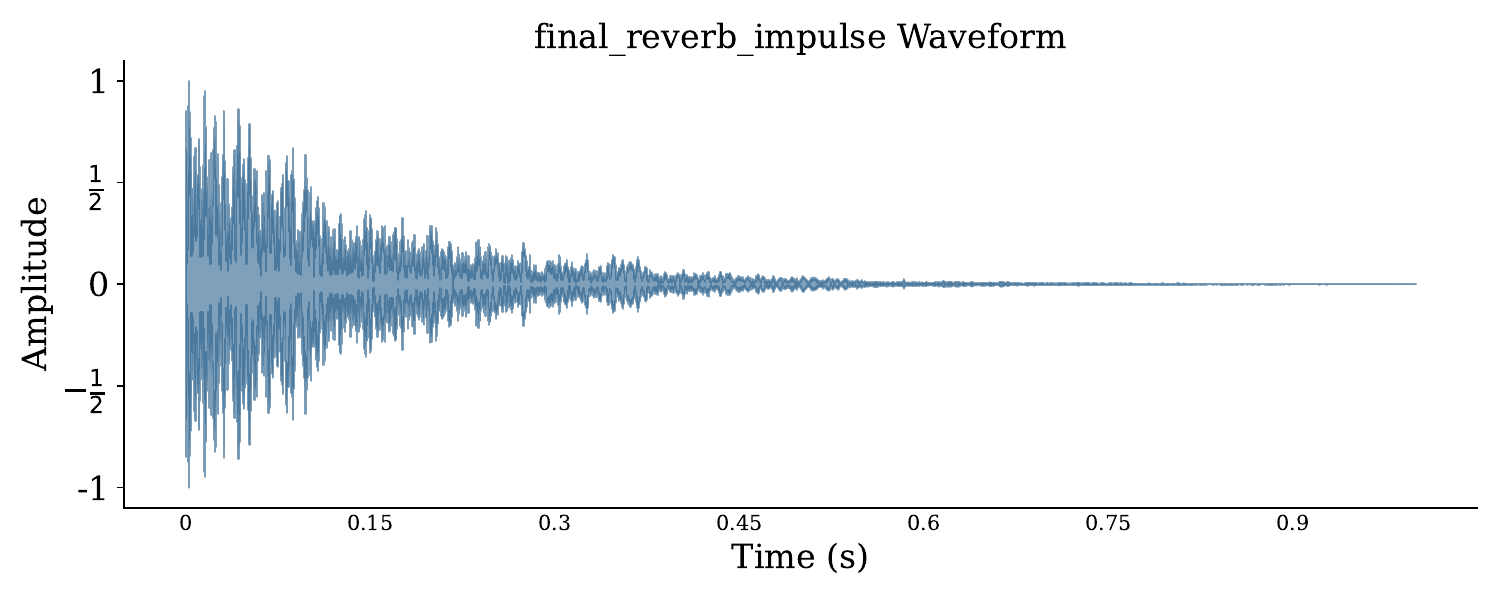}};

                \node (star) [above=of img21, node distance=0cm, xshift=-0.1cm, yshift=-1.2cm,  font=\color{black}]{\Huge{$\star$}};
                
                \node (img31)[right=of star, xshift=1.3cm]{\includegraphics[trim={2.5cm 1.75cm 4.7cm 1.0cm}, clip, width=.4\linewidth]{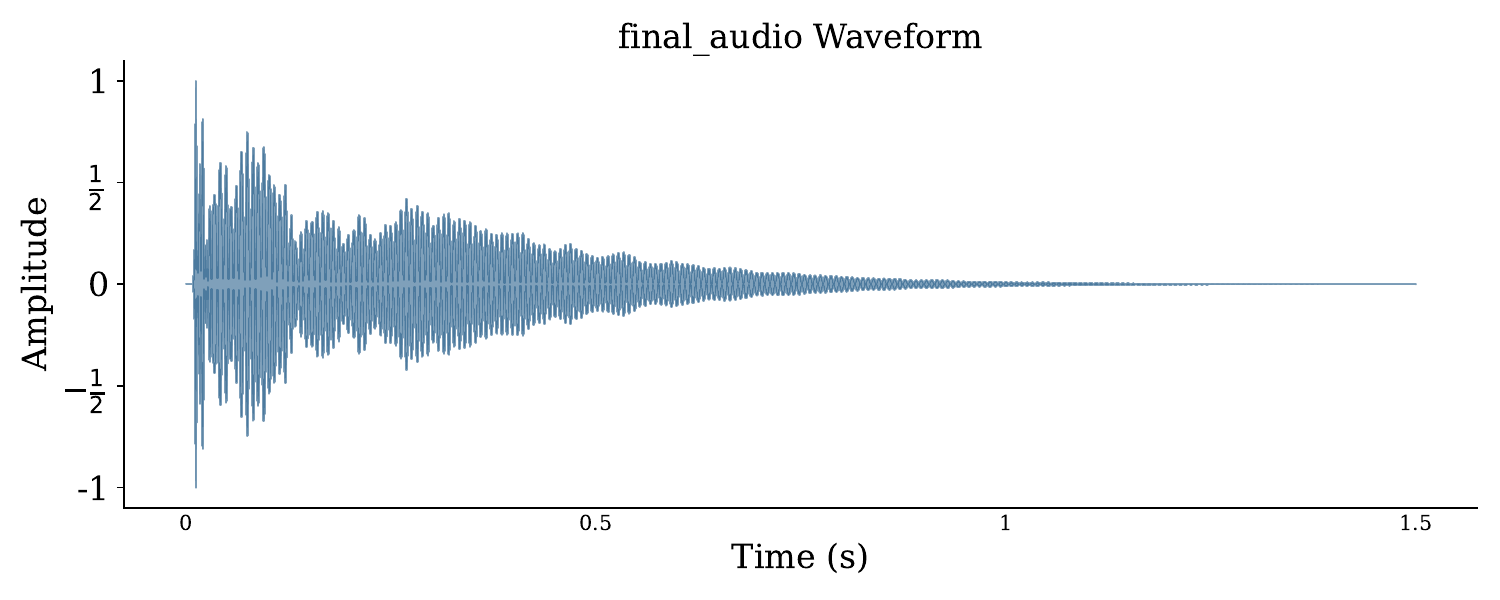}};

                \node (star) [above=of img21, node distance=0cm, xshift=-0.1cm, yshift=-1.2cm,  font=\color{black}]{\Huge{$\star$}};
                \node[right=of star, node distance=0cm, xshift=0.1cm,  yshift=-0.08cm, font=\color{black}]{\Huge{$=$}};

                \node[left=of img11, node distance=0cm, rotate=90, xshift=0.75cm, yshift=-0.9cm,  font=\color{black}]{\tiny{Obj. Impulse $\impulseObj^{\optParams}$}};
                \node[left=of img21, node distance=0cm, rotate=90, xshift=1cm, yshift=-.9cm,   font=\color{black}]{\tiny{Rev. Impulse $\impulseReverb^{\optParams}$}};
                \node[below=of img31, node distance=0cm, xshift=0.1cm, yshift=1cm,  font=\color{black}]{\tiny{Rendered Audio $\render \!=\! \renderFunc(\!\optParams\!) \!\!=\!\! \impulseObj^{\optParams} \!\!\star \!\impulseReverb^{\optParams}$}};
            \end{tikzpicture}
            \vspace{-0.03\textheight}
            \caption{
                \textbf{Impact Synthesis Overview.}
                Learned components and final audio for the impact synthesis problem (\secref{sec:method_diff_impact}) with prompt $\prompt = $ ``\texttt{kick drum, bass, reverb}''. Parameters $\optParams$ control object and reverb impulses $\impulseObj^{\optParams}$,  $\impulseReverb^{\optParams}$ (\eqref{eq:object_impulse} and \eqref{eq:reverb_impulse}), whose convolution is the rendered audio $\render = \impulseObj^{\optParams} \!\!\star \!\impulseReverb^{\optParams}$.
                The optimized output (\href{https://drive.google.com/file/d/1ekCge0x2-JSKsw3Rjm1w7F69al81qxkw/view?usp=sharing}{{\color{nvidiagreen}listen here}}) provides {\color{darkgreen}$+\num{0.1}$ CLAP} over the initialization (\href{https://drive.google.com/file/d/1j-KS2OFdtptFngSh372canj0NMc1wRtQ/view?usp=sharing}{{\color{nvidiagreen}listen here}}) illustrating improved alignment with the prompt.
            }
            \label{fig:diff_impact_overview}
            \vspace{-0.01\textheight}
        \end{figure}

        \vspace{-0.01\textheight}
        \subsubsection{Physically Informed Impact Synthesis}\label{sec:method_diff_impact}
        \vspace{-0.01\textheight}
            We now apply Audio-SDS to impact sounds, overviewed in \figref{fig:diff_impact_overview}. We use a simplified version of \citep{langlois2014eigenmode}, where the object impulse is parametrized by a sum of $\numImpulseComponent$ damped sinusoids:
            \begin{equation}\label{eq:object_impulse}
            \mathclap{
                \smash{
                \impulseObj^{\optParams}[\audioSampleIndex] = 
                \sum\nolimits_{\impulseComponentIndex=1}^\numImpulseComponent 
                    \amplitude_\impulseComponentIndex \;
                    \exp \bigl(-\decay_\impulseComponentIndex \,\audioSampleIndex\bigr) 
                    \;\cos\bigl(\frequency_\impulseComponentIndex \,\audioSampleIndex\bigr)
                }
            }
            \end{equation}
            The reverb impulse is a sum of bandpassed white noise $\mathcal{W}$:
            \begin{equation}\label{eq:reverb_impulse}:
            \mathclap{
                \smash{
                \!\!\impulseReverb^{\optParams}[\audioSampleIndex] =\! 
                \sum\nolimits_{\impulseComponentIndex=1}^\numImpulseComponent 
                    \!\tilde{\amplitude}_\impulseComponentIndex \!
                    \exp\bigl(-\tilde{\decay}_\impulseComponentIndex \,\audioSampleIndex\bigr) \;
                    \bandpassFunc(\mathcal{W}(\audioSampleIndex),\tilde{\frequency}_\impulseComponentIndex)
                }
            }
            \end{equation}
            The final waveform is the convolution of these impulses $\render \!=\! \renderFunc(\optParams) \!=\! \impulseImpact \!\star\! \impulseObj^{\optParams} \!\star\! \impulseReverb^{\optParams}$, where we choose $\impulseImpact$ to simply be a delta function. By running SDS with text prompts like ``\texttt{sharp metallic clang}'', the model guides $\optParams = \{\frequency_\impulseComponentIndex, \decay_\impulseComponentIndex,\amplitude_\impulseComponentIndex, \tilde{\frequency}_\impulseComponentIndex, \tilde{\decay}_\impulseComponentIndex, \tilde{\amplitude}_\impulseComponentIndex\}_{\impulseComponentIndex=1}^{\numImpulseComponent}$ to produce waveforms deemed plausible for that descriptor.             

        \begin{figure}[ht]
            \vspace{-0.015\textheight}
            \centering
            \scalebox{.92}{
            \begin{tikzpicture}[
              every node/.style={font=\sffamily},
              >={Latex[length=3mm]}, %
              node distance=1cm and 1cm %
            ]
            
            \node[minimum width=0.25\linewidth, minimum height=0.1\linewidth, align=center] (saxophone) {\includegraphics[width=0.22\linewidth, trim={2.5cm 1.75cm 4.7cm .95cm}, clip]{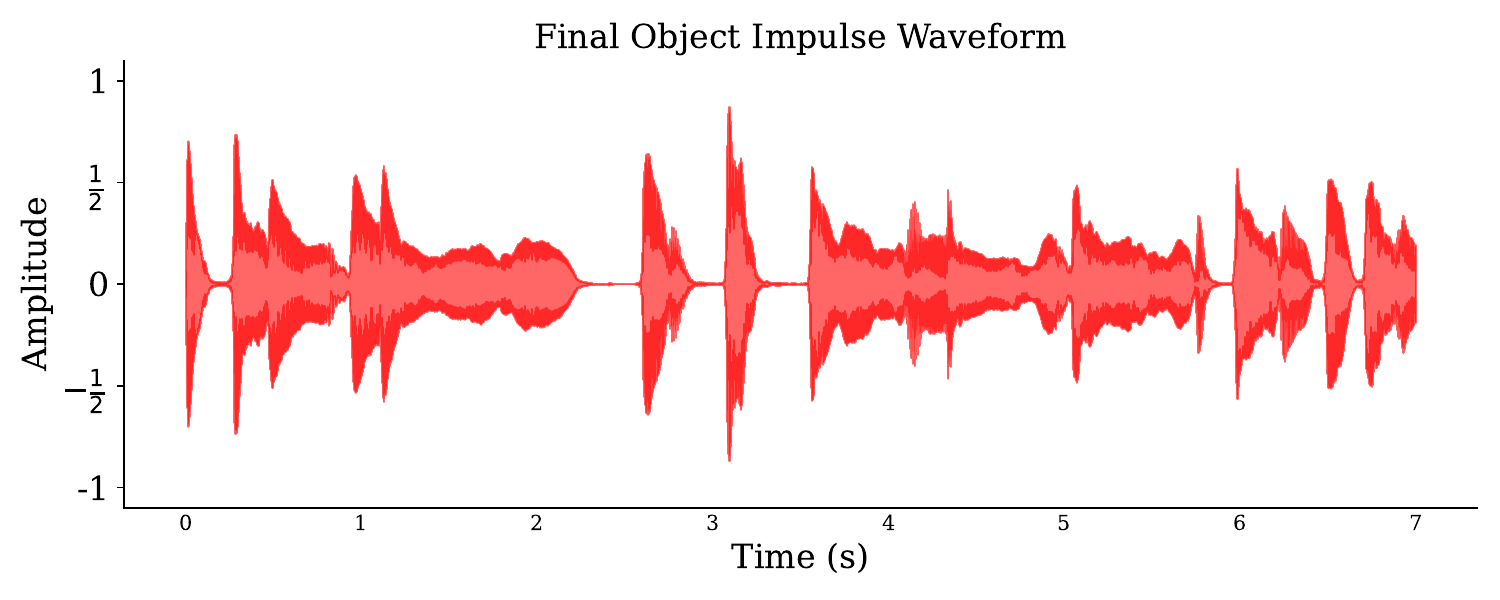}};
            \node[below=of saxophone, yshift=1.2cm, minimum width=0.25\linewidth, align=center] (saxophone_label) {\tiny{Saxophone Solo}};
            \node[below=of saxophone_label, yshift=1.2cm, minimum width=0.25\linewidth, align=center] (saxophone_label2) {\tiny{(Ground-truth source)}};

            \node[below=of saxophone_label2, yshift=0.9cm, minimum width=0.25\linewidth, align=center] (groundtruth_label) {\tiny{Note: Ground-truth audio}};
            \node[below=of groundtruth_label, yshift=1.2cm, minimum width=0.25\linewidth, align=center] (groundtruth_label2) {\tiny{sources are not available in}};
            \node[below=of groundtruth_label2, yshift=1.2cm, minimum width=0.25\linewidth, align=center] (groundtruth_label3) {\tiny{real-world recordings}};
            
            \node[below=of saxophone, yshift=-1.0cm, minimum width=0.25\linewidth, minimum height=0.1\linewidth, align=center] (traffic) {\includegraphics[width=0.22\linewidth, trim={2.5cm 1.75cm 4.7cm .95cm}, clip]{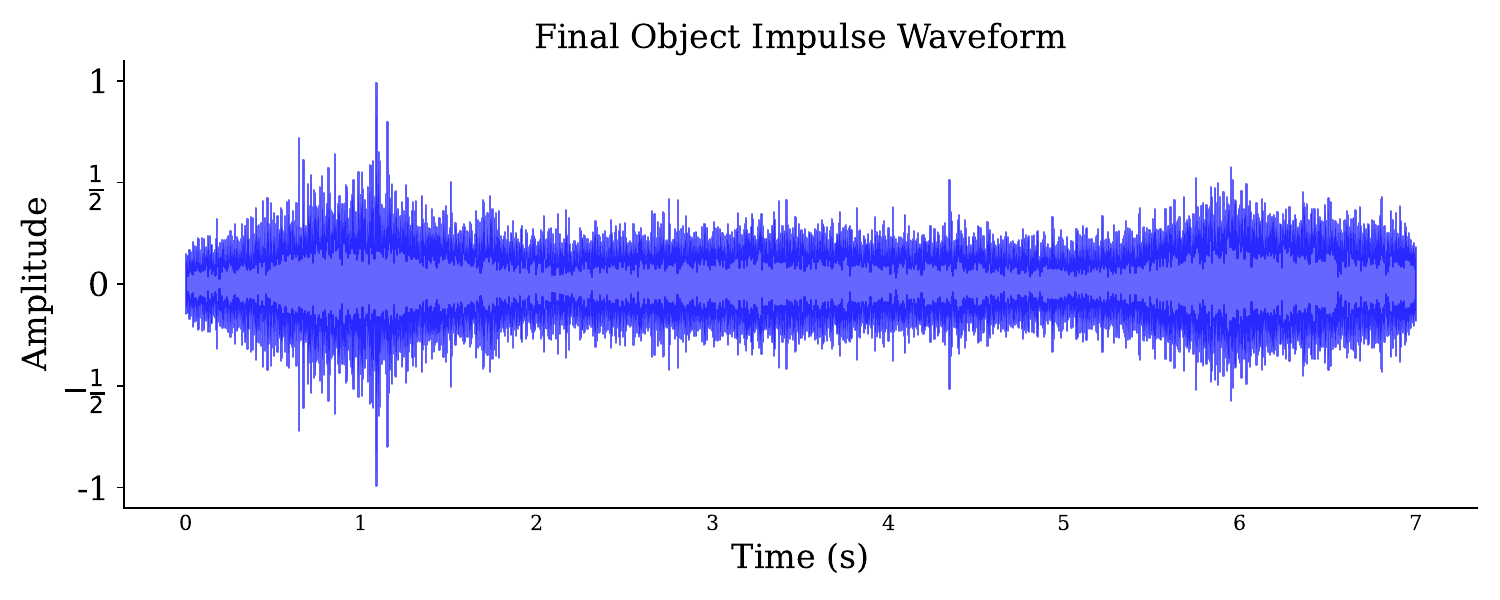}};
            \node[below=of traffic, yshift=1.2cm, minimum width=0.25\linewidth, align=center] (traffic_label) {\tiny{Traffic Noise}};
            \node[below=of traffic_label, yshift=1.2cm, minimum width=0.25\linewidth, align=center] (traffic_label2) {\tiny{(Ground-truth source)}};
            
            \node[right=of $(saxophone)!0.5!(traffic)$, xshift=1cm, minimum width=0.25\linewidth, minimum height=0.1\linewidth, align=center] (mixture) {\includegraphics[width=0.22\linewidth, trim={2.5cm 1.9cm 4.7cm .95cm}, clip]{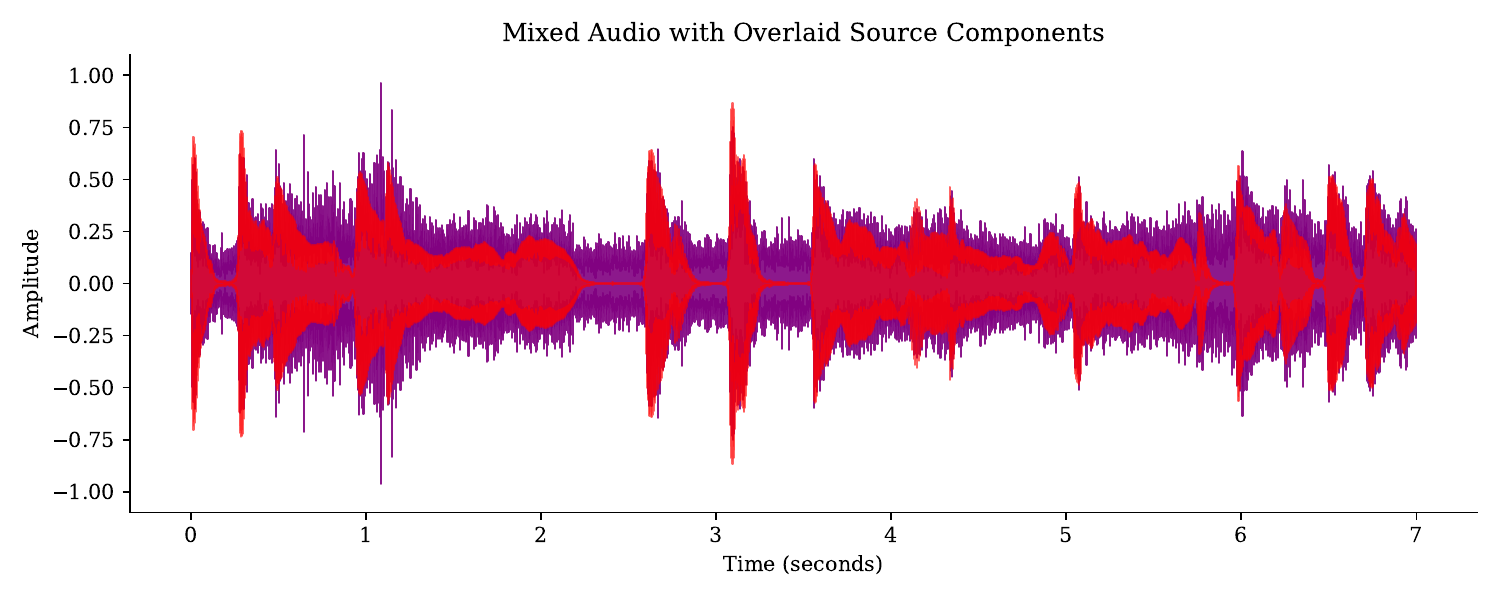}};

            \node[above=of mixture, xshift=-.0cm, yshift=-.2cm, minimum width=0.25\linewidth, align=center] (mixture_text_line1) {\tiny{Our provided audio $\targetAudio$ is a mixture}};
            \node[below=of mixture_text_line1, yshift=1.2cm, minimum width=0.25\linewidth, align=center] (mixture_text_line2) {\tiny{of various (unknown) sources, which}};
            \node[below=of mixture_text_line2, yshift=1.2cm, minimum width=0.25\linewidth, align=center] (mixture_text_line3) {\tiny{we want to separate}};
            
            \node[below=of mixture, xshift=-.0cm, yshift=1.0cm, minimum width=0.25\linewidth, align=center] (separation_text_line1) {\tiny{Separate $\targetAudio$ by training with}};
            \node[below=of separation_text_line1, yshift=1.2cm, minimum width=0.25\linewidth, align=center] (separation_text_line2) {\tiny{mix of reconstruction loss and}};
            \node[below=of separation_text_line2, yshift=1.2cm, minimum width=0.25\linewidth, align=center] (separation_text_line3) {\tiny{Audio-SDS with prompt $\smash{\prompt_{\sourceIndex}}$}};
            \node[below=of separation_text_line3, yshift=1.2cm, minimum width=0.25\linewidth, align=center] (separation_text_line4) {\tiny{per-component $\sourceIndex$ as in Eq.~\eqref{eq:separation_combined_loss}}};
            \node[below=of separation_text_line4, yshift=1.2cm, minimum width=0.25\linewidth, align=center] (separation_text_line5) {\tiny{($\totalNumSources=2$ here)}};
            
            \node[right=of saxophone, xshift=3cm, minimum width=0.25\linewidth, minimum height=0.1\linewidth, align=center] (saxophone_out) {\includegraphics[width=0.22\linewidth, trim={2.5cm 1.75cm 4.7cm .95cm}, clip]{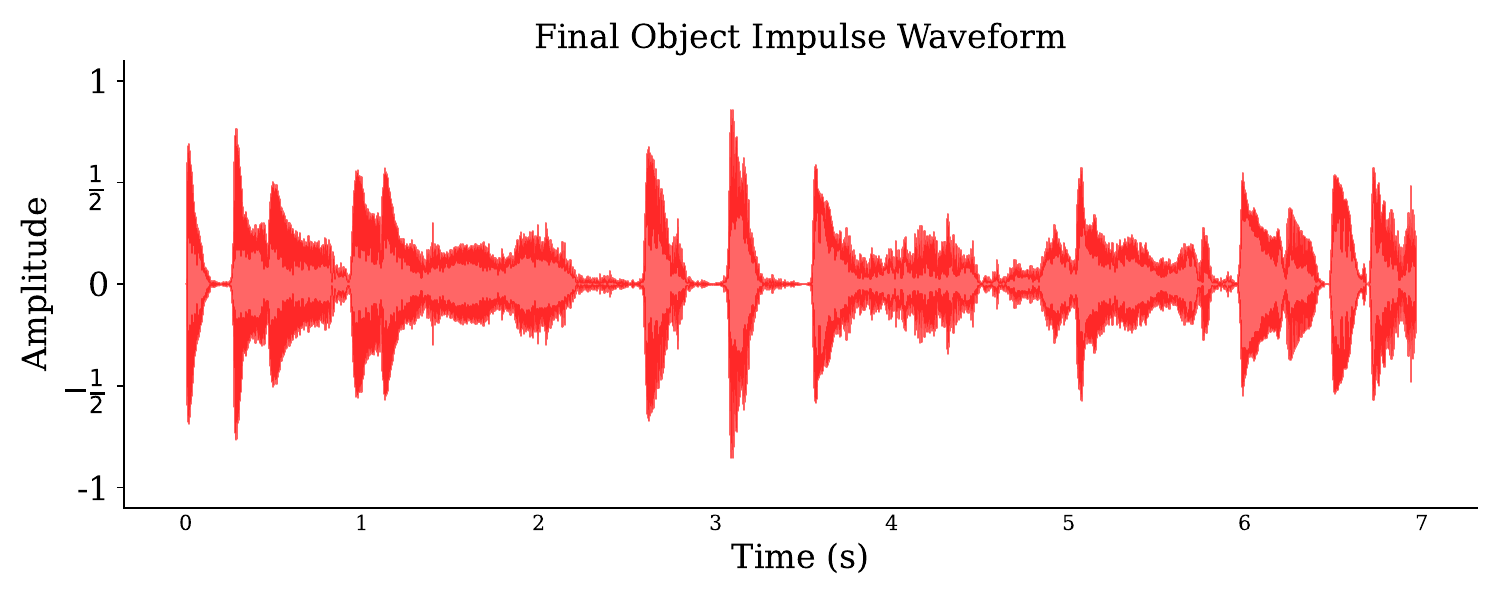}};
            \node[below=of saxophone_out, yshift=1.2cm, minimum width=0.25\linewidth, align=center] (saxophone_text_line1) {\tiny{``\texttt{sax playing melody,}}};
            \node[below=of saxophone_text_line1, yshift=1.2cm, minimum width=0.25\linewidth, align=center] (saxophone_text_line2) {\tiny{\texttt{jazzy, modal,}}};
            \node[below=of saxophone_text_line2, yshift=1.2cm, minimum width=0.25\linewidth, align=center] (saxophone_text_line3) {\tiny{\texttt{interchange, post bop}''}};
            
            \node[below=of saxophone_out, yshift=-1.0cm, minimum width=0.25\linewidth, align=center] (traffic_out) {\includegraphics[width=0.22\linewidth, trim={2.5cm 1.75cm 4.7cm .95cm}, clip]{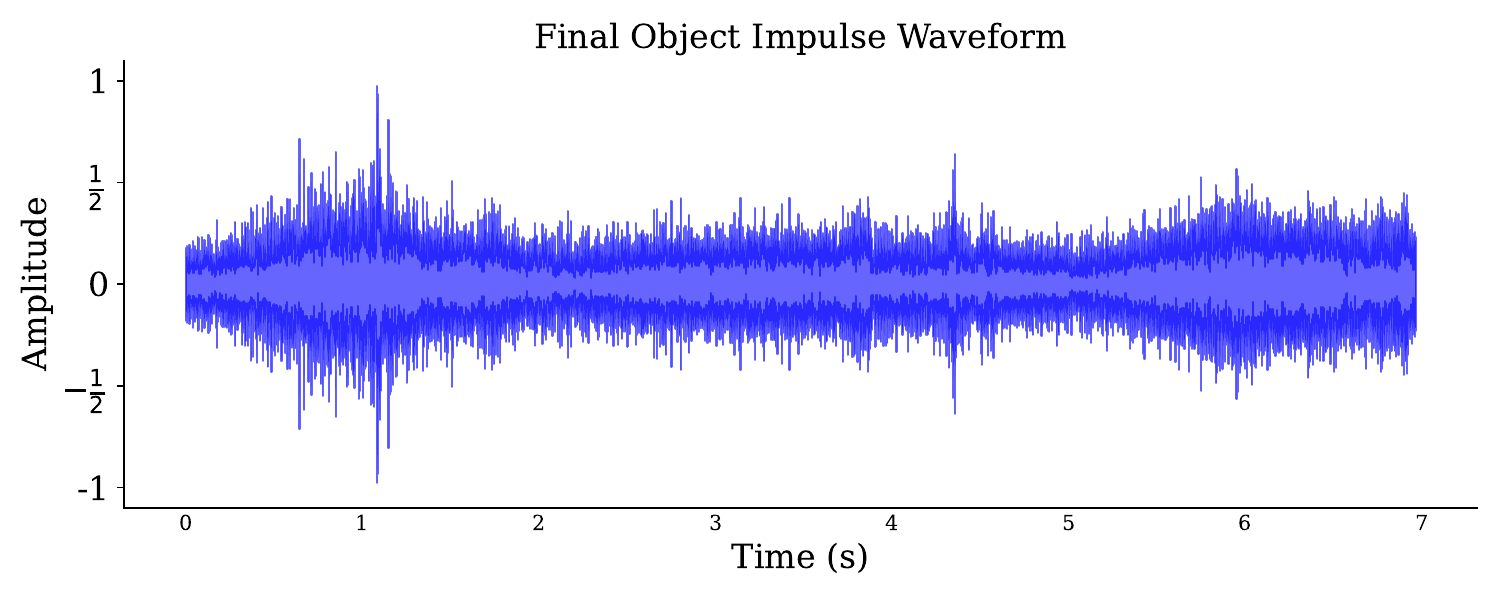}};
            \node[below=of traffic_out, yshift=1.2cm, minimum width=0.25\linewidth, align=center] (traffic_text_line1) {\tiny{``\texttt{cars passing by on}}};
            \node[below=of traffic_text_line1, yshift=1.2cm, minimum width=0.25\linewidth, align=center] (traffic_text_line2) {\tiny{\texttt{ a busy street,}}};
            \node[below=of traffic_text_line2, yshift=1.2cm, minimum width=0.25\linewidth, align=center] (traffic_text_line3) {\tiny{\texttt{traffic, road noise}''}};
            
            \draw[thick, ->] (saxophone.east) -- (mixture.west);
            \draw[thick, ->] (traffic.east) -- (mixture.west);
            \draw[thick, ->] (mixture.east) -- (saxophone_out.west);
            \draw[thick, ->] (mixture.east) -- (traffic_out.west);
            
            \end{tikzpicture}
            }
            \vspace{-0.035\textheight}
            \caption{
                \textbf{Prompt-Driven Source Separation.} 
                Decomposing a single audio mixture of a saxophone and traffic noise into two separate waveforms, each guided by its text prompt. The sum of these separated waveforms reconstructs the original mixture.
                Audio links:
                    \href{https://drive.google.com/file/d/1rcfa1Er4TGLcXvG7cdPyn0uVrNF4fKaN/view?usp=sharing}{{\color{nvidiagreen}sax source}}, \href{https://drive.google.com/file/d/1UNJp4--Ceypc8lxy6EMameyUpCZ_zOoG/view?usp=sharing}{{\color{nvidiagreen}traffic source}},
                    \href{https://drive.google.com/file/d/1GW-DEzex9tj73BeF1nnmj4qU8iiKhsyU/view?usp=sharing}{{\color{nvidiagreen}mixed audio $\targetAudio$}},
                    \href{https://drive.google.com/file/d/1iF1B7_Mg6zFqMMfI2IfYnApjJBgIi1N5/view?usp=sharing}{{\color{nvidiagreen}recovered ``\texttt{sax}\dots''}},
                    \href{https://drive.google.com/file/d/197MVw23v2a3fbaK0-1BWecd6P-el5Jev/view?usp=sharing}{{\color{nvidiagreen}recovered ``\texttt{cars}\dots''}},
                    \href{https://drive.google.com/file/d/1D0tE30sxPDTK3HguzSqE93YyVqW0PQW0/view?usp=sharing}{{\color{nvidiagreen}reconstructed mixed audio}}.
            }
            \label{fig:source_separation_overview}
            \vspace{-0.01\textheight}
        \end{figure}
        \subsubsection{Source Separation}\label{sec:method_source_separation}
             Overviewed in \figref{fig:source_separation_overview}, we tackle prompt-guided source separation as a key application of our Audio-SDS framework, where a user seeks to disentangle a mixed audio signal into separate, discrete audio sources with different semantic meanings. For example, disentangling a real-world recording of a saxophone player on a busy street into the saxophone audio and traffic audio. We also propose a method to automate the source decomposition for users. 
            
            Specifically, we assume a single mixture signal $\targetAudio$ composed of $\totalNumSources$ unknown sources and a set of text prompts $\prompt_k$ for $k=1\dots\totalNumSources$, where each prompt semantically describes a source. Denoting each source's waveform $\render_{\sourceIndex} = \renderFunc(\optParams_{\sourceIndex})$ with parameters $\optParams_{\sourceIndex}$, the mixture should (approximately) satisfy:
            \begin{equation}\label{eq:mixture_equation}
            \mathclap{
                \targetAudio \approx \smash{\sum\nolimits_{\sourceIndex=1}^{\totalNumSources}} \render_{\sourceIndex} = \smash{\sum\nolimits_{\sourceIndex=1}^{\totalNumSources}} \renderFunc_{\sourceIndex}(\optParams_{\sourceIndex}) = \renderFunc(\optParams)
            }
            \end{equation}
            Since~\eqref{eq:mixture_equation} is underdetermined, we will regularize the decomposition via text prompts $\{\prompt_{\sourceIndex}\}_{\sourceIndex=1}^\totalNumSources$$\{\render_{\sourceIndex}\}_{\sourceIndex=1}^\totalNumSources$, one for each source (e.g., ``\texttt{saxophone}'', ``\texttt{traffic}'', ``\texttt{rain}'').
            We first impose that the sum of the separated sources recovers $\targetAudio$. We define the reconstruction loss:
            \begin{equation} \label{eq:l_recons}
            \mathclap{
                \smash{
                \!\!\!\!\!\!\!\!\lossRecons\!(\optParams) \!=\! \sum\nolimits_{\STFTIndex=1}^{\totalNumSTFT} \!\Big| \STFT_{\STFTIndex}(\targetAudio) \!-\! \STFT_{\STFTIndex}\Big(\!\sum\nolimits_{\sourceIndex=1}^{\totalNumSources} \!\renderFunc_{\sourceIndex}(\optParams_{\sourceIndex})\!\Big) \!\Big|_{2}^{2}
                }
            }
            \end{equation}                
            using the set of STFT magnitudes $\{\STFT_\STFTIndex\}_{\STFTIndex = 1}^{ \totalNumSTFT}$ described in \secref{sec:method_spectrogram_updates}. 
            Informally, we seek solutions $\optParams^{*}$ (approximately) minimizing reconstruction loss while getting high probability for their prompt via an audio diffusion model:
            \begin{align*}
                \smash{\optParams^{*}} &= \smash{\{\optParams_{\sourceIndex}^{*}\}_{\sourceIndex=1}^{\totalNumSources} \in \argmin\nolimits_{\optParams} \lossRecons(\optParams),}\\
                \textnormal{where render } \render^{*}_{\sourceIndex} &= \renderFunc(\optParams_{\sourceIndex}^{*}) \textnormal{ high-prob. under generative}\\
                \textnormal{model } &p_{\diffusionParams}(\cdot | \prompt_{\sourceIndex}) \textnormal{ for prompts } \sourceIndex=1\dots\totalNumSources
            \end{align*}
            We find such solutions by using updates combining the reconstruction gradient and SDS update in a weighted sum:
            \begin{equation}\label{eq:separation_combined_loss}
            \mathclap{
                \smash{
                \!\!\!\!\!\!\!\!\!\!\!\updateSourceSep\!(\!\optParams; \!\{\prompt_{\sourceIndex}\}_{\sourceIndex=1}^{\totalNumSources}\!) \!=\! \nabla_{\!\optParams}\!\lossRecons\!(\!\optParams\!) \!+ \lossScaleSourceSep\!\! \sum\nolimits_{\sourceIndex=1}^{\totalNumSources}\! \updateSDS^{\STFT\!, \decodeFunc}\!(\!\optParams_{\sourceIndex};\! \prompt_{\sourceIndex}\!)
                }
            }
            \end{equation}
            Where $\lossScaleSourceSep \!>\! 0 $ trades off between exact reconstruction (i.e., matching $\targetAudio$) and semantic alignment with prompts $\prompt_{\sourceIndex}$, minimized by gradient descent on $\optParams \!=\! \{\optParams_{\sourceIndex}\}$. This encourages the sources to sum to $\targetAudio$ while aligning with their prompt. In practice, $\lossScaleSourceSep$ is a hyperparameter set sufficiently small so the reconstructions are high-quality throughout optimization while progressing on the SDS objective.

            \textbf{Optimizing Latents vs.\ Waveforms:}
                When using a latent diffusion model (e.g., Stable Audio Open), each source can be parametrized directly in latent space $\latentRender_\sourceIndex \in \stableAudioLatentDomain$, significantly reducing dimensionality relative to raw waveforms and often accelerating convergence. We apply SDS updates in the latent domain, while the reconstruction loss \eqref{eq:l_recons} is computed in audio space by decoding $\latentRender_\sourceIndex$ through $\decodeFunc$. Optimizing waveforms instead of latents is easier to perfectly reconstruct the mixture (e.g., by adding residual error each iteration), but requires differentiating through the encoder for SDS updates, which we found brittle.

            \textbf{Automating Source Decomposition with LLMs and Audio-Captioning models:}
                The preceding formulation introduces the set of $\totalNumSources$ prompts as a tool for the user to use, which can be straightforward to select for simple audio. Yet, for more complex audio or users with less expertise, we aim to generate different potential relevant source decompositions for the user automatically. For this, we propose using a captioning model on our audio $\targetAudio$ to get a textual description to give to an LLM, which suggests potential prompt sets given the caption, detailed in App. \secref{app:sec_method_automating_source_decomposition}.

\vspace{-0.01\textheight}
\section{Experiments}\label{sec:experiments}
\vspace{-0.01\textheight}
    We now present quantitative and qualitative results demonstrating the effectiveness of our proposed Audio-SDS framework on three audio tasks: (1) FM synthesis, (2) impact synthesis, and (3) source separation. 
    Our general experimental setup is explained in \secref{sec:exp_setup}, including metrics and baseline comparisons. We then dive into results in \secref{sec:exp_fm_synthesis}--\ref{sec:exp_source_separation}. We conclude with ablations in \secref{sec:exp_ablation}.
    
    \vspace{-0.01\textheight}
    \subsection{Experimental Setup}\label{sec:exp_setup}
    \vspace{-0.01\textheight}
        
        \textbf{Reproducibility Details:}
            We include a summary of our supplementary files in App. \secref{sec:reproducibility_details}, full implementation details in App. \secref{sec:app_implementation}, and complete experimental details in App. \secref{sec:app_experimental_details}, including compute requirements (\secref{sec:app_compute}), all hyperparameters (\secref{sec:app_hyperparameters}), and additional ablations (\secref{sec:app_ablations}). We summarize only the relevant details to understand our experimental results.

        \vspace{-0.01\textheight}
        \textbf{Pretrained Audio Diffusion Model:} 
            We use the publicly available Stable Audio Open checkpoint~\citep{evans2024stable}, which uses a latent diffusion architecture, as this is the only sufficient quality open-source text-to-audio diffusion model for general audio. The diffusion parameters are frozen, and we do not differentiate the score network itself (see \secref{sec:audio_sds_method}). 

        \vspace{-0.01\textheight}
        \textbf{Metrics and Evaluations:}
            We report both subjective (listening tests) and objective metrics:
                (1) \emph{CLAP Score:} We measure audio-text alignment using the Contrastive Language-Audio Pretraining (CLAP) model~\citep{elizalde2023clap}. Higher values indicate better agreement with the given prompt. 
                (2) \emph{Distance to Ground Truth:} For tasks with a known ground-truth (e.g., synthetic source separation), we measure the $\ell_2$ error or spectral loss (\ref{sec:method_spectrogram_updates}) between the recovered signal and ground truth. 
                (3) \emph{Source Separation Metrics:} For source separation with a ground truth, we also report (scale-invariant) Signal-to-Distortion Ratio (SDR)~\citep{vincent2006performance} to quantify separation performance.
                (4) \emph{Qualitative Results:} Results are included in the supplementary for qualitative comparison.
            We do not use distribution-based or user-study-centric metrics, like Fréchet Audio Distance (FAD) or large-scale opinion-score tests, because they are prohibitively expensive when multiple text prompts must be tested repeatedly.

        \vspace{-0.01\textheight}
        \textbf{Baselines:}
            Neither FM nor impact sound generators have large (or any) curated datasets. There are no other specialized, pretrained baselines that can be adapted to general text-guided synthesis without extensive re-training on domain-specific data.
            For source separation, we emphasize that classical methods (e.g., DiffSep~\citep{scheibler2023diffusion}, ConvTasNet~\citep{luo2019conv}) focus on specific separation tasks and require either domain-specific training or domain labels. In contrast, our pipeline uses the same pretrained text-to-audio diffusion for all tasks, making direct numerical comparisons less straightforward.

    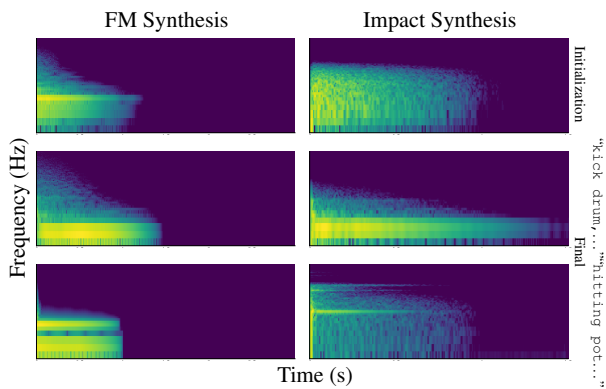
\begin{figure}[t]
        \vspace{-0.01\textheight}
        \centering
        \scalebox{.94}{
        \begin{tikzpicture}
        \centering
            \node (img11){\includegraphics[trim={2.25cm 1.45cm 4.7cm 2.0cm}, clip, width=.45\linewidth]{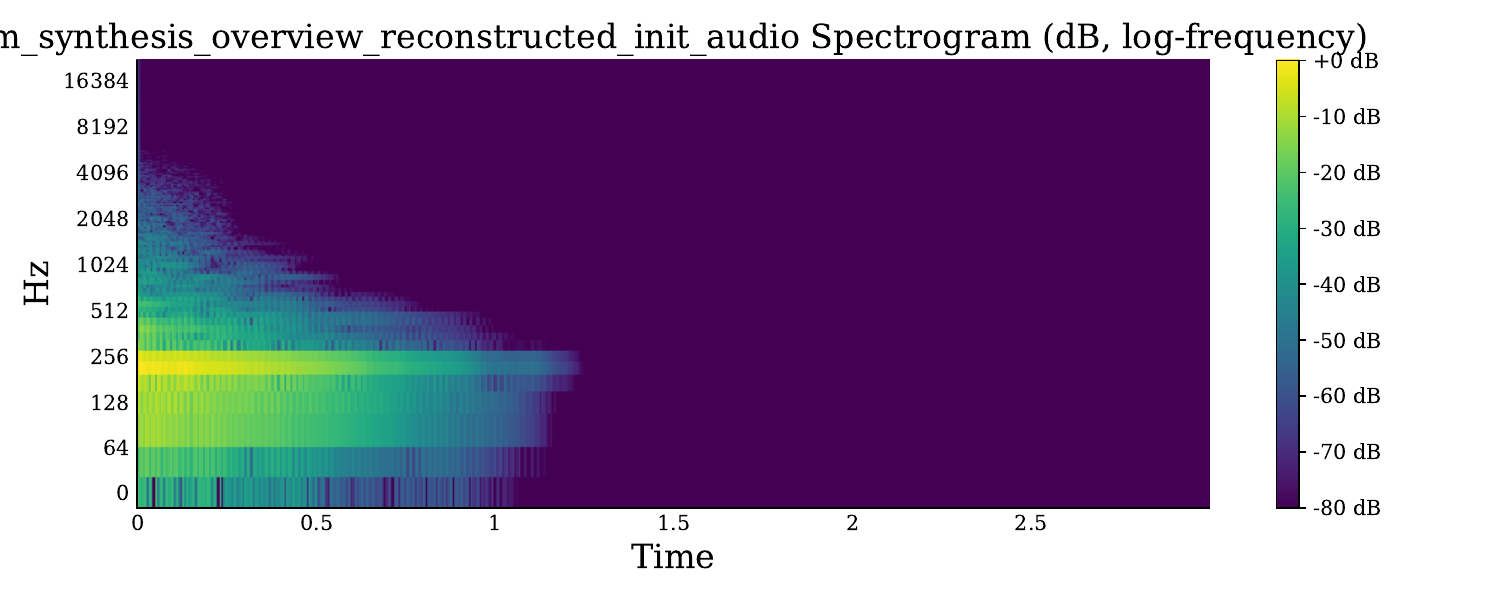}};
            \node (img21)[below=of img11, yshift=1.0cm]{\includegraphics[trim={2.25cm 1.45cm 4.7cm 2.0cm}, clip, width=.45\linewidth]{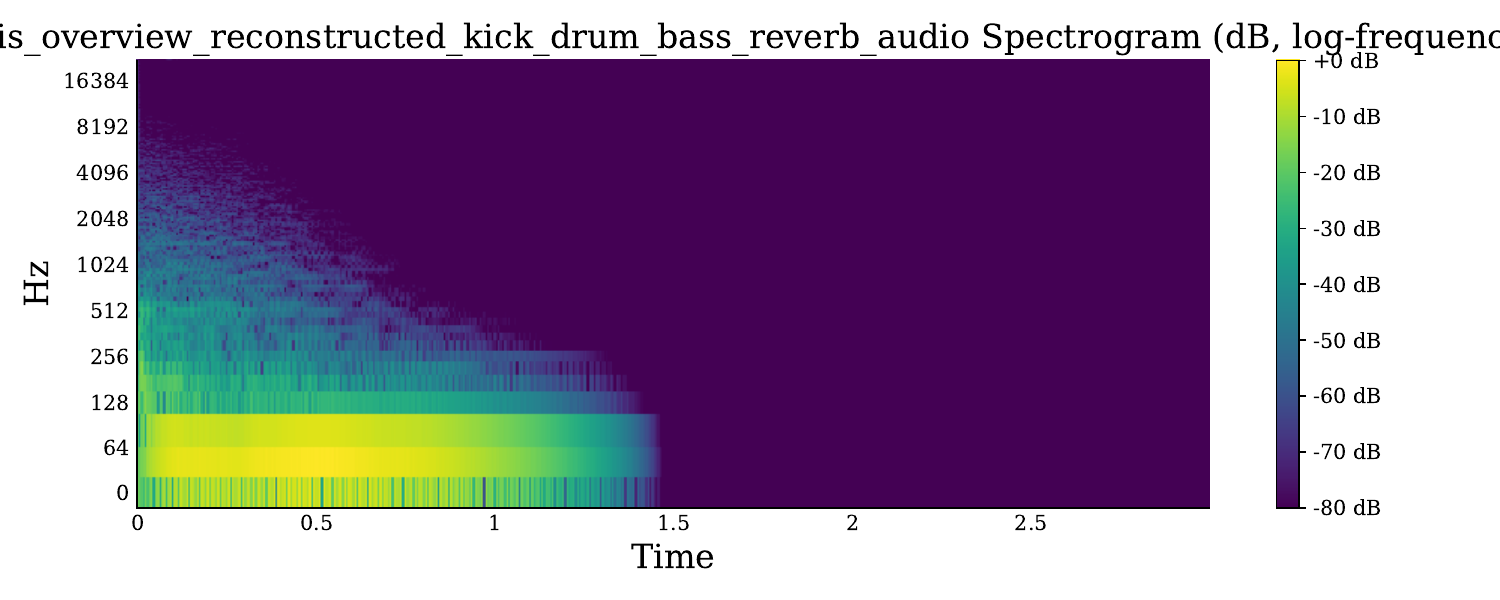}};
            \node (img31)[below=of img21, yshift=1.0cm]{\includegraphics[trim={2.25cm 1.45cm 4.7cm 2.0cm}, clip, width=.45\linewidth]{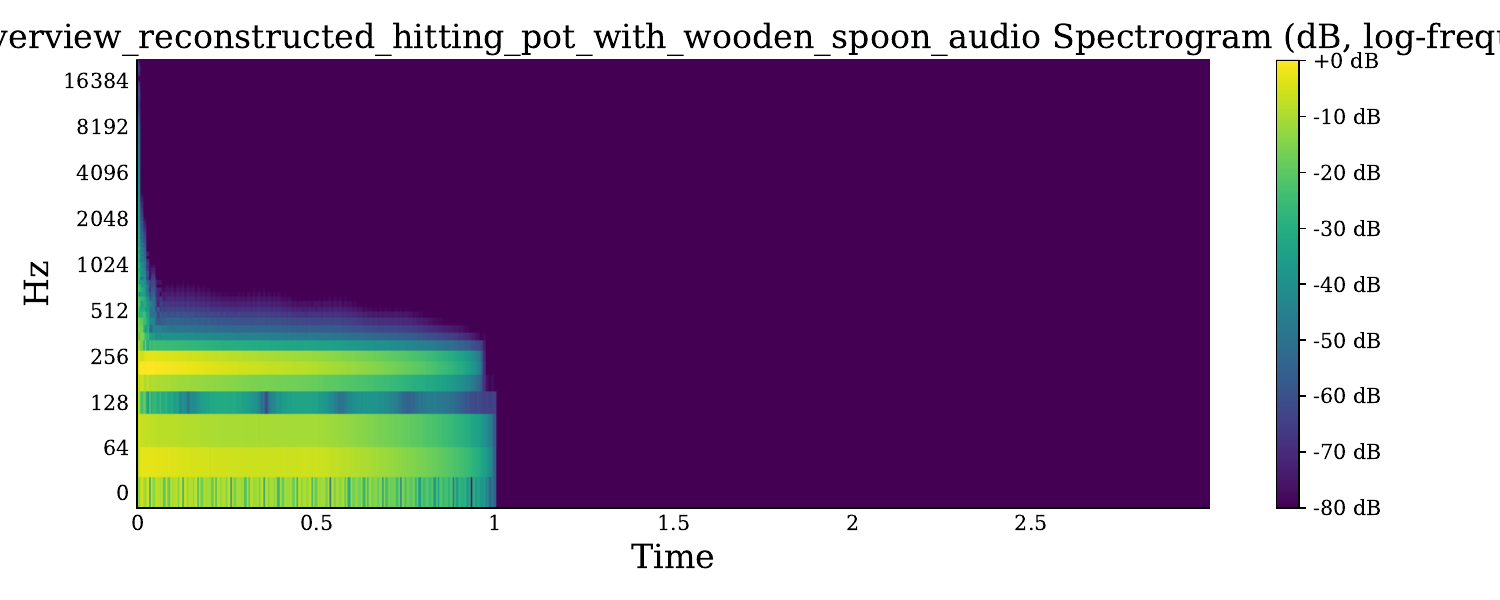}};

            \node (img12)[right=of img11, xshift=-1.1cm] {\includegraphics[trim={2.25cm 1.45cm 4.7cm 2.0cm}, clip, width=.45\linewidth]{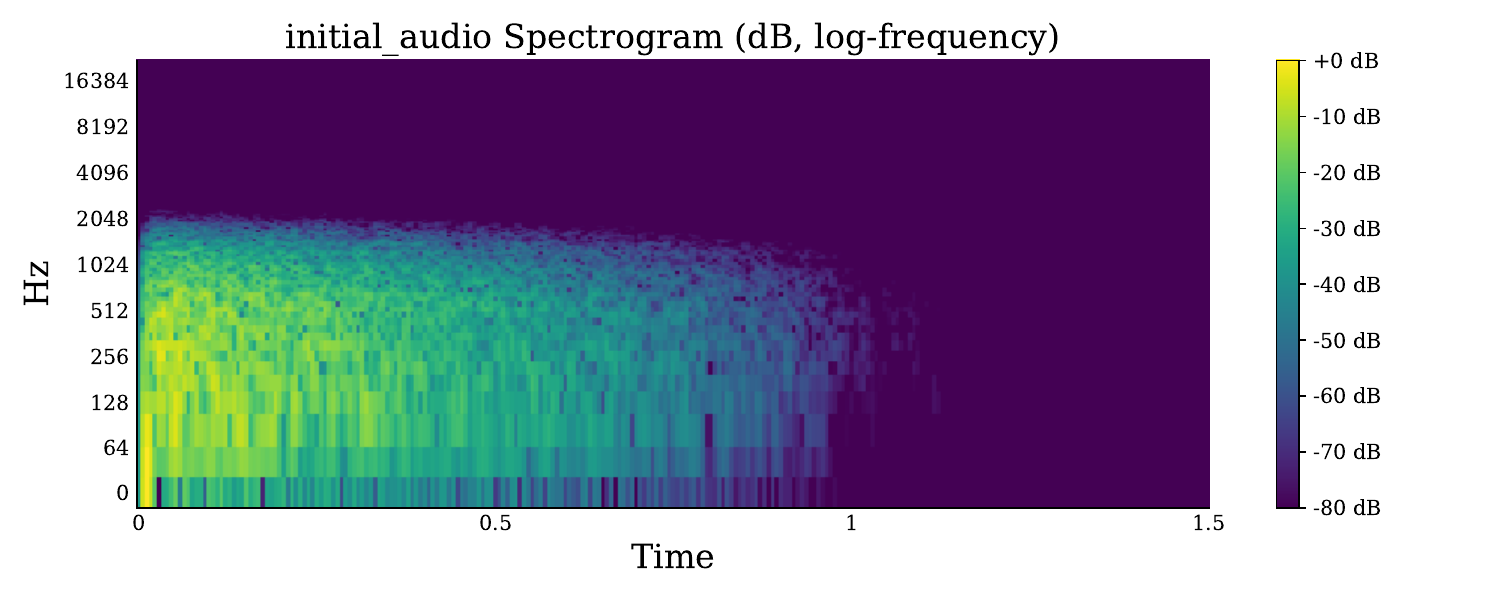}};
            \node (img22)[below=of img12, yshift=1.0cm]{\includegraphics[trim={2.25cm 1.45cm 4.7cm 2.0cm}, clip, width=.45\linewidth]{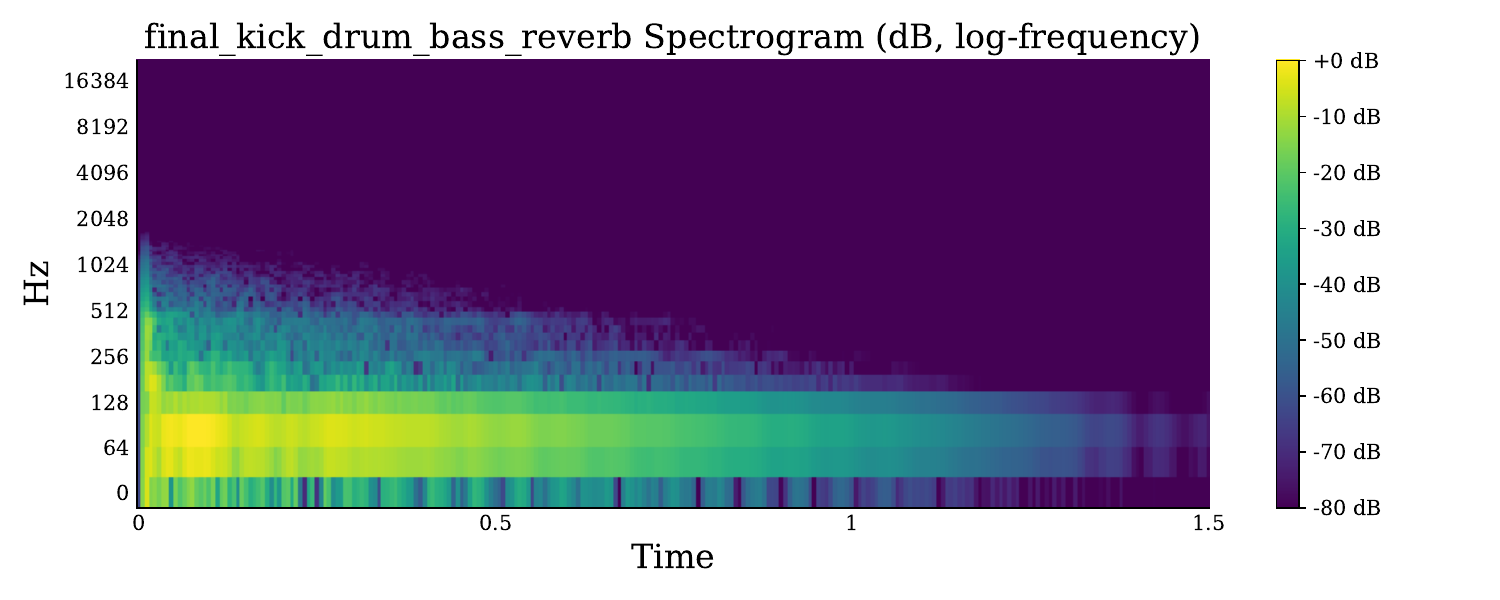}};
            \node (img32)[below=of img22, yshift=1.0cm]{\includegraphics[trim={2.25cm 1.45cm 4.7cm 2.0cm}, clip, width=.45\linewidth]{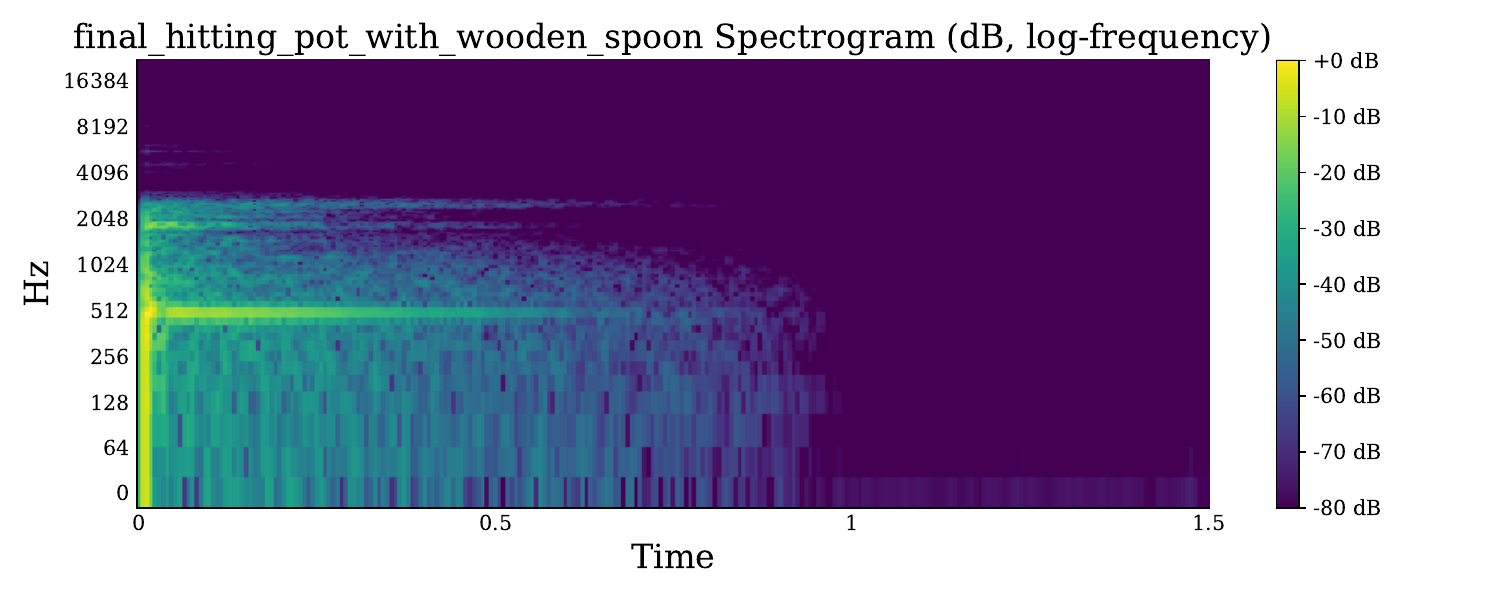}};

            \node[right=of img12, node distance=0cm, xshift=-1cm, yshift=.7cm, rotate=270, font=\color{black}]{\tiny{Initialization}};
            
            \node[right=of img22, node distance=0cm, xshift=-1cm, yshift=-.4cm, rotate=270, font=\color{black}]{\tiny{Final}};
            \node[right=of img22, node distance=0cm, xshift=-.75cm, yshift=1.0cm, rotate=270, font=\color{black}]{\tiny{``\texttt{kick drum,}\dots''}};
            \node[right=of img32, node distance=0cm, xshift=-.75cm, yshift=.9cm, rotate=270, font=\color{black}]{\tiny{``\texttt{hitting pot}\dots''}};

            \node[left=of img21, node distance=0cm, rotate=90, xshift=1.0cm, yshift=-.9cm,  font=\color{black}]{\footnotesize{Frequency (Hz)}};
            \node[below=of img31, node distance=0cm, xshift=2.1cm, yshift=1.15cm,  font=\color{black}]{\footnotesize{Time (s)}};

            \node[above=of img11, node distance=0cm, xshift=0cm, yshift=-1.15cm,  font=\color{black}]{\footnotesize{FM Synthesis}};
            \node[above=of img12, node distance=0cm, xshift=0cm, yshift=-1.15cm,  font=\color{black}]{\footnotesize{Impact Synthesis}};
        \end{tikzpicture}
        }
        \vspace{-0.03\textheight}
        \caption{
            \textbf{FM and Impact Synthesis: Qualitative Results.} Spectrograms of the initialization and final result after optimization for two prompts. %
            \emph{Takeaway:} Outputs separate into distinct results according to prompts, reflecting quantitative results (\figref{fig:impact_synthesis_clap_vs_time}, end of caption). FM Synthesis fits the ``\texttt{kick drum}\dots'', but fails for more challenging  ``\texttt{hitting pot}\dots''. However, the more complex impact synthesis fits both.
            Audio links:
                \href{https://drive.google.com/file/d/11JMwLdoQDhML08GbuKjlaCIjamVM1eDa/view?usp=sharing}{{\color{nvidiagreen}init. FM}},
                \href{https://drive.google.com/file/d/1j-KS2OFdtptFngSh372canj0NMc1wRtQ/view?usp=sharing}{{\color{nvidiagreen}init. impact}}, 
                \href{https://drive.google.com/file/d/182ecHVg32w641PLL3XFoQHKBn8Z_BdJW/view?usp=sharing}{{\color{nvidiagreen}final FM ``\texttt{kick drum, bass, reverb}''}} ({\color{darkgreen} $+\num{0.13}$ CLAP} vs. init.),
                \href{https://drive.google.com/file/d/192gAAiXWN3d7dxbzcZwTG68FYGUyh0qz/view?usp=sharing}{{\color{nvidiagreen}final FM ``\texttt{hitting pot with wooden spoon}''}} ({\color{black} $+\num{0.01}$ CLAP} vs. init.)
                \href{https://drive.google.com/file/d/1ekCge0x2-JSKsw3Rjm1w7F69al81qxkw/view?usp=sharing}{{\color{nvidiagreen}final impact ``\texttt{kick drum,}\dots''}} ({\color{darkgreen} $+\num{0.10}$ CLAP} vs. init., {\color{black} $-\num{0.01}$ CLAP} vs. FM), 
                \href{https://drive.google.com/file/d/1zeMHLsd8e-EFUg1beLfZNQxrUBf0DdH7/view?usp=sharing}{{\color{nvidiagreen}final impact ``\texttt{hitting pot}\dots''}} ({\color{darkgreen} $+\num{0.18}$ CLAP} vs. init., {\color{darkgreen} $+\num{0.30}$ CLAP} vs. FM).
        }
        \label{fig:fm_and_impact_synthesis_overview_qualitative}
        \vspace{-0.03\textheight}
    \end{figure}
        
    \subsection{FM Synthesis}\label{sec:exp_fm_synthesis}
    \vspace{-0.01\textheight}
        First, we implement a simple FM synthesizer (see \secref{sec:method_synthesizer}) for text-based parameter tuning as a proof of concept. A simple attack/decay envelope parametrizes the synthesizer for each oscillator, a frequency modulation matrix encoding each oscillator's modulation on the others, which are rendered to produce an audio waveform. Our SDS update scheme then adjusts these parameters to better align with a user-specified text prompt.
        We experiment with in-domain prompts that are amenable to FM synthesis (e.g., ``\texttt{kick drum, bass, reverb}''), as well as out-of-domain prompts which are more difficult for FM synthesis (e.g., ``\texttt{hitting pot with a wooden spoon}'').
        
        \textbf{Results:} 
            Figure~\ref{fig:fm_and_impact_synthesis_overview_qualitative} shows an example of the synthesized spectrogram and audio. Audio-SDS successfully tunes the FM parameters for a text-consistent result on in-domain prompts, improving CLAP. Our optimization enhances the generated audio's prompt alignment. However, this toy model suffers from more complex prompts, motivating the more sophisticated models in the following section.

    \vspace{-0.01\textheight}
    \subsection{Physically Informed Impact Synthesis}\label{sec:exp_impact_synthesis}
    \vspace{-0.01\textheight}
        Next, we consider a differentiable impact sound simulator based on a modal resonator model (\secref{sec:method_diff_impact}), which includes $\numImpulseComponent$ damped sinusoids for direct impact and additional bandpassed noise for reverberation. We initialize frequencies and damping coefficients randomly and optimize via SDS to align with the textual description while preserving physically plausible decay patterns.
        We test impact-oriented prompts like ``\texttt{Kick drum, bass, reverb}'', ``\texttt{Hitting pot with wooden spoon}'', and ``\texttt{Bass drum, low pitched}''.
        
\begin{figure}[t]
            \vspace{-0.01\textheight}
            \centering
            \begin{tikzpicture}
            \centering
                \node (img11){\includegraphics[trim={.9cm 1.0cm 1.25cm 2.0cm}, clip, width=.95\linewidth]{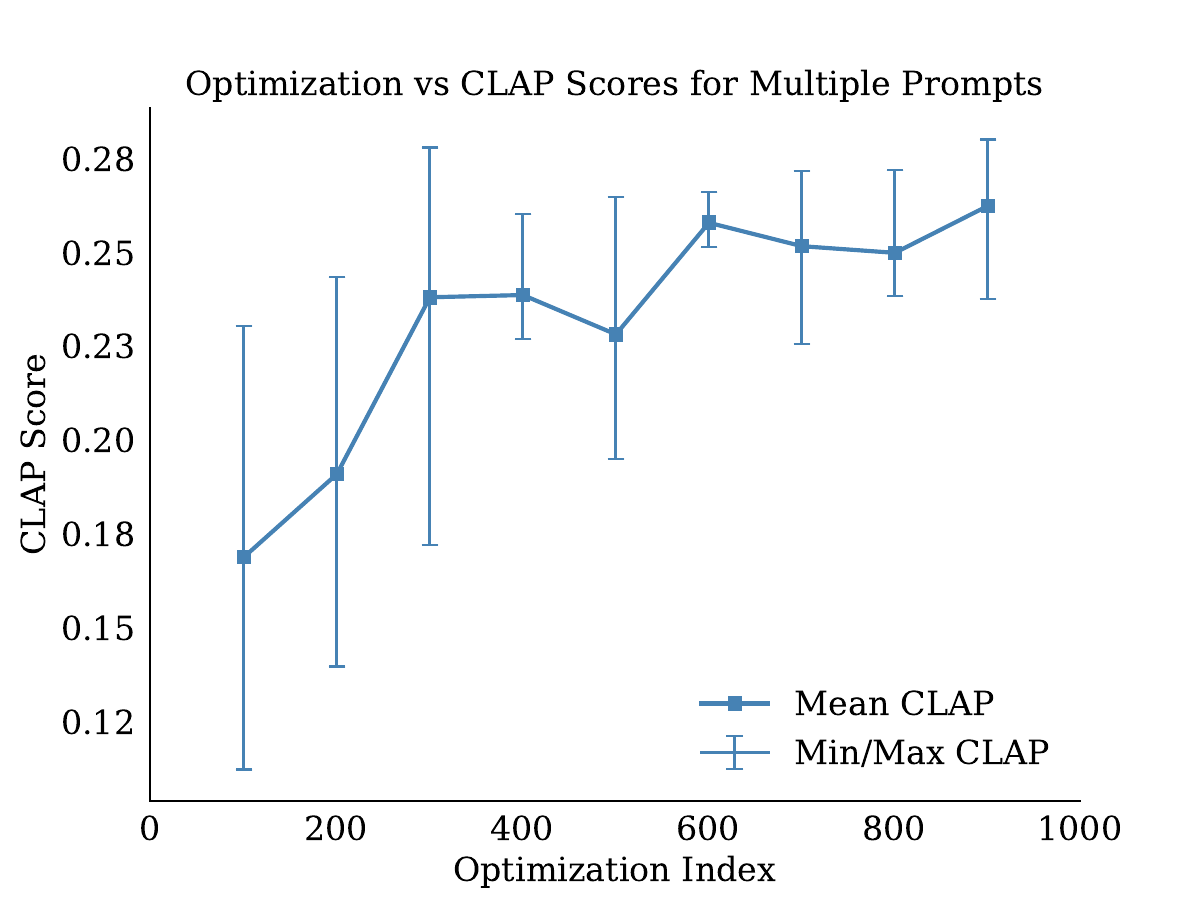}};

                \node[left=of img11, node distance=0cm, rotate=90, xshift=1.0cm, yshift=-.9cm,  font=\color{black}]{CLAP Score};
                \node[below=of img11, node distance=0cm, xshift=0.25cm, yshift=1.15cm,  font=\color{black}]{Optimization Step};
            \end{tikzpicture}
            \vspace{-0.03\textheight}
            \caption{
                \textbf{Impact Synthesis: CLAP Over Time.} We track the CLAP alignment score from initialization to convergence for various prompts. \emph{Takeaway:} The generated audio's alignment with the prompt increases during training, showing our optimization is successful, along with qualitative results shown in \figref{fig:fm_and_impact_synthesis_overview_qualitative}.
            }
            \label{fig:impact_synthesis_clap_vs_time}
            \vspace{-0.02\textheight}
        \end{figure}

        \textbf{Results:}
            \figref{fig:diff_impact_overview} overviews the impact synthesis pipeline, illustrating the learned components and final audio for a prompt.
            \figref{fig:impact_synthesis_clap_vs_time} shows the CLAP score during optimization, demonstrating a steady increase, meaning that our optimization successfully enhances the generated audio's alignment with the prompt.
            We show qualitative results in \figref{fig:fm_and_impact_synthesis_overview_qualitative}, including visualizations of (and links to) the learned audio.

    \subsection{Source Separation}\label{sec:exp_source_separation}
    \vspace{-0.005\textheight}
        Finally, as visualized in \figref{fig:source_separation_overview}, we apply Audio-SDS to the longstanding challenge of source separation (\secref{sec:method_source_separation}). Given a mixture $\targetAudio$ (e.g., a \SI{10}{\second} clip of street noise and saxophone), we then optimize the latent waveforms (e.g., using $\totalNumSources=2$ gives $\optParams = (\latentRender_1, \latentRender_2)$ to align to their respective prompts (e.g., ``\texttt{traffic noise}'' and ``\texttt{saxophone}''). Optimization is done with the update $\updateSourceSep(\optParams)$ as defined in Eq.~\eqref{eq:separation_combined_loss}, which is a mixture of a per-prompt/source SDS update and a reconstruction loss update.
        
        \textbf{Setups and Prompts:} 
            We primarily focus on mixtures where the audio has a simple, known structure of prompts, detailed in App. \secref{sec:app_experiment_details_prompt_selections_source_separation}, to circumvent the issue -- separate from our Audio-SDS investigations -- of selecting prompts for the mixed audio $\targetAudio$. We use a simple baseline for comparisons by using the mixed audio $\targetAudio$ for each source. We report SDR to assess improvement over baselines for synthetic mixtures where ground-truth isolated tracks are available. For real-world recordings, we rely on CLAP scores and qualitative analysis.

        \vspace{-0.005\textheight}
        \textbf{Quantitative Results:} 
            Table~\ref{tab:sep_sdr_table} contains the SDR of our separated sources and reconstruction vs.\ ground truth for our synthetic mixtures, showing Audio-SDS achieves effective separations while closely reconstructing the mixed audio.
            However, we do not have ground-truth sources for SDR in real setups. Table~\ref{tab:sep_clap_table} shows the CLAP for our separated sources on non-synthetic mixtures without ground truth, indicating that we more successfully align with prompts, while SDR to the mixed audio indicates a strong reconstruction.

        \vspace{-0.005\textheight}
        \textbf{Qualitative Observations:} 
            As illustrated in \figref{fig:source_separation_overview}, the separated channels strongly align with their prompts, even for background or soft sounds. We also confirm that adding the channels almost perfectly reconstructs the original mixture. All results are in the supplement.

        \vspace{-0.005\textheight}
        \textbf{Automated Prompt Generation for Real Audio:}
            To automatically generate prompt decompositions given mixed audio $\targetAudio$, as a preliminary proof of concept in App. \figref{fig:automatic_real_source_separation}, we show results of taking real audio, captioning it, and then using an LLM to suggest sources automatically.
            We show that the captioner provides sufficient context for the LLM to suggest a reasonable source decomposition, which, after training, yields a reasonable source separation.

        \begin{table}[t]
            \centering
            \caption{
                \textbf{Source Separation Distance to Ground-truth.} 
                We show Signal-to-Distortion Ratio (SDR) in dB (higher is better) for each separated source of the full $\num{10}$s clips.
                Our method improves significantly over the baseline, boosting mean SDR to the source prompts from $\num{-2.5} \!\to\! \num{2.2} ({\color{darkgreen}+4.7})$.
                Most recovery artifacts are at the audio's end, which we hypothesize is due to the diffusion model preferring audio with silence at the end. Comparing recovery for only the first half, we see the mean SDR goes from $\num{-2.3} \!\to\! \num{9.4} ({\color{darkgreen}+11.7})$.
            }
            \label{tab:sep_sdr_table}
            \scalebox{0.83}{
            \begin{tabular}{lccc}
                \toprule
                Mixture & \multicolumn{2}{c}{\scriptsize{SDRs for reconstructed $(\textnormal{source}_1, \textnormal{source}_2, \textnormal{mixture})$}} \\
                    \cmidrule(lr){2-3}
                    & Initialization & Us \\
                \midrule
                Traffic + Sax &  $(\num{-0.7}, \num{-5.2}, \num{3.2})$ & $({\bf{8.5}}, {\bf{8.1}}, {\bf{13.1}})$ \\
                Bongo + Waves & $(\num{1.2}, \num{-6.4}, \num{3.9})$ & $({\bf{1.2}}, {\bf{-2.1}}, {\bf{8.7}})$ \\
                Pipes + Glass & $(\num{-4.2}, \num{0.0}, \num{2.8})$ & $({\bf{1.5}}, {\bf{6.5}}, {\bf{7.7}})$ \\
                Clock + Bongo & $(\num{-15.8}, {\bf{10.4}}, \num{11.8})$ & $({\bf{-10.4}}, {\num{3.6}}, {\bf{13.9}})$ \\
                Wind + Pipes & $(\num{-6.1}, \num{1.9}, \num{5.8})$ & $({\bf{-0.5}}, {\bf{8.6}}, {\bf{7.3}})$ \\
                \bottomrule
            \end{tabular}
            }
            \vspace{-0.015\textheight}
        \end{table}

        \begin{table}[t]
            \centering
            \caption{
                \textbf{Source Separation Prompt Alignment.} 
                We show CLAP scores (higher is better) for each source and reconstruction loss (lower is better) to the mixture $\targetAudio$. %
                We achieve better CLAP scores, maintaining low reconstruction loss, indicating better source separation than baselines, boosting mean/min/max CLAP to the prompts from $(\num{0.18}, \num{0.02}, \num{0.27})\!\to\! (\num{0.2} ({\color{darkgreen}+0.02}), \num{0.05} ({\color{darkgreen}+0.03}), \num{0.34} ({\color{darkgreen}+0.10}))$.
            }
            \label{tab:sep_clap_table}
            \scalebox{0.83}{
                \begin{tabular}{lccc}
                \toprule
                Mixture & \multicolumn{2}{c}{\scriptsize{CLAPs for reconstructed $\textnormal{source}_1$ and  $\textnormal{source}_2$, then SDR for $\textnormal{mixture}$}} \\
                    \cmidrule(lr){2-3}
                    & Initialization & Us \\
                \midrule
                Traffic + Sax &  $(\num{0.17}, \num{0.02}, \num{3.2})$ & $({\bf{0.2}}, {\bf{0.05}}, {\bf{13.1}})$ \\
                Bongo + Waves & $(\num{0.15}, {\bf{0.14}}, \num{3.9})$ & $({\bf{0.16}}, {\num{0.08}}, {\bf{8.7}})$ \\
                Pipes + Glass & $({\bf{0.25}}, \num{0.27}, \num{2.8})$ & $({\num{0.21}}, {\bf{0.3}}, {\bf{7.7}})$ \\
                Clock + Bongo & $(\num{0.22}, \num{0.24}, \num{11.8})$ & $({\bf{0.30}}, {\bf{0.34}}, {\bf{13.9}})$ \\
                Wind + Pipes &  $(\num{0.25}, {\num{0.06}}, \num{5.8})$ & $({\num{0.25}}, {\bf{0.09}}, {\bf{7.3}})$\\
                \bottomrule
            \end{tabular}
            }
            \vspace{-0.025\textheight}
        \end{table}

    \vspace{-0.01\textheight}
    \subsection{Ablation Studies on Audio-SDS}\label{sec:exp_ablation}
    \vspace{-0.01\textheight}
        We conclude our experiments with ablation studies measuring the impact of key design choices for Audio-SDS, including (a) Decoder-SDS from \secref{sec:avoiding_instabilities}, (b) Multiscale Spectrogram Emphasis from \secref{sec:method_spectrogram_updates}, and (c) using multistep denoising as in \secref{sec:method_multi_step_audio_sds}. Additional ablations on problem design choices are included in App. \secref{sec:app_ablations}.

        \vspace{-0.005\textheight}
        \textbf{Our Decoder-SDS vs.\ Classic Encoder-SDS:}
            We found instabilities differentiating through the encoder for audio diffusion models, as done with the original SDS update \citet{pooledreamfusion} in \eqref{eq:sds_update}. We use our Decoder-SDS update from \eqref{eq:decoder_sds_update} to avoid encoder differentiation, which shows improved quality as demonstrated in App. \figref{fig:ablation_decoder_sds}.

        \textbf{Spectrogram Emphasis vs.\ Pure Time-Domain:} 
            While na\"ive time-domain $\ell_2$ losses suffice for less percussive audio, the multiscale spectrogram approach can improve quality for transients, as App. \figref{fig:ablation_spectrogram_emphasis} demonstrates.

        \textbf{Single-Step vs.\ Multistep Denoising:} 
            Our multi-step partial denoising procedure (\secref{sec:method_multi_step_audio_sds}) consistently improved convergence stability across all tasks. In particular, we observed fewer artifacts (e.g., chirps, glitchy transients) when running even a small number (2--5) of partial denoising steps per iteration, as demonstrated in App. \figref{fig:ablation_multistep_denoising}.

    \vspace{-0.005\textheight}
    \subsection{Remark on Performance vs.\ Flexibility}\label{sec:exp_other_notes}
    \vspace{-0.005\textheight}
        App. \secref{sec:app_exp_failures} explores failure cases of our model. 
        We stress that our goal is not to outperform SOTA methods finetuned for single tasks (e.g., wavelet-based impact solvers or deep source-separation systems). Instead, these setups show Audio-SDS is a unified, prompt-driven interface for many generative audio problems. Each parametric model -- be it an FM synthesizer or a source decomposition -- is guided by the \emph{same} pretrained text-to-audio backbone without dataset-specific training, opening new avenues for rapid prototyping, creative exploration, and downstream tasks where labeled data or domain expertise may be scarce.

\vspace{-0.005\textheight}
\section{Discussion}\label{sec:discussion}
\vspace{-0.005\textheight}
    Due to space constraints, our related works are in App. \secref{sec:related_works} and our ideas for future exploration are in \secref{sec:future_work}.

    \vspace{-0.01\textheight}
    \subsection{Limitations}\label{sec:limitations}
    \vspace{-0.005\textheight}
        \textbf{Model Coverage:}
            Because our method relies on a pretrained text-to-audio diffusion model, it inherits any gaps or biases within its training distribution. In particular, handling spoken dialogue or highly specialized environments (e.g., ``\texttt{a singer in a cathedral}'') remains challenging, as current checkpoints often lack robust representations of nuanced spatial cues or certain rare prompt classes.

        \textbf{Latent Encoding Artifacts:}
            Our experiments use a latent diffusion approach whose encoder-decoder occasionally introduces audible discrepancies between the synthesized waveform and its autoencoded counterpart. These reconstruction gaps can degrade the final quality of complex or highly transient sounds, although improvements in future versions of the audio autoencoder may mitigate such issues.

        \textbf{Optimization Sensitivities:}
            We focus on relatively short audio clips; extending to significantly longer clips can strain the diffusion model’s receptive field and computational resources. Also, SDS optimization can be sensitive to hyperparameter tuning and initialization. Although broadly applicable, convergence can be slow or unstable for specific prompts if parameter settings are not carefully selected.

    \subsection{Conclusion}\label{sec:conclusion}
        By applying Score Distillation Sampling (SDS) to pretrained text-to-audio diffusion models, we reveal a flexible framework unifying data-driven priors with user- or physics-based parametric representations. In tasks ranging from modal impact synthesis to prompt-guided source separation, we show that SDS can tune various audio parameters to match high-level textual descriptions. This capability unlocks new possibilities for creative audio design, research on physically grounded sound generation, and semantic audio editing -- all without reliance on large-scale domain-specific datasets. We anticipate that these findings will serve as a springboard for broader multimodal research, fostering tighter integration between audio, vision, and beyond through the lens of distillation-based generative methods.

\section*{Impact Statement}
    Using large-scale diffusion models, our Score Distillation Sampling (SDS) approach introduces high-fidelity, prompt-driven audio synthesis and separation, eliminating the need for massive curated datasets. While this innovation benefits creators and researchers by streamlining tasks such as physically informed synthesis, it raises concerns about unauthorized audio manipulation or copyright infringement. Ensuring transparency, robust detection measures, and clear ethical guidelines is crucial to the safe and equitable adoption of SDS-based audio systems.

\section*{Acknowledgements}
    We want to thank Kevin Shih, Rafael Valle, Zhifeng Kong, Alex Keller, Doug James, Sanja Fidler, and Vismay Modi for their insightful discussions, and Sanggil Lee for their help with performance metrics.

\newpage

{\small{
\bibliography{main}
\bibliographystyle{icml2025}
}}

\newpage
\appendix
\onecolumn

\begin{table*}[h]\caption{Glossary and notation}
    \begin{center}
    \scalebox{0.8}{
        \begin{tabular}{c c}
            \toprule
            SDS & Score Distillation Sampling\\
            VR/AR & Virtual Reality / Augmented Reality\\
            SFX & Sounds Effects\\
            FM & Frequency-modulation\\
            LLM & Large Language Model\\
            SOTA & State of the art\\
            MSE & Mean-squared Error\\
            CLAP & Contrastive Language-Audio Pretraining \\
            SDR & Signal-to-Distortion Ratio\\
            FAD & Fréchet Audio Distance\\
            CFG & Classifier-Free Guidance\\
            STFT & Short-Time Fourier Transform\\
            $\identity$ & The identity matrix\\
            $\mathcal{U}, \mathcal{N}$ & The Uniform and Normal distributions respectively\\
            $\optParams \in \optParamsDomain$ & The parameters to be optimized and their domain\\
            $\loss:\optParamsDomain \to \R$ & An arbitrary loss function\\
            $\render \in \renderDomain$ & A rendered audio (or image) and its domain\\
            $\camera \in \cameraDomain$ & Sampled render parameters and their domain (e.g., camera location in text-to-3D)\\
            $\renderFunc: \optParamsDomain \times \cameraDomain \to \renderDomain$ & The rendering function, where $\render = \renderFunc(\optParams, \camera)$\\
            $\noise \sim \mathcal{N}(0, \identity)$ & The (Gaussian) noise added to the diffusion model input\\
            $\timestep \sim \mathcal{U}[\timestepMin, \timestepMax]$ & The sampled diffusion timestep (small is low noise)\\
            $\signalScale_\timestep, \noiseScale_\timestep \in \R$ & The diffusion model's signal and noise scale at timestep $\timestep$\\
            $\noisedRender(\optParams, \camera) = \signalScale_\timestep \renderFunc(\optParams, \camera) + \noiseScale_\timestep \noise$ & The noised data at diffusion timestep $\timestep$ for optimizable parameters $\optParams$\\
            $\diffusionParams$ & The (frozen) diffusion models weights\\
            $\prompt$ & A text prompt for conditioning the diffusion model\\
            $\noisePrediction(\noisedRender,\timestep,\prompt)$ & The predicted noise for noised render $\noisedRender$ and timestep $\timestep$ for prompt $\prompt$\\
            $\guidanceScale \in \R$ & The classifier-free guidance (CFG) scale\\
            $\guidanceNoisePrediction(\noisedRender, \timestep, \prompt) = (1 + \guidanceScale)\noisePrediction(\noisedRender, \timestep, \prompt) - \guidanceScale \noisePrediction(\noisedRender, \timestep)$ & The noise prediction with CFG\\
            $\timestepWeight \in \R$ & A scaling weight at diffusion timestep $\timestep$\\
            $\lossSDS(\optParams; \prompt) = \E_{\timestep\!\!,\noise,\camera} [ \timestepWeight || \noisePrediction ( \noisedRender(\optParams, \!\camera), \!\timestep, \prompt ) - \noise ||^{2} ]$ & The ``SDS loss'', whose approximate gradient gives us our updates\\
            $\updateSDS \!=\! \E_{\timestep\!\!,\noise,\camera} [\timestepWeight(\!\guidanceNoisePrediction \!( \noisedRender(\!\optParams\!, \!\camera\!)\!, \!\timestep\!, \prompt \!) \!-\! \noise)\!\nabla_{\!\optParams} \noisedRender(\!\optParams\!,\! \camera)\!]$ & The SDS update, which is an approximation of $\nabla_{\optParams}\lossSDS$\\
            $\encodeFunc, \decodeFunc$ & The encoding and decoding functions for our latent diffusion model\\
            $\latentRender = \encodeFunc(\render)$ & The encoded latent for audio $\render$\\
            $\denoisedNoisedLatent_{\diffusionParams}(\optParams, \timestep, \noise, \prompt)$ & The denoised, noised latent of audio $\render$\\
            $\noisedDenoisedRender_{\diffusionParams}(\optParams, \timestep, \noise, \prompt) = \decodeFunc(\denoisedNoisedLatent_{\diffusionParams}(\optParams, \timestep, \noise, \prompt))$ & The decoded, denoised, noised latent of audio $\render$\\
            $\updateSDS^{\decodeFunc} \!=\! (\E[\noisedDenoisedRender] \!-\! \render)\nabla_{\!\optParams}\render$ & The Decoder-SDS update, avoiding differentiating the encoder\\
            $\audioSampleIndex \in \{1, \dots, \totalAudioSamples\}$ & The total number of samples in an audio waveform\\
            $\audioRender = \audioRenderFunc(\optParams) \in \audioRenderDomain$ & The rendered, stereo audio waveform\\
            $\STFTIndex \in \{1, \dots, \totalNumSTFT \}$ & The total number of STFT magnitudes to use\\
            $\STFT_{\STFTIndex}(\cdot)$ &  A STFT magnitude with set window size\\
            $\specRender = \sum_{\STFTIndex=1}^{\totalNumSTFT}\STFT_{\STFTIndex}(\render),\noisedDenoisedSpecRender = \sum_{\STFTIndex=1}^{\totalNumSTFT}\STFT_{\STFTIndex}(\noisedDenoisedRender)$ & The spectrogram emphasized render and denoised, decoded render respectively\\
            $\updateSDS^{\STFT\!, \decodeFunc} \!\!=\! (\E[\noisedDenoisedSpecRender] \!-\! \specRender)\!\nabla_{\!\optParams}\specRender$ & The spectrogram-emphasized Decoder-SDS update\\
            $\fmMatrixIndex \in \{1, \dots , \fmMatrixSize \}$ & The size of the FM Matrix\\
            $\fmMatrix \in \R^{\fmMatrixSize \times \fmMatrixSize+1}$ & The FM matrix\\
            $\fmState \in \R^{\fmMatrixSize \times \totalAudioSamples}$ & The oscillator state at each timestep $\audioSampleIndex$ defined in Eq.\eqref{eq:fm_state}\\
            $\fmAttackDecayFunc_{\fmMatrixIndex}(\audioSampleIndex)$ & The envelope controlling attack and decay for the $\fmMatrixIndex$th oscillator \\
            $\{\fmFrequency_\fmMatrixIndex,\fmAttackDecayA_\fmMatrixIndex, \fmAttackDecayB_\fmMatrixIndex\}_{\fmMatrixIndex=1}^{\fmMatrixSize}$ & The envelope function $\fmAttackDecayFunc_{\fmMatrixIndex}$'s parameters\\
            $\impulseImpact, \impulseObj^{\optParams}, \impulseReverb^{\optParams}$ & The impact, object, and reverberation impulses respectively\\
            $\star$ & The convolution operator\\
            $\impulseComponentIndex \in \{1, \dots, \numImpulseComponent\}$ & The number of impulse components\\
            $\{\frequency_\impulseComponentIndex, \decay_\impulseComponentIndex,\amplitude_\impulseComponentIndex, \tilde{\frequency}_\impulseComponentIndex, \tilde{\decay}_\impulseComponentIndex, \tilde{\amplitude}_\impulseComponentIndex\}_{\impulseComponentIndex=1}^{\numImpulseComponent}$ & The frequency, decay, and amplitude of each impulse sine component\\
            $\bandpassFunc(\cdot, \cdot)$ & A function bandpassing the first argument centered at the second's frequency\\
            $\targetAudio \in \audioRenderDomain$ & A target, mixed audio waveform for source separation\\
            $\sourceIndex \in \{1, \dots, \totalNumSources\}$ & The number of sources to decompose the target audio into\\
            $\prompt_{\sourceIndex}, \optParams_{\sourceIndex}, \render_{\sourceIndex}$ & The text prompt, optimizable parameters and audio for the $\sourceIndex$th source\\
            $\!\!\!\!\!\!\!\!\lossRecons\!(\optParams) \!=\! \sum\nolimits_{\STFTIndex=1}^{\totalNumSTFT} \!| \STFT_{\STFTIndex}(\targetAudio) \!-\! \STFT_{\STFTIndex}(\sum\nolimits_{\sourceIndex=1}^{\totalNumSources} \!\renderFunc_{\sourceIndex}(\optParams_{\sourceIndex})) |_{2}^{2}$ & A spectrogram-emphasized reconstruction loss for audio $\targetAudio$ and source sum $\render_{\sourceIndex}(\optParams_{\sourceIndex})$\\
            $\lossScaleSourceSep \in \R$ & The weighting between our reconstruction and SDS updates\\
            $\updateSourceSep(\optParams; \{\prompt\}_{\sourceIndex=1}^{\totalNumSources}) = \nabla_{\optParams}\lossRecons(\optParams) + \lossScaleSourceSep \sum\nolimits_{\sourceIndex=1}^{\totalNumSources}\! \updateSDS^{\STFT\!, \decodeFunc}(\optParams_{\sourceIndex}; \prompt_{\sourceIndex})$ & The source separation update, a mixture of reconstruction gradient and SDS update\\
            $\optParams^{*} \in \argmin\nolimits_{\optParams} \loss(\optParams)$ & A loss-minimizing parameter value\\
            \bottomrule
        \end{tabular}
        }
    \end{center}
    \label{tab:TableOfNotation}
\end{table*}

\section{Reproducibility Details}\label{sec:reproducibility_details}
    Additional details and future updates will be available at our \href{https://research.nvidia.com/labs/toronto-ai/Audio-SDS/}{{\color{nvidiagreen}project website}}.
    Table~\ref{tab:TableOfNotation} summarizes all notation used throughout the paper.
    All core implementation details are found in Appendix~\ref{sec:app_implementation}.
    Without additional fine-tuning or proprietary data, we rely on pretrained diffusion checkpoints (Stable Audio Open~\citep{evans2024stable}). Full details on prompt selection (FM synthesis, impact synthesis, and source separation) appear in Appendix~\ref{sec:app_experiment_details_prompt_selections}. 
    Our optimization protocols, including batch sizes, learning rates, and guidance scales, are exhaustively listed in Appendix~\ref{sec:app_hyperparameters}. 
    All experiments were run on a single NVIDIA A100 GPU; Appendix~\ref{sec:app_compute} details the total runtimes and memory usage for each experiment.

\section{Related Works}\label{sec:related_works}
    We review key literature that informs our approach, organized by Score Distillation Sampling (SDS), as well as general machine learning techniques for audio, FM synthesis, impact sound modeling, and source separation. 
    
    \textbf{Score Distillation Sampling (SDS)}
        was originally introduced by \citet{pooledreamfusion} and had multiple variants such as VSD~\citep{wang2024prolificdreamer}, SDI~\citep{lukoianov2024score}, and beyond~\citep{ma2025scaledreamer, wang2023score} developed for properties such as improved quality, robustness, and convergence speed.
        SDS used with image-conditional diffusion models~\citep{liu2023zero} and text-conditional image diffusion models for 3D generation of NeRFs~\citep{pooledreamfusion, lorraine2023att3d}, meshes~\citep{lin2023magic3d, xie2024latte3d}, Gaussian splats~\citep{chen2023text, tangdreamgaussian}, relightable materials~\citep{deng2025flashtex}, or editing generated assets~\citep{zhuang2023dreameditor}.
        SDS has also been combined with video diffusion models for generating 4D~\citep{ling2024align, bahmani20244d} and elastic material property fields~\citep{zhang2025physdreamer}.
        However, to our knowledge, no prior works use SDS with audio diffusion models.

    \textbf{Machine Learning for Audio Generation}
        has advanced rapidly with deep models for speech, music, and general sound effects \citep{van2016wavenet, yamamoto2020parallel, engel2019gansynth}. Autoregressive approaches~\citep{van2016wavenet} introduced fine-grained sample-level synthesis, while GAN-based~\citep{donahueadversarial, binkowski2019high} and transformer-based~\citep{fugatto2025} methods improved efficiency. More recently, diffusion-based techniques \citep{kongdiffwave, liu2023audioldm} have taken center stage for text-to-audio generation, as shown in various recent works \citep{kreukaudiogen, agostinelli2023musiclm, huang2023make}. Our work leverages these advances by integrating SDS updates with pretrained diffusion models, enabling flexible parametric audio synthesis and editing without massive task-specific datasets.

    \textbf{Frequency Modulation (FM) Synthesis}
        \citep{chowning1973synthesis} revolutionized digital sound design by allowing one oscillator’s frequency to be modulated by another’s output, generating rich spectra with relatively few parameters. Its commercial success, facilitated by instruments like the Yamaha DX7, standardized FM techniques \citep{lavengood2019makes}, while later efforts explored phase modulation and hybrid approaches \citep{roads1985micro}. More recent neural audio frameworks integrated FM into differentiable pipelines \citep{engel2020ddsp}, enabling the automated learning of parameters from data. By applying SDS on a pretrained text-to-audio diffusion model, our method unlocks text-driven optimization of FM parameters, bridging classic signal processing with modern generative priors.

    \textbf{Impact Sounds}
        arise from short transients and complex resonance behaviors, making them challenging to model. Traditional work used parametric or modal decompositions \citep{obrien2002synthesizing, langlois2014eigenmode} and physically guided resonators \citep{dendoel2001foleyautomatic}. Inverse solvers and differentiable simulators \citep{langlois2014inverse, su2023physics, zhang2025physdreamer} have more recently enabled parameter fitting from recordings, yielding a realistic reproduction of collisions or material interactions. We integrate these physically informed models with SDS, using a pretrained diffusion model as a high-level ``critic'' allowing the underlying physical parameters (e.g., oscillator frequencies or damping) to be optimized to match prompt-driven criteria, rather than relying solely on domain-specific data.

    \textbf{Source Separation}
        is critical to music demixing, speech enhancement, and environmental sound analysis \citep{defossez2019music, hennequin2020spleeter, stoter2019open}, and was traditionally tackled with signal processing (e.g., non-negative matrix factorization) or direct neural networks \citep{jansson2017singing, stoller2018wave}. Recently, text-based guidance has enabled more flexible or zero-shot scenarios, where users specify which source to extract~\citep{liu2024separate, tesch2023multi}. Building on this paradigm, we propose using SDS updates from a pretrained diffusion model to enforce that separated components align with particular text prompts. Our method ``assigns'' each mixture component to audio consistent with the user’s description, while a reconstruction loss ensures each source sums back to the original mixture.

\newpage
\section{Implementation Details}\label{sec:app_implementation}
\vspace{-0.005\textheight}
    Our experiments are implemented in JAX~\citep{jax2018github}, though we expect PyTorch~\citep{paszke2019pytorch} to work similarly, albeit with slightly more complexity due to our careful use of JAX's automatic differentiation interface.
    
    A significant implementation hurdle is encountered when trying to differentiate through spectrogram losses. Using the default JAX implementation, differentiating through the Short-Time Fourier Transform (STFT) incurs what appears to be significant memory overhead, as it traces through the windowing procedure. We were able to mitigate this issue by carefully linearizing the function before computing the vector-Jacobian product; however, it remains slower in runtime and consumes more memory than expected. Future work might implement an adjoint method for the STFT, which avoids any memory overhead by using the Fourier transform's unitary property.

\section{Experimental Details}\label{sec:app_experimental_details}
\vspace{-0.005\textheight}
    \subsection{Compute Requirements}\label{sec:app_compute}
    \vspace{-0.005\textheight}
        Our experiments were carried out on an A100 GPU, each using a single one. 
        As an upper bound on compute utilization: Each FM synthesis experiment (\figref{fig:fm_and_impact_synthesis_overview_qualitative}) used a single GPU for at most $10$ hours. Each impact synthesis experiment (\figref{fig:diff_impact_overview}) run used a single A100 for at most $12$ hours. Each source separation run used a GPU for at most $4$ hours.
        For the synthesis results, as an upper bound on those shown in \figref{fig:fm_and_impact_synthesis_overview_qualitative} and \figref{fig:impact_synthesis_clap_vs_time}, $20 + 24 + 12 \times 6 = 60$ A100 GPU hours.
        We show $5$ source separation results, which use at most $4 \times 5 = 20$ A100 GPU hours collectively.

    \subsection{Hyperparameters}\label{sec:app_hyperparameters}
    \vspace{-0.005\textheight}
        \subsubsection{Optimization Hyperparameters}        
            We use Adam~\citep{kingma2014adam} as our optimizer, with the specific parameters varying by experiment and set to their defaults unless otherwise specified.
            For FM synthesizer experiments, we use a batch size of $8$, a learning rate between $5 \times 10^{-3}$ and $3 \times 10^{-2}$, and $\num{1000}$ total iterations.
            For the impact synthesis experiments, we use a batch size of $8$, a learning rate between $5 \times 10^{-3}$ and $ 10^{-2}$, and a total of $\num{1000}$ to $\num{5000}$ iterations.
            We use a batch size of $10$ for the source separation experiments, a learning rate of $5 \times 10^{-2}$ for $\num{1000}$ total iterations.

        \subsubsection{Loss Function Details}
        \vspace{-0.005\textheight}
            We use a clip length of $3$ seconds and $\ell_2$ emphasis for FM synthesizer and impact synthesis experiments.
            For source separation experiments, we use a clip length of $10$ seconds during optimization, although we report enhanced results using only the first $5$ seconds for metrics. Source separation also uses a spectrogram emphasis on reconstruction and SDS, with $3$ bin sizes of $\num{1024}$, $\num{2048}$, and $\num{4096}$.

        \subsubsection{Guidance Hyperparameters}
        \vspace{-0.005\textheight}
            For FM synthesizer experiments, we use min and max timesteps of $[\timestepMin, \timestepMax] = [0.6, 1.0]$, a guidance scale of $15$, and $3$--$5$ denoising steps.
            For impact synthesis experiments, we use min and max timesteps of $[\timestepMin, \timestepMax] = [0.7, 1.0]$, a guidance scale of $15$, and $10$--$25$ denoising steps.
            For source separation experiments, we use min and max timesteps of $[\timestepMin, \timestepMax] = [0.025, 0.875]$, a guidance scale of $60$, and $2$ denoising steps.

        \subsubsection{Problem Specific Hyperparameters}
        \vspace{-0.005\textheight}
            \textbf{FM Synthesis:}
                We use an FM matrix $\fmMatrix$ of size $5\times 4$ (i.e., $\fmMatrixSize = 4$ operators), as well as a logarithmic parametrization of the matrix itself -- so the parameter matrix was exponentiated before being used. This is typical for audio since volume is perceived logarithmically. The attack/decay envelope parameters are similarly mapped through a sigmoid to constrain their range from $0$ to $1$ (in seconds), and each operator has its own envelope. The output row of $\fmMatrix$ is initialized to be close to $[1,0,0,0]$ after exponentiation, so the synthesizer starts with (approximately, due to a small Gaussian noise perturbation) only one oscillator audible.

            \textbf{Impact Synthesis:}
                We use $\num{2048}$ sinusoids for the object impulse and $2048$ bandpass filters for the reverb impulse. For both, the frequencies are initialized linearly between $100$Hz--$18$kHz, biasing them towards the lower end of the spectrum. Gaussian noise with a standard deviation of $10^{-4}$ is added to perturb these initial frequencies.

            \textbf{Source Separation:}
               We use an SDS update scaling of $\lossScaleSourceSep = \num{0.02}$ and a latent parametrization of the sources.

        \subsubsection{Hyperparameter Selection Strategies}
            We initially selected default hyperparameter values from the threestudio~\citep{threestudio2023} open-source re-implementation of DreamFusion. Of these, for our setup, multiple values had to be changed from their defaults; notably, the guidance scale $\guidanceScale$ for the FM and impact synthesis experiments are considerably lower values of $\sim\!15-25$, rather than the $\sim\!100$ used in \cite{pooledreamfusion}.

            Source separation required moderate tuning of the SDS weight $\lossScaleSourceSep$. A value that is too large can distort results, overly matching the prompt at the expense of fidelity, while a value that is too small yields minimal semantic guidance. A range of $[0.01, 0.1]$ typically worked well in practice.

        \subsection{Prompt Selections}\label{sec:app_experiment_details_prompt_selections}

            \subsubsection{Source Separation Prompts}\label{sec:app_experiment_details_prompt_selections_source_separation}

                We construct the source audio to mix for separation using Stable Audio Open by generating $10$s clips using standard settings of $100$ denoising steps with a CFG scale of $\guidanceScale = 7$.
                We use the following prompts for each source in our experiments:
                \begin{itemize}
                    \item Traffic: ``\texttt{cars passing by on a busy street, traffic, road noise}''
                    \item Sax: ``\texttt{saxophone playing melody, jazzy, modal interchange, post bop}''
                    \item Bongo: ``\texttt{bongo drum playing a rhythmic beat}''
                    \item Waves: ``\texttt{waves crashing on a rocky shore}''
                    \item Pipes: ``\texttt{clanging metal pipes dropping on concrete}''
                    \item Glass: ``\texttt{glass breaking and shattering}''
                    \item Clock: ``\texttt{rhythmic ticking of an old grandfather clock}''
                    \item Wind: ``\texttt{gentle rustling of leaves in the wind}''
                    \vspace{0.05\textheight}
                \end{itemize}
                
            \newpage
            \subsubsection{Automating Source Separation Prompt Generation with Audio Captioners and LLMs}\label{sec:app_experiment_details_prompt_selections_automated}
            
            For LLM-assisted source separation, we use the following prompt template:
            \begin{tcolorbox}[
              enhanced,
              title=LLM Prompt Template,
              colframe=gray!60!black,
              colback=gray!5,
              fonttitle=\bfseries,
              varwidth upper,
              boxsep=2pt,
              arc=4pt
            ]
            \textbf{System Prompt / Instruction with parameter \texttt{AUDIO\_CAPTION}:} 
\begin{verbatim}
I have an audio file (e.g., a .wav) that contains various real-world sounds 
(for example, cars driving by, music playing in the background, cicadas
chirping, and water flowing). I process this audio with a model that generates
a natural language caption describing the sounds present in the audio. I will
provide you with this caption.

I also have a separate source-separation tool that works as follows: I give it
the original audio, along with a list of N text descriptions—one description 
per channel. This tool uses a text-to-audio diffusion model to separate the 
original audio into N channels, each channel corresponding to one of the text 
prompts. The resulting channels, when mixed together, should reconstruct the 
original audio. Each channel’s prompt should describe a distinct element or 
sound source present in the recording.

Your task is to take the caption I provide and generate several examples of 
sets of N text prompts suitable for the source-separation tool. Each example 
should present a plausible partitioning of the audio into N distinct channels, 
reflecting different ways to break down the overall scene. Provide a few 
different examples, each with a unique approach to organizing the sounds into 
N text prompts.

Here is the output caption for my audio: <AUDIO_CAPTION>
\end{verbatim}
            \end{tcolorbox}
\newpage
            \begin{tcolorbox}[
              enhanced,
              title=LLM Output,
              colframe=gray!60!black,
              colback=gray!5,
              fonttitle=\bfseries,
              varwidth upper,
              boxsep=2pt,
              arc=4pt
            ]
            \textbf{Example LLM Output where \texttt{AUDIO\_CAPTION} = ``\texttt{Someone is clicking on a keyboard and talking.}''}
\begin{verbatim}
Below are several example sets of prompts (each set corresponds to one way of
partitioning the audio into N channels). Every example reflects a different
approach to organizing the sounds described by the caption “Someone is
clicking on a keyboard and talking.”

Example 1 (N=2)
This example matches your previous usage with two distinct channels:

Channel 1 Prompt: “music playing quietly with indiscernible talking”
Channel 2 Prompt: “clicking on a keyboard”
Example 2 (N=2)
A simpler two-channel breakdown emphasizing voice versus keyboard:

Channel 1 Prompt: “clear speaking voice in a room”
Channel 2 Prompt: “typing sounds on a keyboard”
Example 3 (N=3)
A three-channel separation, isolating each element more explicitly:

Channel 1 Prompt: “soft background ambience with distant echoes”
Channel 2 Prompt: “person talking or speaking softly”
Channel 3 Prompt: “repetitive keyboard clicking sounds”
Example 4 (N=3)
Another three-channel separation, focusing on different qualities of speech
and background noise:

Channel 1 Prompt: “whispered conversation or murmuring”
Channel 2 Prompt: “tapping and clicking from a computer keyboard”
Channel 3 Prompt: “subtle room tone with very faint background music”
Feel free to modify any of these examples to suit your specific needs or the
capabilities of your source-separation tool. Each prompt should guide the
model to isolate the corresponding element in the audio.
\end{verbatim}
            \end{tcolorbox}

\newpage
\subsection{Extended Experimental Results}\label{sec:app_experiments}
    \subsubsection{Synthesis}
\begin{figure}[t]
                \centering
                \scalebox{1.0}{
                \begin{tikzpicture}
                \centering
                    \node (img11){\includegraphics[trim={2.25cm 1.65cm 4.7cm 1.5cm}, clip, width=.45\linewidth]{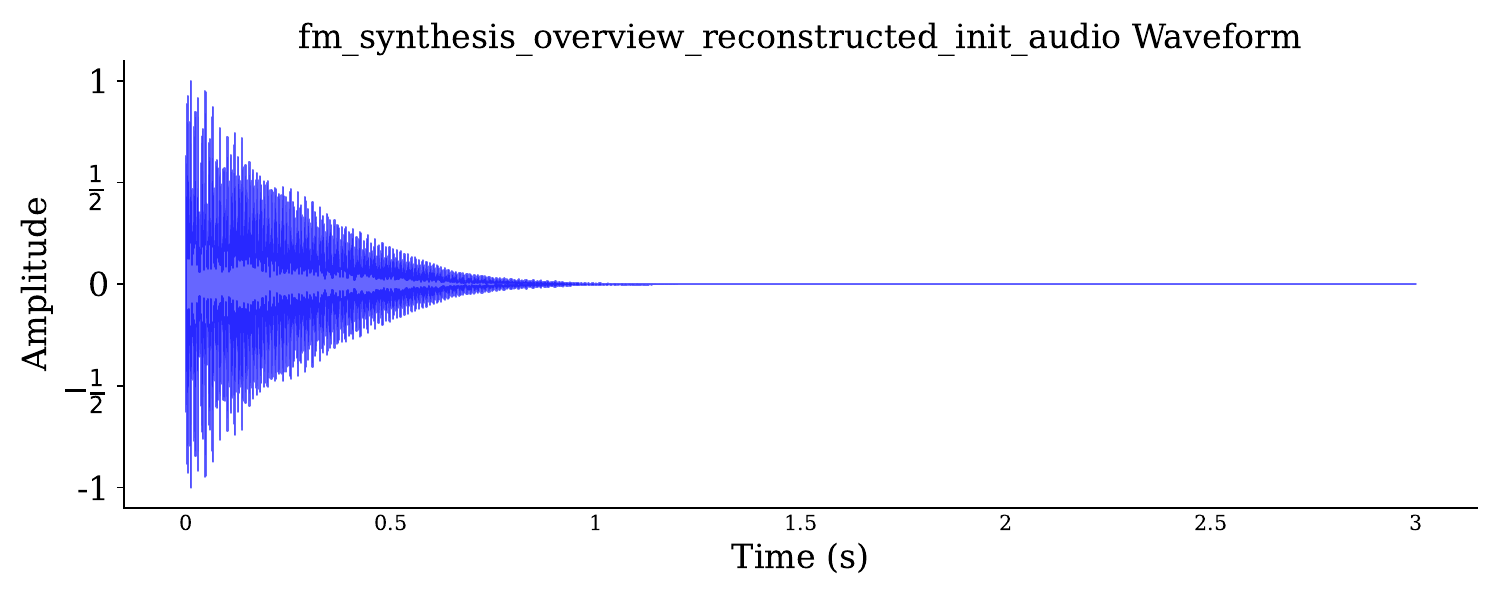}};
                    \node (img21)[below=of img11, yshift=1.0cm]{\includegraphics[trim={2.25cm 1.65cm 4.7cm 1.5cm}, clip, width=.45\linewidth]{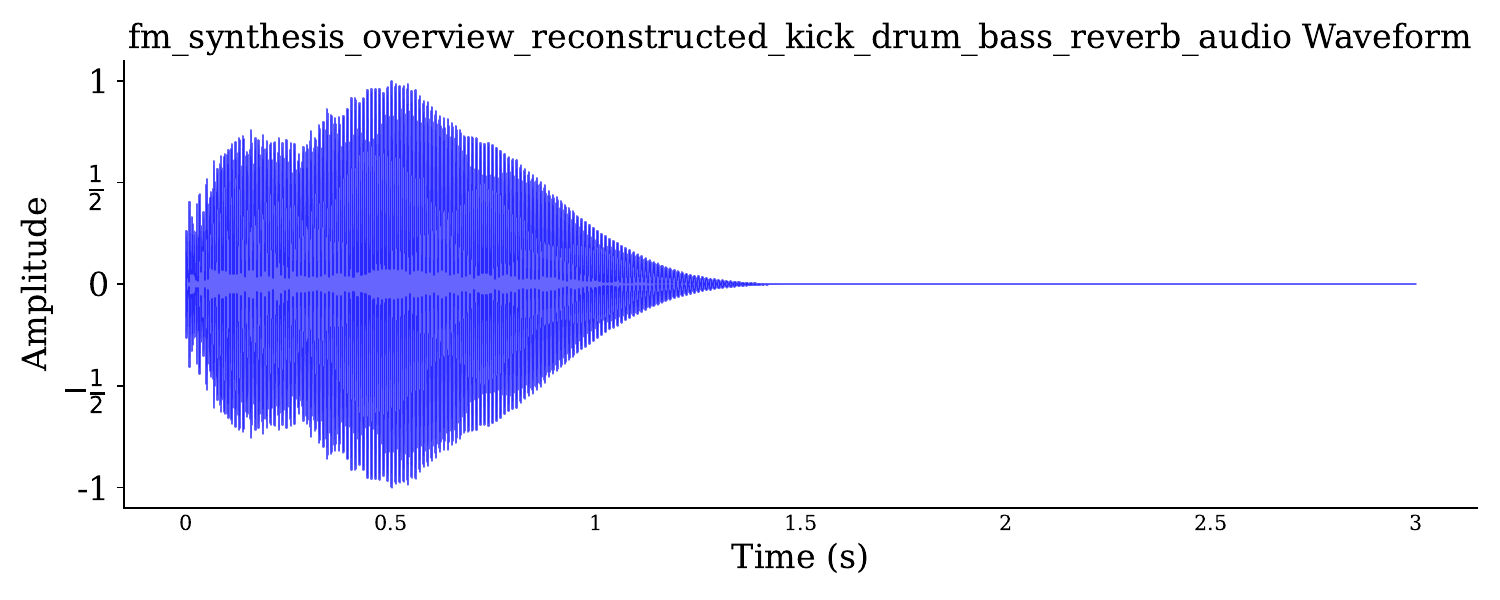}};
                    \node (img31)[below=of img21, yshift=1.0cm]{\includegraphics[trim={2.25cm 1.65cm 4.7cm 1.5cm}, clip, width=.45\linewidth]{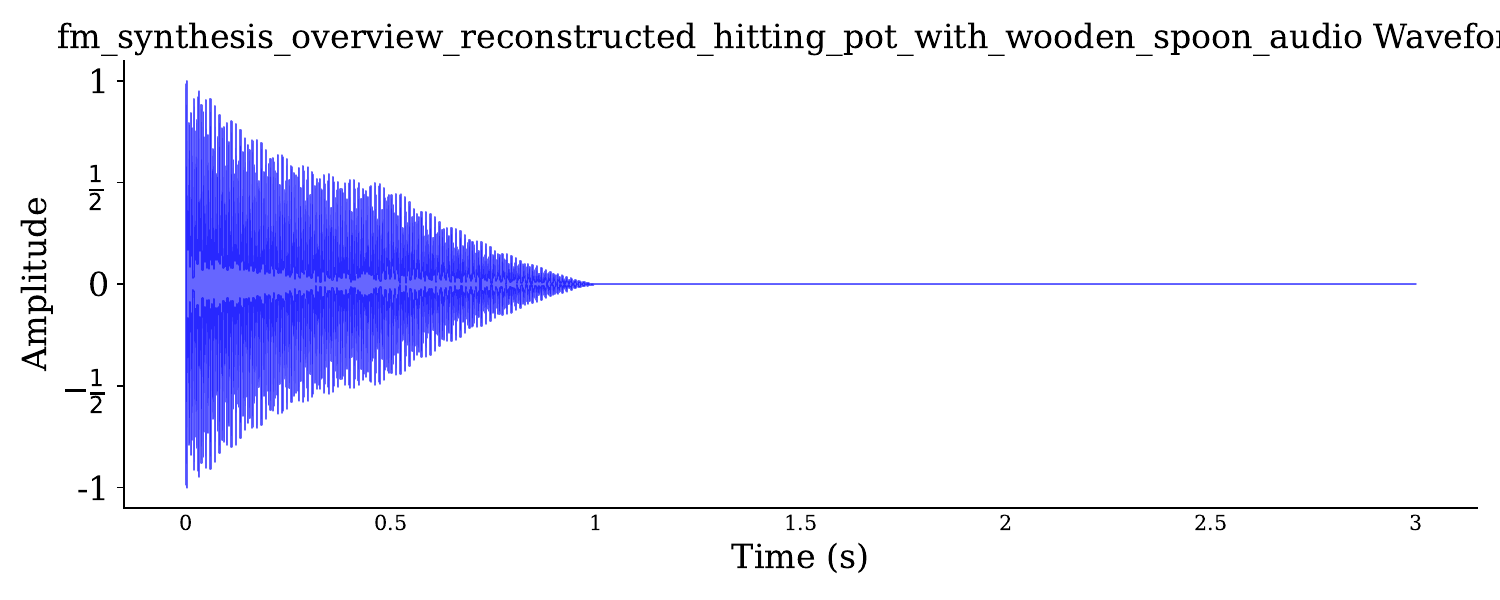}};
    
                    \node (img12)[right=of img11, xshift=-1.0cm] {\includegraphics[trim={2.25cm 1.65cm 4.7cm 1.5cm}, clip, width=.45\linewidth]{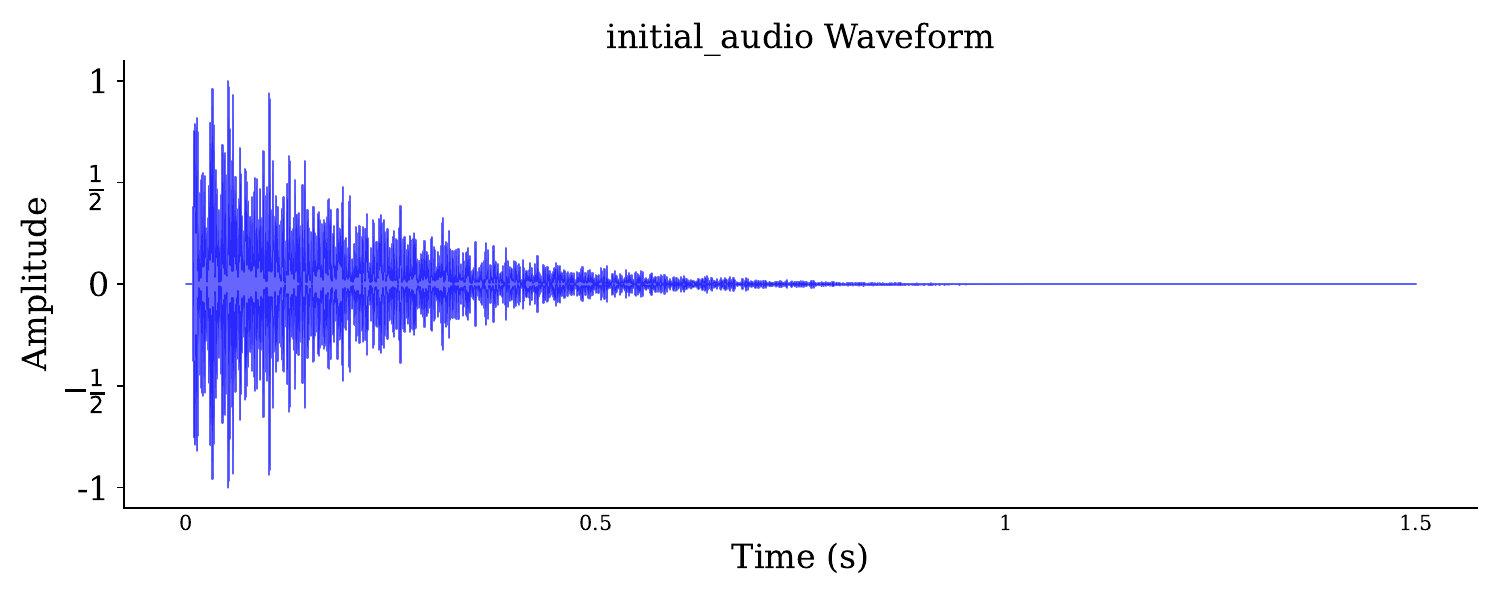}};
                    \node (img22)[below=of img12, yshift=1.0cm]{\includegraphics[trim={2.25cm 1.65cm 4.7cm 1.5cm}, clip, width=.45\linewidth]{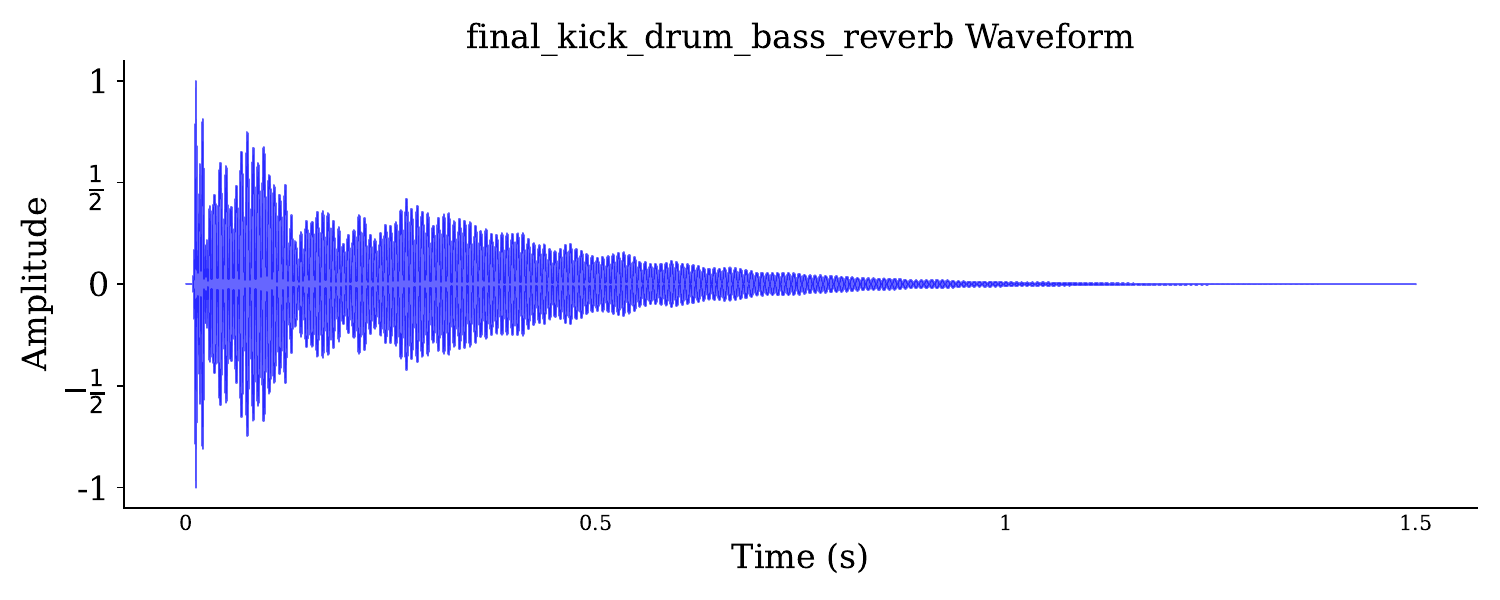}};
                    \node (img32)[below=of img22, yshift=1.0cm]{\includegraphics[trim={2.25cm 1.65cm 4.7cm 1.5cm}, clip, width=.45\linewidth]{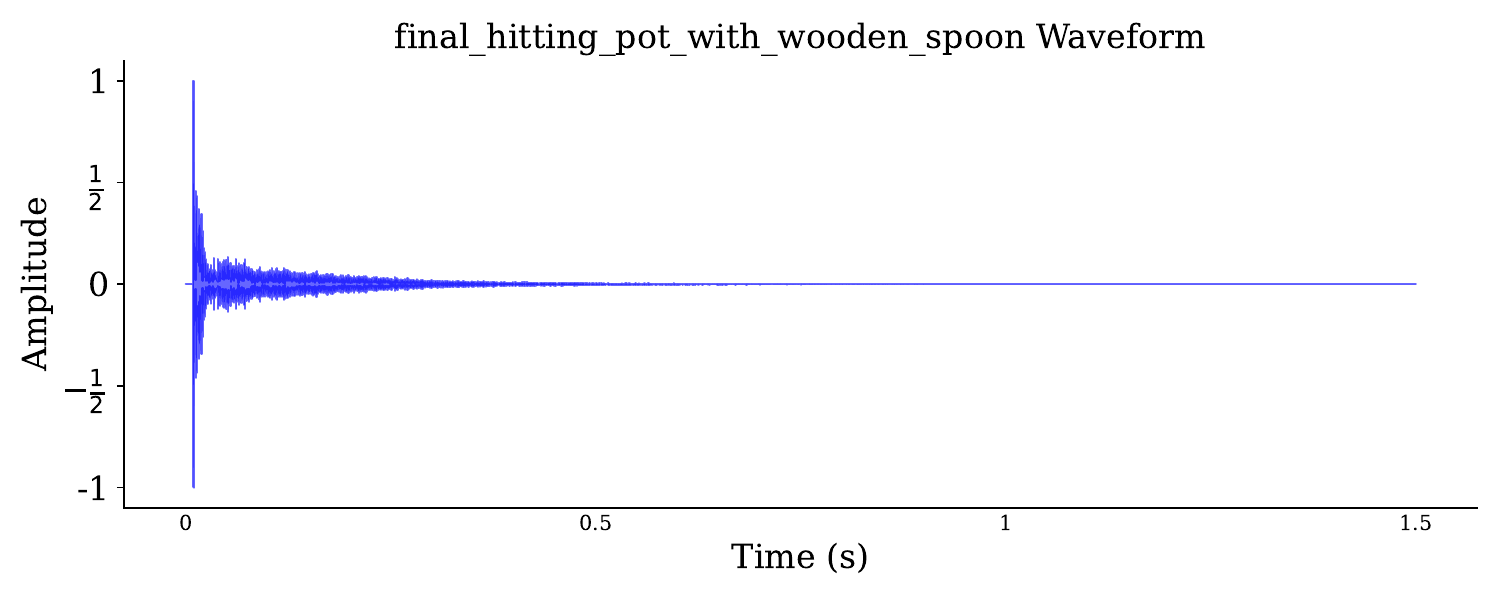}};
    
                    \node[right=of img12, node distance=0cm, xshift=-.8cm, yshift=.6cm, rotate=270, font=\color{black}]{\normalsize{Initialization}};
                    
                    \node[right=of img22, node distance=0cm, xshift=-.8cm, yshift=-1cm, rotate=270, font=\color{black}]{\normalsize{Final}};
                    \node[right=of img22, node distance=0cm, xshift=-.5cm, yshift=1.0cm, rotate=270, font=\color{black}]{\tiny{``\texttt{kick drum,}\dots''}};
                    \node[right=of img32, node distance=0cm, xshift=-.5cm, yshift=.9cm, rotate=270, font=\color{black}]{\tiny{``\texttt{hitting pot}\dots''}};
    
                    \node[left=of img21, node distance=0cm, rotate=90, xshift=1.0cm, yshift=-.9cm,  font=\color{black}]{\normalsize{Amplitude}};
                    \node[below=of img31, node distance=0cm, xshift=4.5cm, yshift=1.15cm,  font=\color{black}]{\normalsize{Time (s)}};
    
                    \node[above=of img11, node distance=0cm, xshift=0cm, yshift=-1.15cm,  font=\color{black}]{\normalsize{FM Synthesis}};
                    \node[above=of img12, node distance=0cm, xshift=0cm, yshift=-1.15cm,  font=\color{black}]{\normalsize{Impact Synthesis}};
                \end{tikzpicture}
                }
                \vspace{-0.02\textheight}
                \caption{
                    We repeat the visualization of \figref{fig:fm_and_impact_synthesis_overview_qualitative}, with waveforms instead of spectrograms.
                }
                \label{fig:impact_synthesis_qualitative_waveform}
                \vspace{-0.01\textheight}
            \end{figure}

        See \figref{fig:impact_synthesis_qualitative_waveform}.

    \subsubsection{Automated Prompt Generation for Source Separation with Real Audio}\label{sec:app_exp_automatic_source_separation}
\begin{figure}[t]
                \vspace{-0.01\textheight}
                \centering
                \scalebox{.94}{
                \begin{tikzpicture}
                \centering
                    \node (img11){\includegraphics[trim={2.25cm 1.45cm 4.7cm 2.0cm}, clip, width=.45\linewidth]{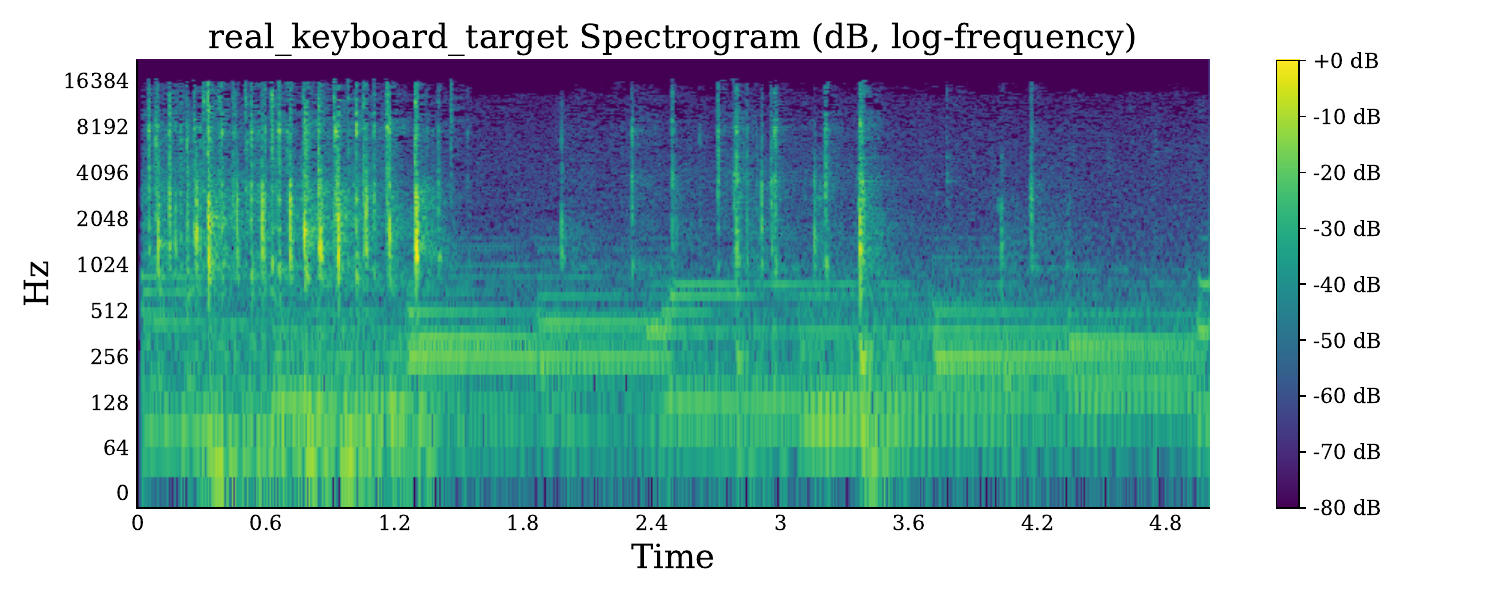}};
                    \node (img21)[below=of img11, yshift=1.0cm]{\includegraphics[trim={2.25cm 1.45cm 4.7cm 2.0cm}, clip, width=.45\linewidth]{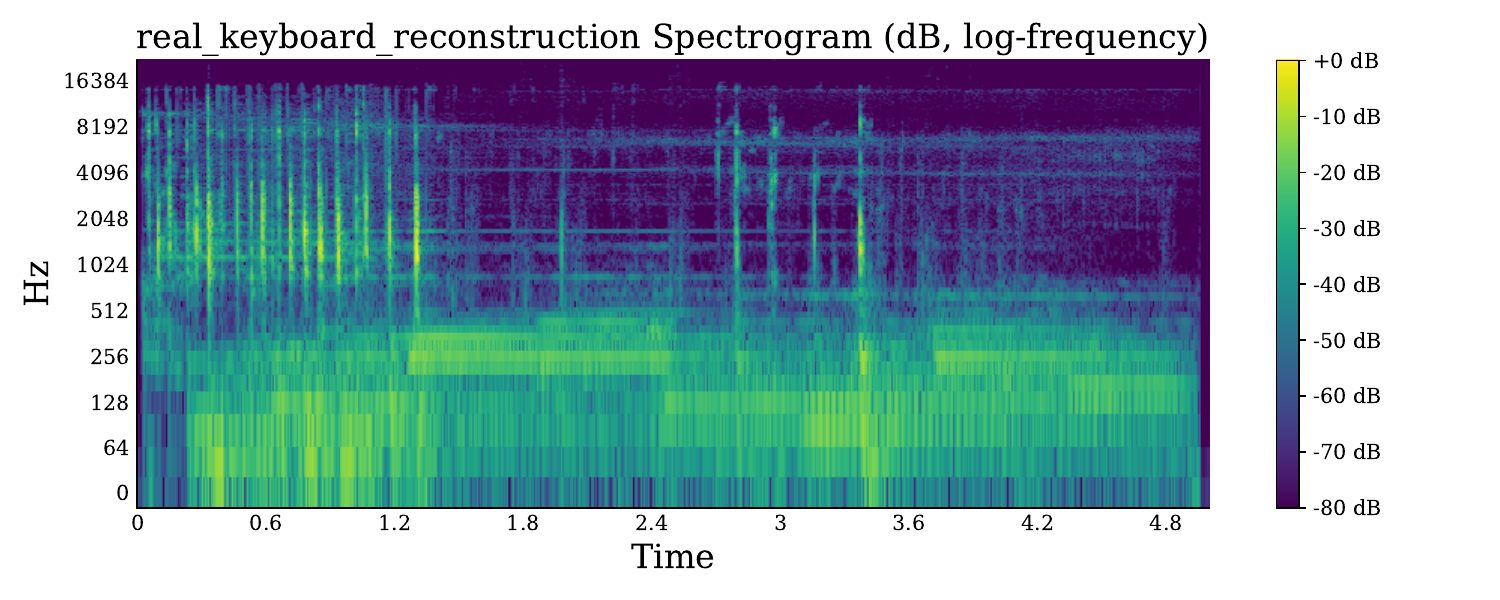}};
                    \node (img31)[below=of img21, yshift=1.0cm]{\includegraphics[trim={2.25cm 1.45cm 4.7cm 2.0cm}, clip, width=.45\linewidth]{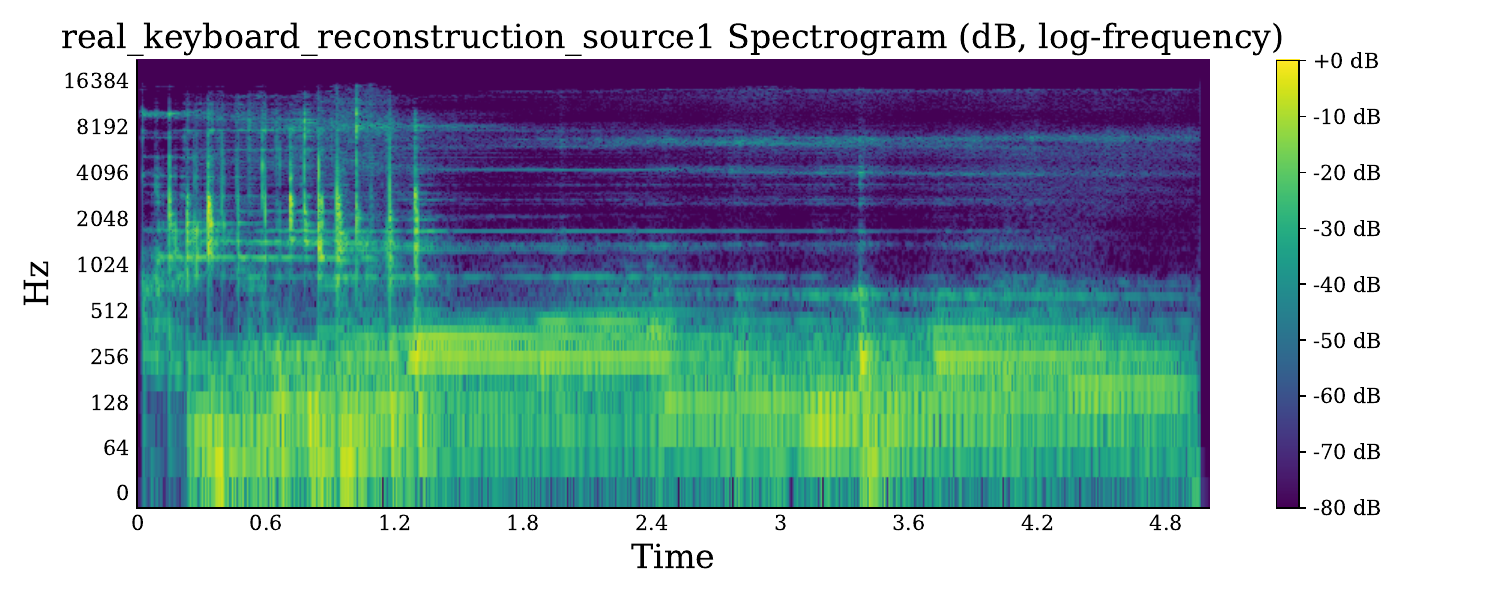}};
                    \node (img41)[below=of img31, yshift=1.0cm]{\includegraphics[trim={2.25cm 1.45cm 4.7cm 2.0cm}, clip, width=.45\linewidth]{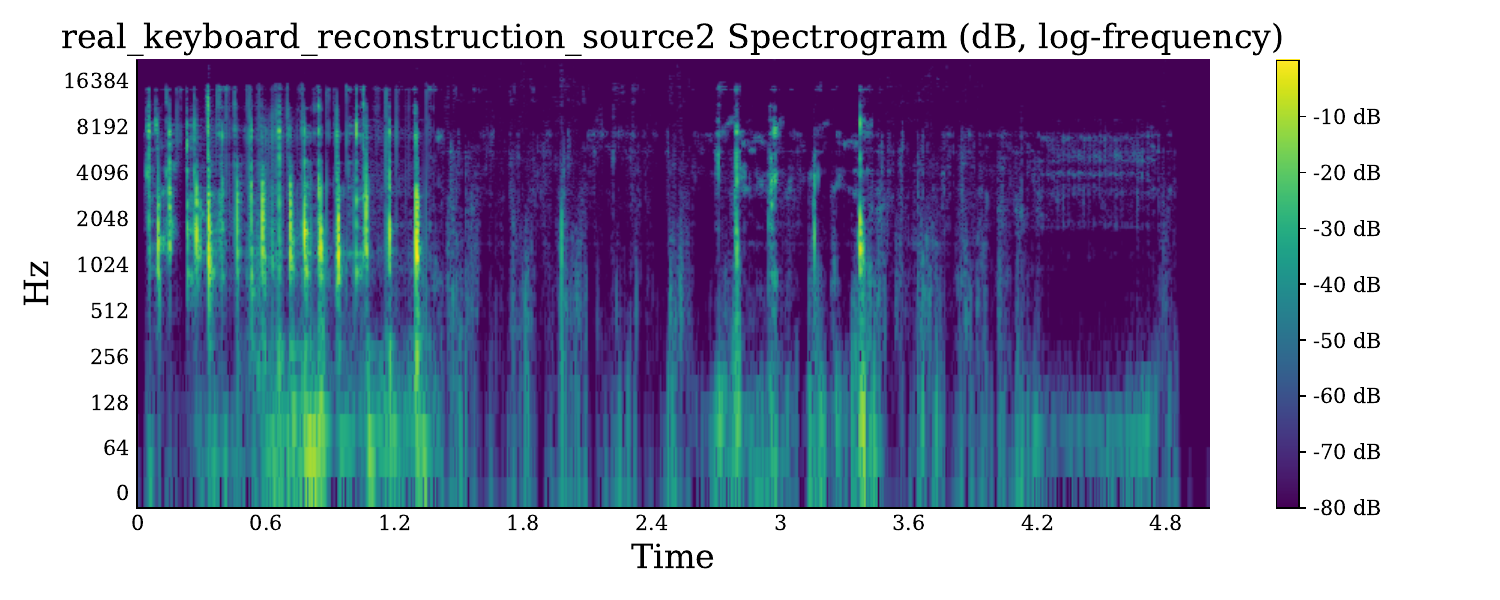}};

                    \node[right=of img11, node distance=0cm, xshift=-1cm, yshift=1.1cm, rotate=270, font=\color{black}]{\scriptsize{Target Audio $\targetAudio$}};
                    
                    \node[right=of img21, node distance=0cm, xshift=-1cm, yshift=1.2cm, rotate=270, font=\color{black}]{\scriptsize{Reconstructed Audio}};
                    \node[right=of img31, node distance=0cm, xshift=-1cm, yshift=1.5cm, rotate=270, font=\color{black}]{\scriptsize{Reconstructed ``\texttt{music}\dots''}};
                    \node[right=of img41, node distance=0cm, xshift=-1cm, yshift=1.6cm, rotate=270, font=\color{black}]{\scriptsize{Reconstructed ``\texttt{clicking}\dots''}};
    
                    \node[left=of img21, node distance=0cm, rotate=90, xshift=-.5cm, yshift=-.9cm,  font=\color{black}]{\normalsize{Frequency (Hz)}};
                    \node[below=of img41, node distance=0cm, xshift=0.5cm, yshift=1.15cm,  font=\color{black}]{\normalsize{Time (s)}};
                \end{tikzpicture}
                }
                \vspace{-0.02\textheight}
                \caption{
                    \textbf{Automated Prompt Generation for Source Separation with Real Audio}
                    We run the audio captioning model on the target audio $\targetAudio$, getting the caption ``\texttt{Someone is clicking on a keyboard and talking}''.
                    Providing this caption to the LLM in the template prompt from \secref{sec:app_experiment_details_prompt_selections_automated} generates a dialogue (also in \secref{sec:app_experiment_details_prompt_selections_automated}) which contains the prompt decomposition ``\texttt{clicking on a keyboard}'' and ``\texttt{music playing quietly with indiscernible talking}''. 
                    \emph{Takeaway:} After running our method, the reconstructed sources better fit their prompts qualitatively while closely reconstructing the target audio, all without requiring a person to specify the prompts.
                    Audio links:
                        \href{https://drive.google.com/file/d/1PKv-ok220wKjnKamaB6UlKlQPHP2Khdz/view?usp=sharing}{{\color{nvidiagreen}target audio $\targetAudio$}}, 
                        \href{https://drive.google.com/file/d/1BXrPg6sW3eaPYoQRxpxMq17adEDLNkGQ/view?usp=sharing}{{\color{nvidiagreen}reconstructed audio}}, 
                        \href{https://drive.google.com/file/d/1tPoFHY0R0UxHQnaZ7qRDrZI2R66blTTi/view?usp=sharing}{{\color{nvidiagreen}recovered ``\texttt{clicking}\dots''}}, 
                        \href{https://drive.google.com/file/d/10r9ViJM_1m1QTtQdSXvSa-1juhPk747h/view?usp=sharing}{{\color{nvidiagreen}recovered ``\texttt{music}\dots''}}.
                }\label{fig:automatic_real_source_separation}
                \vspace{-0.02\textheight}
            \end{figure}

        See \figref{fig:automatic_real_source_separation}.

    \subsubsection{Failure Cases}\label{sec:app_exp_failures}
        Despite the versatility of Audio-SDS, we observed several notable failure modes:

        \textbf{Out-of-distribution prompts:}
            Because the method relies on a pretrained text-to-audio diffusion model, prompts far outside the model’s training distribution often yield unrealistic or collapsed audio. For instance, highly specific or unusual prompts such as \emph{``a singing whale that transitions into a squeaky door''} may cause the output to degenerate into repetitive noise or faint chirping artifacts. In these scenarios, the diffusion prior struggles to provide coherent guidance, resulting in SDS updates failing to converge to a stable solution.
        
        \textbf{Overly complex or lengthy audio:}
            The latent diffusion model can miss fine temporal details when generating or separating longer clips (e.g., beyond $10$–$15$s) or highly layered soundscapes. This sometimes results in abrupt sonic transitions or partial ``silences'' in the final audio. Extending the receptive field or applying hierarchical approaches may mitigate such issues, but it remains a challenging area for future research.
        
        \textbf{Parametric-model mismatch:}
            Tasks such as FM synthesis or physically informed impact sounds require domain-appropriate parameter ranges. The resulting audio can become unstable if the learned updates push parameters beyond plausible physical limits (e.g., negative frequency or extreme decay). Similarly, a simple FM synthesizer struggles to realize prompts describing percussive hits on unconventional surfaces (\emph{``wooden pot with metallic ring''}) because those timbres lie outside the synthesizer’s expressivity.
        
        \textbf{High guidance scale instability:}
            Excessively large classifier-free guidance (CFG) scales $\guidanceScale$ sometimes force the optimizer into pathological local minima, introducing issues such as buzzes or clicks. Moderate guidance values (e.g., $5$--$25$) and partial denoising steps generally reduce these artifacts, but ideal hyperparameters can vary across prompts.
        
        Addressing these limitations likely involves improved prompt-to-parameter alignment, negative prompt guidance, or future checkpoints with broader text–audio coverage. More advanced scheduling schemes and domain-specific priors may also help stabilize SDS for out-of-domain prompts and extended audio durations.

    \subsection{Ablations}\label{sec:app_ablations}

        \subsubsection{Our Decoder-SDS vs.\ Classic Encoder-SDS:}\label{sec:app_ablations_decoder_sds}
\begin{figure}[t]
                \vspace{-0.01\textheight}
                \centering
                \scalebox{.94}{
                \begin{tikzpicture}
                \centering
                    \node (img11){\includegraphics[trim={2.25cm 1.45cm 4.7cm 2.0cm}, clip, width=.45\linewidth]{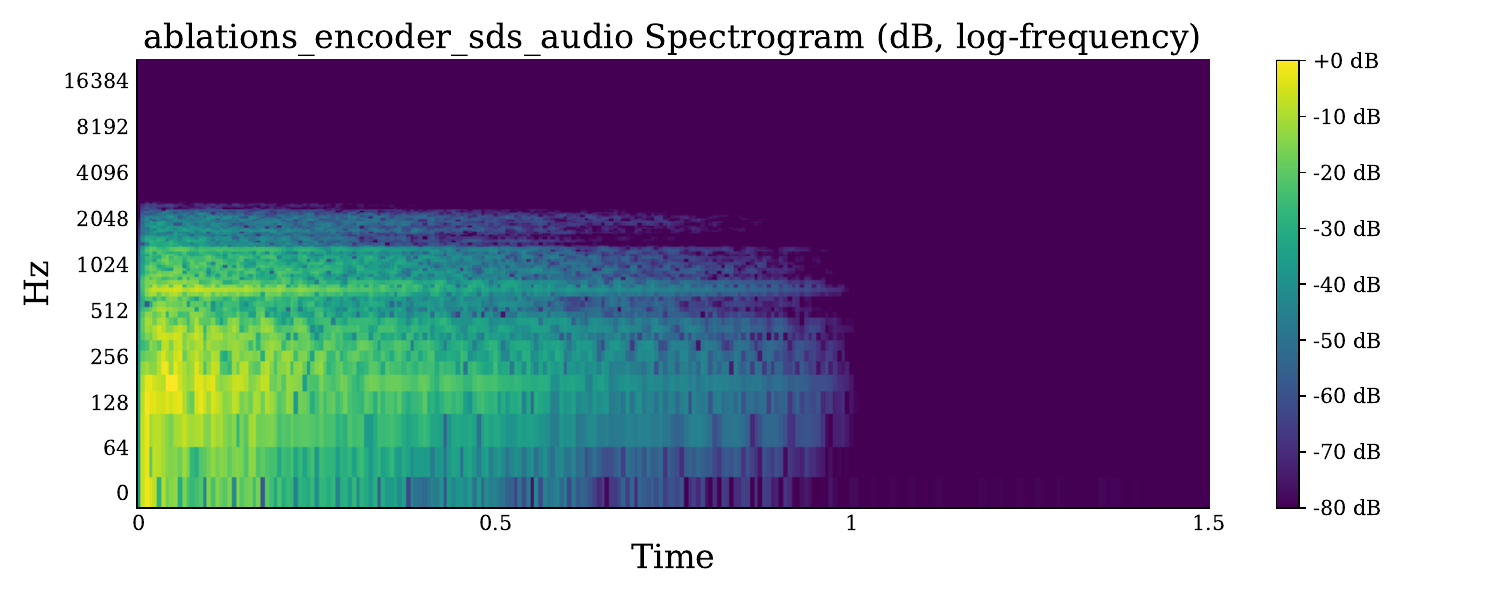}};
                    \node (img21)[below=of img11, yshift=1.0cm]{\includegraphics[trim={2.25cm 1.45cm 4.7cm 2.0cm}, clip, width=.45\linewidth]{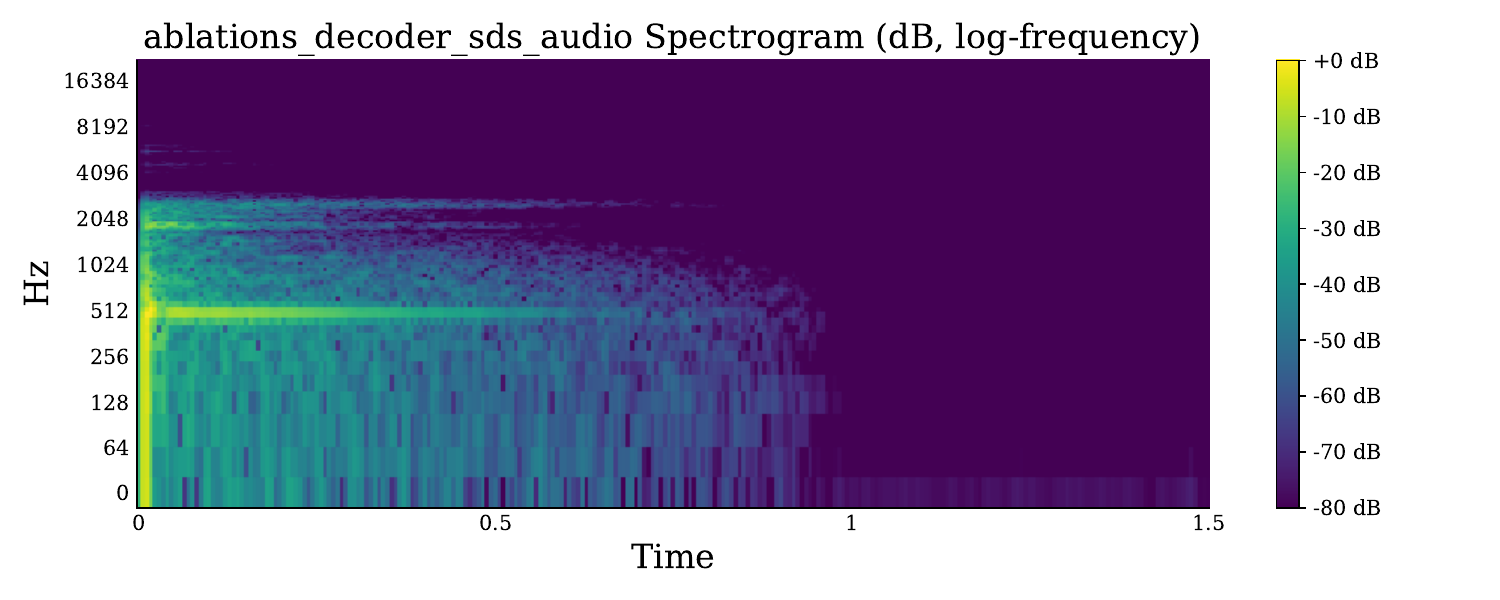}};

                    \node[right=of img11, node distance=0cm, xshift=-1cm, yshift=1.3cm, rotate=270, font=\color{black}]{\scriptsize{Encoder-SDS (Baseline)}};
                    
                    \node[right=of img21, node distance=0cm, xshift=-1cm, yshift=1.0cm, rotate=270, font=\color{black}]{\scriptsize{Decoder-SDS (Us)}};
    
                    \node[left=of img21, node distance=0cm, rotate=90, xshift=2.5cm, yshift=-.9cm,  font=\color{black}]{\normalsize{Frequency (Hz)}};
                    \node[below=of img21, node distance=0cm, xshift=0.5cm, yshift=1.15cm,  font=\color{black}]{\normalsize{Time (s)}};
                \end{tikzpicture}
                }
                \vspace{-0.02\textheight}
                \caption{
                    \textbf{Decoder-SDS vs. Encoder-SDS} 
                    Optimizing impact synthesis with prompt ``\texttt{hitting pot with wooden spoon}'', comparing our proposed Decoder-SDS variant and the na\"ive Encoder-SDS. 
                    \emph{Takeaway:} Decoder-SDS performs qualitatively and quantitatively better than the Encoder-SDS here.
                    Audio links:
                        \href{https://drive.google.com/file/d/1aeh8jCftYrJPC7zl0qF5ZOFw50m26_-v/view?usp=sharing}{{\color{nvidiagreen}Encoder-SDS}}, 
                        \href{https://drive.google.com/file/d/1IdHzzcwcg9oM-Wb1lPIkCzAgSQACyza1/view?usp=sharing}{{\color{nvidiagreen}Decoder-SDS}} ({\color{darkgreen} $+\num{0.15}$ CLAP} vs. Encoder-SDS).
                }\label{fig:ablation_decoder_sds}
                \vspace{-0.01\textheight}
            \end{figure}
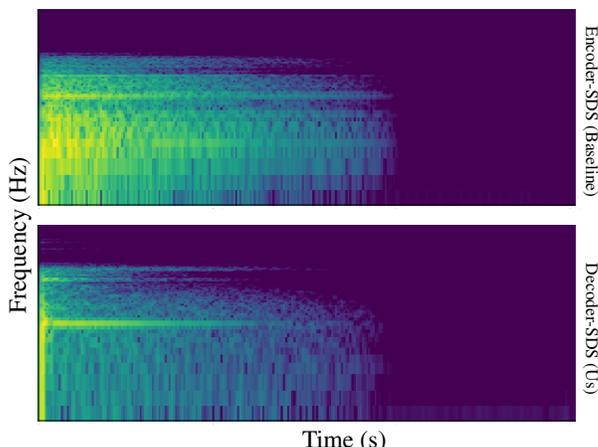

            See \figref{fig:ablation_decoder_sds}.
        
        \subsubsection{Spectrogram Emphasis vs.\ Pure Time-Domain:}\label{sec:app_ablations_spectrogram_emphasis} 
\begin{figure}[t]
                \vspace{-0.01\textheight}
                \centering
                \scalebox{.94}{
                \begin{tikzpicture}
                \centering
                    \node (img11){\includegraphics[trim={2.25cm 1.65cm 4.7cm 2.0cm}, clip, width=.45\linewidth]{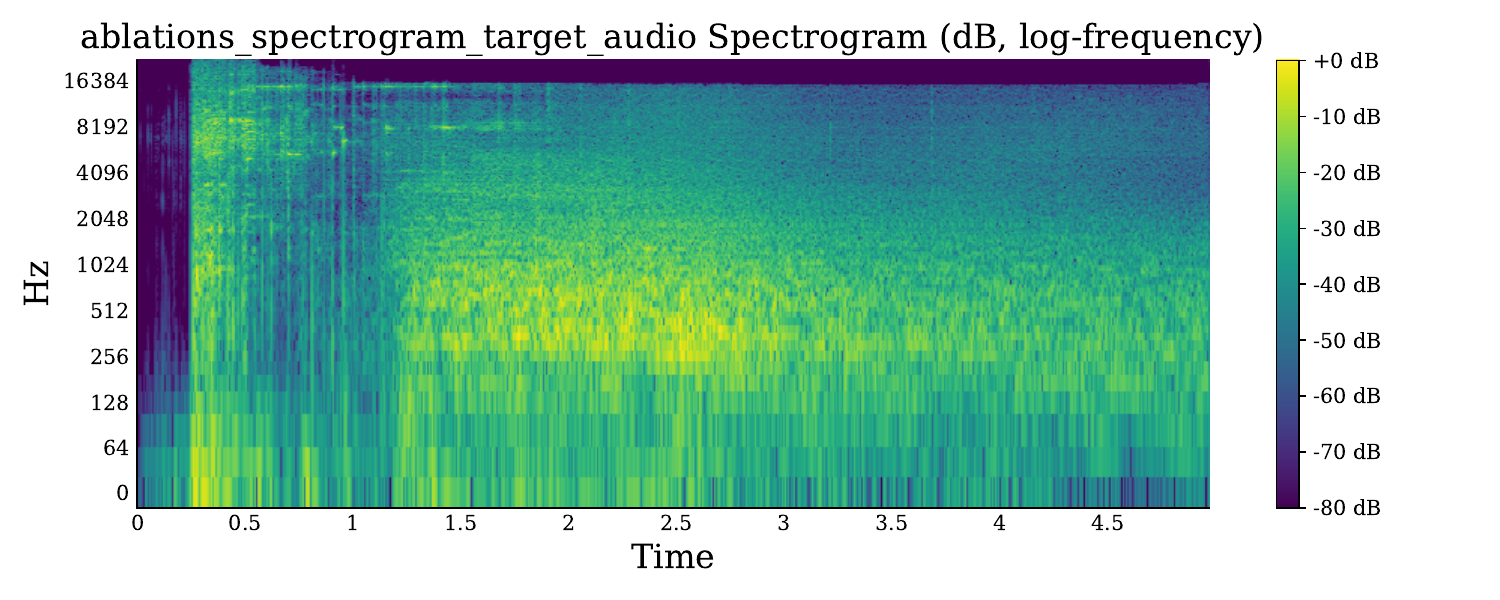}};
                    \node (img21)[below=of img11, yshift=1.0cm]{\includegraphics[trim={2.25cm 1.65cm 4.7cm 2.0cm}, clip, width=.45\linewidth]{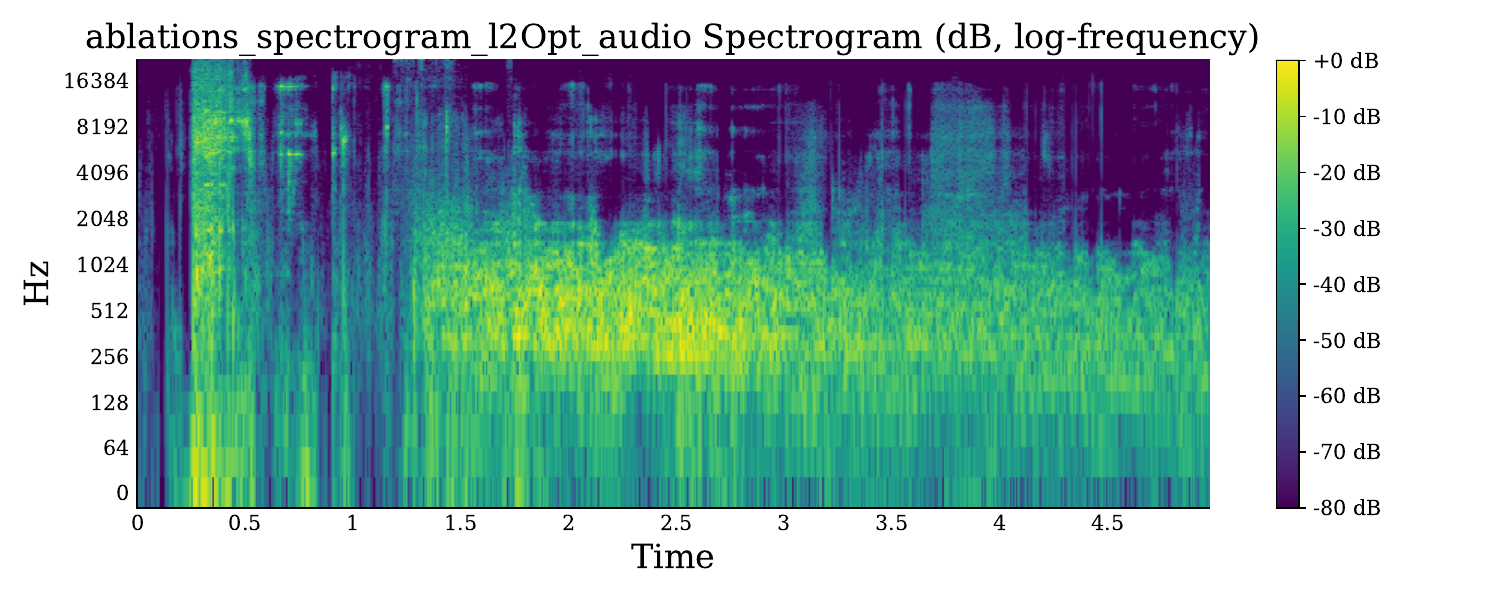}};
                    \node (img31)[below=of img21, yshift=1.0cm]{\includegraphics[trim={2.25cm 1.65cm 4.7cm 2.0cm}, clip, width=.45\linewidth]{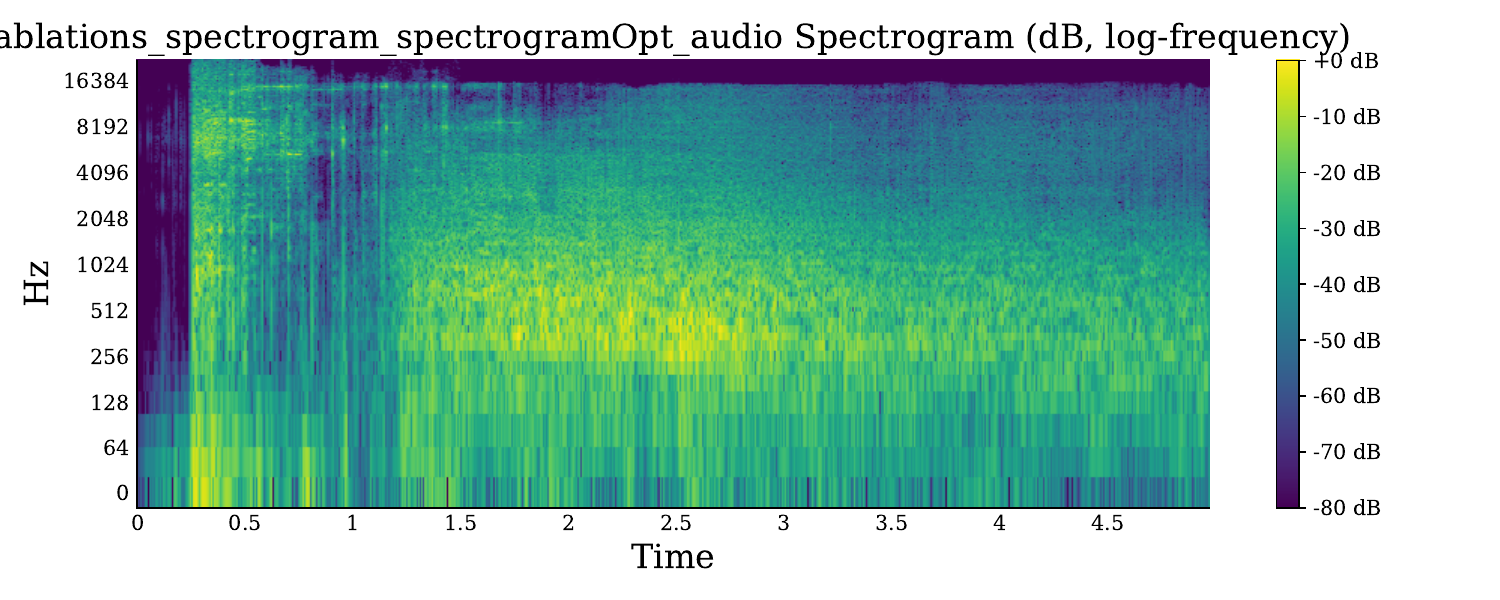}};

                    \node (img12)[right=of img11, xshift=-1cm]{\includegraphics[trim={2.25cm 1.65cm 4.7cm 2.0cm}, clip, width=.45\linewidth]{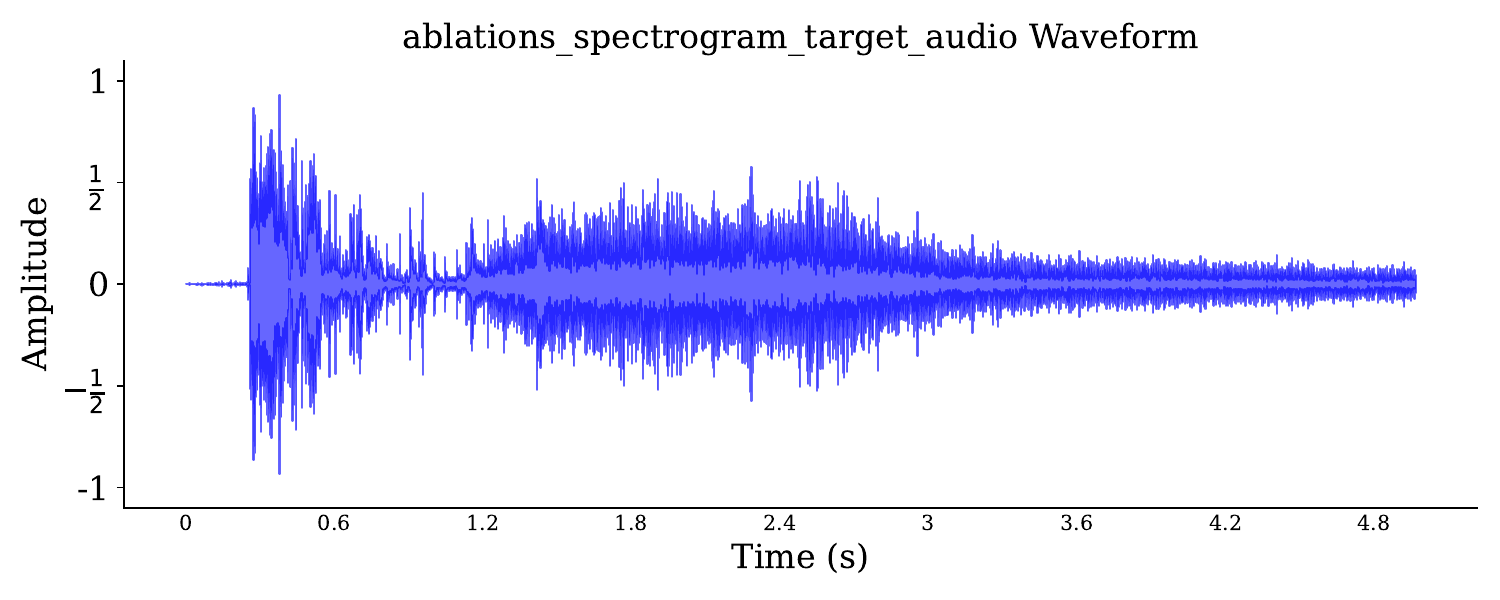}};
                    \node (img22)[below=of img12, yshift=1.0cm]{\includegraphics[trim={2.25cm 1.65cm 4.7cm 2.0cm}, clip, width=.45\linewidth]{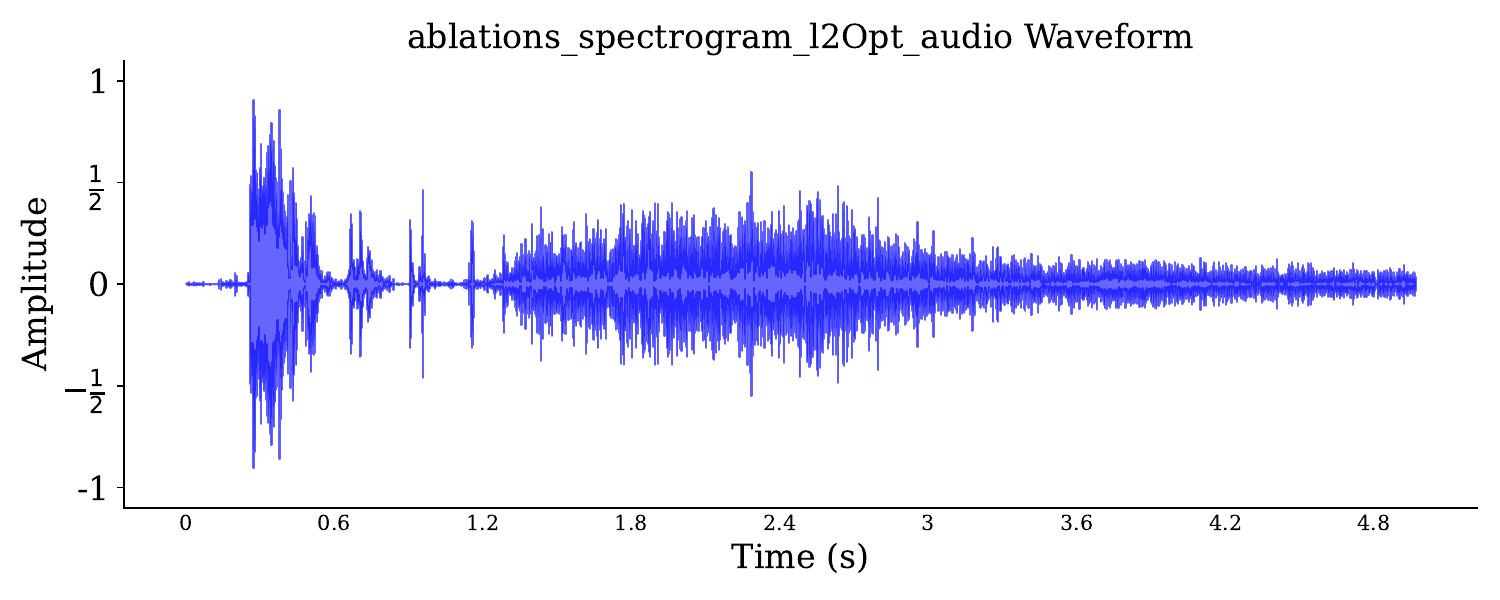}};
                    \node (img32)[below=of img22, yshift=1.0cm]{\includegraphics[trim={2.25cm 1.65cm 4.7cm 2.0cm}, clip, width=.45\linewidth]{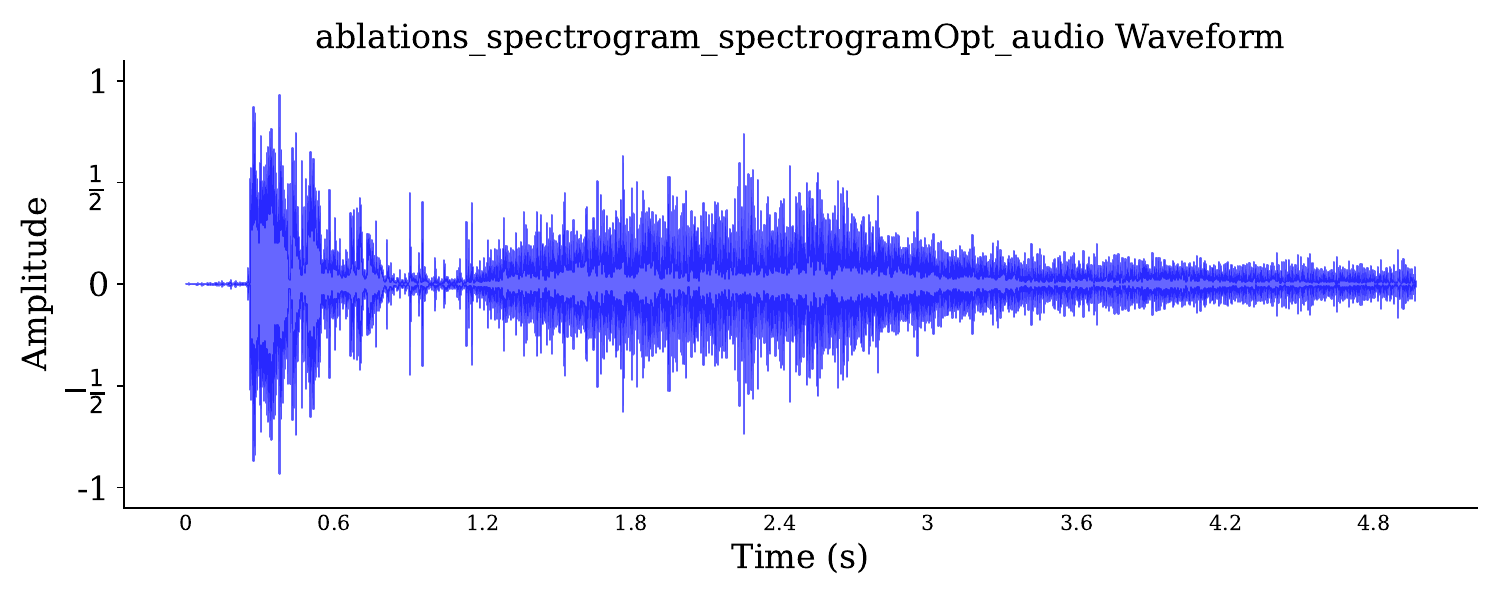}};

                    \node[right=of img12, node distance=0cm, xshift=-.75cm, yshift=.5cm, rotate=270, font=\color{black}]{\scriptsize{Target}};
                    \node[right=of img22, node distance=0cm, xshift=-.75cm, yshift=1.35cm, rotate=270, font=\color{black}]{\scriptsize{$\ell_2$ Emphasis (Baseline)}};
                    \node[right=of img32, node distance=0cm, xshift=-.75cm, yshift=1.5cm, rotate=270, font=\color{black}]{\scriptsize{Spectrogram Emphasis (Us)}};
    
                    \node[above=of img11, node distance=0cm, xshift=0.1cm, yshift=-1.15cm,  font=\color{black}]{\normalsize{Spectrogram}};
                    \node[above=of img12, node distance=0cm, xshift=0.1cm, yshift=-1.15cm,  font=\color{black}]{\normalsize{Waveform}};
                \end{tikzpicture}
                }
                \vspace{-0.015\textheight}
                \caption{
                    \textbf{Spectrogram vs. $\ell_2$ Emphasis:} We visualize spectrograms (in dB) and waveforms of a target audio, and our reconstruction result using a spectrogram and $\ell_2$ emphasis.
                    \emph{Takeaway:} The spectrogram emphasis is qualitatively better than the $\ell_2$ emphasis.
                    Audio links:
                        \href{https://drive.google.com/file/d/1lJmsVYKirawa82pNtTPmK-PLW41KNA1C/view?usp=sharing}{{\color{nvidiagreen}target}}, 
                        \href{https://drive.google.com/file/d/1rVo0RyMB3dcKJx83LAlGVp2CYToOazq6/view?usp=sharing}{{\color{nvidiagreen}$\ell_2$ emphasis}},
                        \href{https://drive.google.com/file/d/1h15o81-qZTAO2SB-9DOjt-hWfleiEtu0/view?usp=sharing}{{\color{nvidiagreen}spectrogram emphasis}}.
                }\label{fig:ablation_spectrogram_emphasis}
            \end{figure}

            See \figref{fig:ablation_spectrogram_emphasis}.

        \subsubsection{Multistep Denoising vs.\ Single-Step:}\label{sec:app_ablations_multistep_denoising}
\begin{figure}[t]
                \vspace{-0.01\textheight}
                \centering
                \scalebox{.94}{
                \begin{tikzpicture}
                \centering
                    \node (img11){\includegraphics[trim={2.25cm 1.45cm 4.7cm 2.0cm}, clip, width=.95\linewidth]{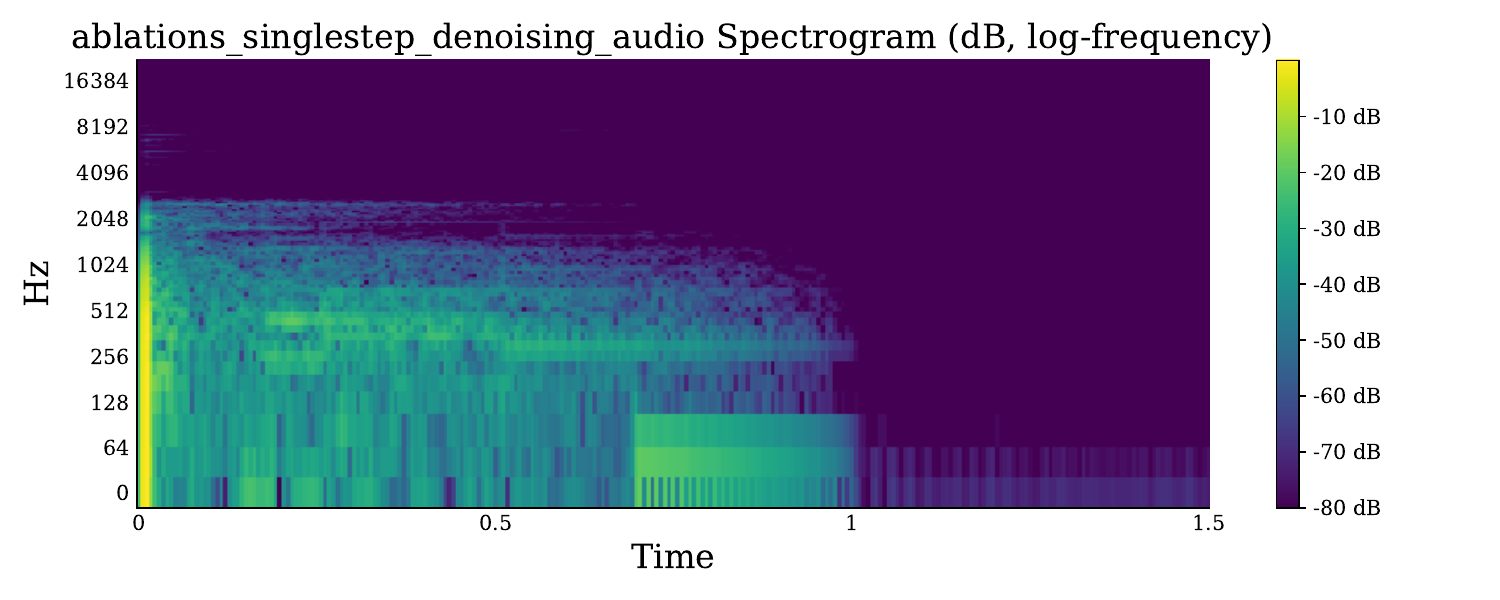}};
                    \node (img21)[below=of img11, yshift=1.0cm]{\includegraphics[trim={2.25cm 1.45cm 4.7cm 2.0cm}, clip, width=.95\linewidth]{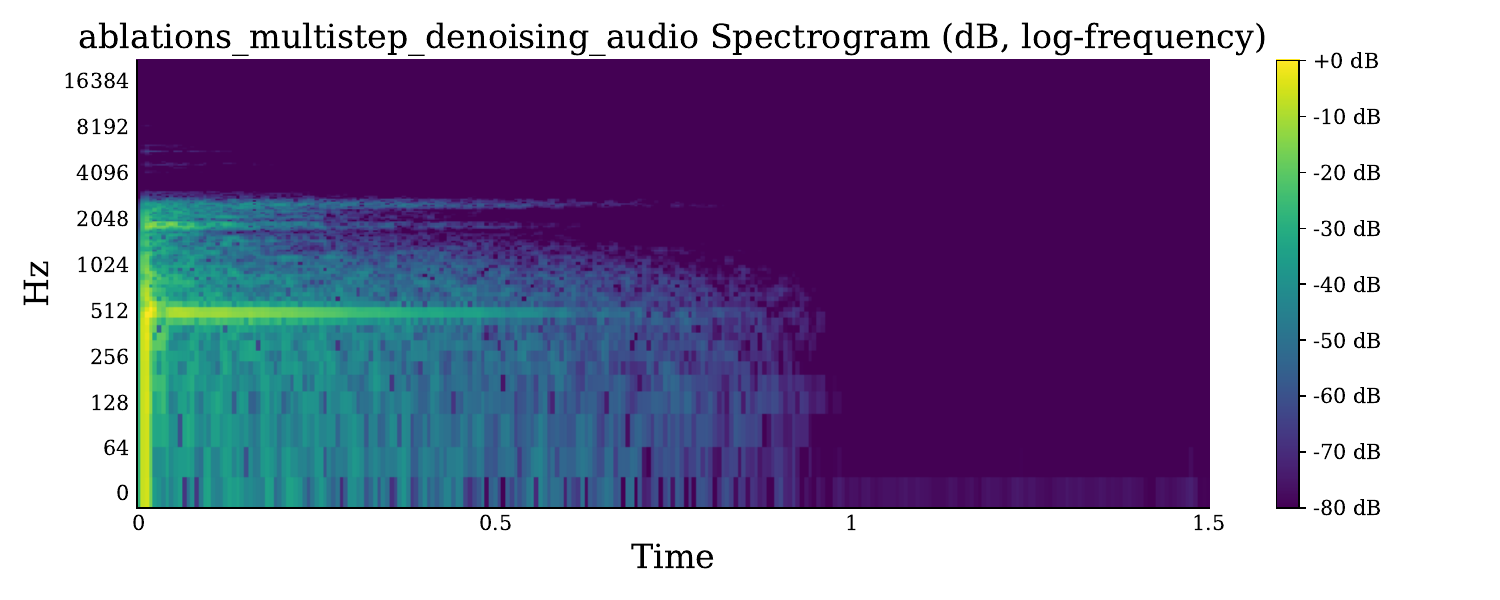}};
    
                    \node[right=of img11, node distance=0cm, xshift=-1.0cm, yshift=1.45cm, rotate=270, font=\color{black}]{Single-step (Baseline)};
                    
                    \node[right=of img21, node distance=0cm, xshift=-1.0cm, yshift=1.2cm, rotate=270, font=\color{black}]{Multi-step (Us)};
    
                    \node[left=of img21, node distance=0cm, rotate=90, xshift=4.25cm, yshift=-.75cm,  font=\color{black}]{Frequency (Hz)};
                    \node[below=of img21, node distance=0cm, xshift=0.5cm, yshift=1.15cm,  font=\color{black}]{Time (s)};
                \end{tikzpicture}
                }
                \caption{
                    \textbf{Single-Step vs. Multi-Step Denoising} We visualize spectrograms in dB of the audio initialization and the final result after optimization for our proposed multi-step denoising and the standard single-step denoising for impact synthesis with the prompt ``\texttt{hitting pot with wooden spoon}''.
                    \emph{Takeaway:} The Decoder-SDS performs qualitatively and quantitatively better than the Encoder-SDS here.
                    Audio links:
                        \href{https://drive.google.com/file/d/1P0gky4rymR27w_qad2nM1XS-0Q2nmu6j/view?usp=sharing}{{\color{nvidiagreen}single-step}}, \href{https://drive.google.com/file/d/1LsiWHlfy7RPG6wPJBLWsQSL3P1DjuVAY/view?usp=sharing}{{\color{nvidiagreen}\num{10}-step}} ({\color{darkgreen} $+\num{0.14}$ CLAP} vs. single-step).
                }\label{fig:ablation_multistep_denoising}
            \end{figure}

            See \figref{fig:ablation_multistep_denoising}.

\section{Extended Methodological Details}

    \subsection{Automating Source Decomposition with LLMs and Audio-Captioning models}\label{app:sec_method_automating_source_decomposition}
        The preceding formulation introduces the set of $\totalNumSources$ prompts as a tool for the user to use, which can be straightforward to select for simple and short mixtures. Yet, for more complex audio or users with less expertise, we seek to automatically generate a set of different potential relevant source decompositions for the user. We mitigate this by proposing a strategy to automate this using an audio captioning model on the audio $\targetAudio$ to get a textual description, which is given to an LLM that suggests potential sets of prompts given the caption.
        The preceding formulation introduces the set of $\totalNumSources$ prompts as a tool for the user to use, which can be straightforward to select for simple and short mixtures. Yet, for more complex audio or users with less expertise, we seek to automatically generate a set of different potential relevant source decompositions for the user. For this, we experiment with:
        \begin{enumerate}[leftmargin=*, nosep]
            \item \textbf{Captioning the audio mixture} using an audio-captioning model (e.g., AudioCaps~\citep{kim2019audiocaps} or WavCaps~\citep{mei2024wavcaps}) to obtain a broad textual description of the mixture’s contents, such as ``\texttt{A saxophone is played on a road where cars are driving past and people walking about.}''
            \item \textbf{Splitting the caption into $\totalNumSources$ prompts} automatically by prompting a large language model (LLM) -- e.g., LLaMA~\citep{dubey2024llama}, GPT-4~\citep{chatgpt2025}, or Claude~\citep{claude2025} --  with the caption and asking it to propose various $\totalNumSources$ sources, each with a distinct semantic label and potential prompt variants. For example, outputting potential decompositions like for $\totalNumSources=2$, ``\texttt{cars driving; music}'' and for $\totalNumSources=3$, ``\texttt{traffic; saxophone; people}''.
            \item \textbf{Optional branching} by re-captioning the resulting audio sources and providing this as input to the LLM to consider if it wants to rebranch, change the prompt, terminate, or otherwise.
        \end{enumerate}

\section{Extended Discussion}
    \subsection{Future Work}\label{sec:future_work}
        \textbf{Deeper Source Separation Investigations:}
            Although our prompt-driven source separation demonstrates promising results, many open directions remain. Future work may combine parametric (e.g., physical or FM synthesis) and latent representations to handle sources varying in their degrees of realism or interpretability. Our source separation formulation can be viewed as a nested optimization problem~\citep{lorraine2024scalable}, motivating exploration of relevant techniques~\citep{gidel2019negative, lorraine2021lyapunov, lorraine2022complex, lorraine2022task, bae2024training} and phenomena~\citep{vicol2022implicit, raghu2021meta}. Integrating negative prompting to exclude specific unwanted components (e.g., background hum) could bolster separation quality. We can look at hierarchical source decompositions. Another intriguing angle is iterative or hierarchical separation, where an initial decomposition is further refined or subdivided based on user feedback or automatic captioning and LLMs.
            One could also incorporate a default ``noise'' channel in a \(\totalNumSources\!+\!1\)-source decomposition to both enable denoising (with a prompt like ``noise'') and serve as a catchall for leftover content not described by other prompts. However, this approach may yield degenerate solutions if the noise channel absorbs too much of the meaningful signal.

        \textbf{Other Audio Tasks for SDS:}
            Extending SDS to dynamic audio scenes represents a promising direction. Instead of modeling sounds as static clips, one could parametrize the positions of mobile sound sources (e.g., cars in motion, footsteps) and iteratively align them with text-based descriptions of movement or location. View-dependent prompting -- borrowed from text-to-3D -- could similarly enhance spatial audio generation, specifying an audio source's relative location to the listener (e.g., a car positioned far away or close up). Another avenue is learning room impulse responses (RIRs) directly from prompts (e.g., ``<PROMPT> \texttt{with a cathedral-like reverb}''), whether using simpler convolution filters or sophisticated methods like neural acoustic fields \citep{luo2022learning} to produce more realistic, location-aware audio suitable for interactive sound design in virtual or augmented reality settings.
            Tasks that leverage the stereo output of the audio diffusion model -- e.g., for spatial tasks -- may prove fruitful.
            Extensions of our impact synthesis experiments to optimize rendering strategies using actual geometry and physical parameters, such as the Young's Modulus, may prove useful for highly-stiff objects where video and image priors fail.
            
        \textbf{Customizing SDS for Audio:}
            While our SDS adaptation yields promising audio results, there is room for improvement. Approaches such as negative prompting, multistep sampling schedules, and other regularization strategies from the image domain may enhance fidelity for challenging prompts (e.g., extremely rare instruments or complex audio scenes). Tighter integration of classifier-free guidance and prompt-specific scheduling is promising, potentially enhancing convergence speed and resolving transient or high-frequency details.
            We expect enhanced results from reduced-variance SDS updates~\citep{lorraine2025training}, enhanced hyperparameter optimization~\citep{jaderberg2017population, lorraine2018stochastic, lorraine2019optimizing, mackay2019self, zhang2023using, mehta2024fmsARXIV} as the training was brittle, alternative parameterizations~\citep{luo2022learning, lorraine2024jacnet, adam2019understanding, limgraph}, or distilled diffusion models~\citep{song2023consistency, song2024multi, sabour2024align} for multi-step update benefits at lower costs.

        \textbf{Extending SDS to Other Output and Conditioning Modalities:}
            SDS originated in text-to-image and text-to-3D pipelines, then extended to video and audio, naturally raising the question of audio+visual fusion. A powerful next step could combine SDS with joint audio-video diffusion models to co-optimize a time-varying 4D representation that looks and sounds right (e.g., learning a material’s appearance and resonance in tandem). Beyond audiovisual tasks, SDS-guided parameter updates could benefit other diffusion-based domains, such as molecular graph generation, physics simulations, or robotic trajectory planning. As diffusion models expand into increasingly specialized modalities and condition types, SDS provides a unifying framework for aligning flexible parametric representations with rich, pretrained priors.

\end{document}